\date{\today}

\documentclass[aps,pra,twocolumn,superscriptaddress,groupedaddress]{revtex4-1}
\usepackage{dcolumn,graphicx,amsmath,amssymb,algorithm,algpseudocode}
\usepackage{todonotes}
\usepackage{subcaption}
\usepackage{fancyvrb}
\usepackage{qcircuit}
\usepackage{placeins}
\usepackage[colorlinks=true,linkcolor=blue,citecolor=blue,urlcolor=blue]{hyperref}
\usepackage[capitalise]{cleveref}
\usepackage{orcidlink}
\usepackage[version=3]{mhchem} 
\usepackage{booktabs}
\usepackage{rotating}

\newcommand{\br}{\mathbf{r}}
\newcommand{\bx}{\mathbf{x}}

\newcommand{\mnhf}{\ce{MnNO\bond{...}HF}}
\newcommand{\mnoh}{\ce{MnNO\bond{...}HOH}}
\newcommand{\mnnh}{\ce{MnNO\bond{...}HNH2}}
\newcommand{\mnch}{\ce{MnNO\bond{...}HCH3}}
\begin{document}
\title{Interaction Energies on Noisy Intermediate-Scale Quantum Computers}
\author{Matthias Loipersberger~\orcidlink{0000-0002-3648-0101}}
 \thanks{These two authors contributed equally to this work.}
\affiliation{QC Ware Corporation, Palo Alto, CA, 94301, USA}
\author{Fionn D.~Malone~\orcidlink{0000-0001-9239-0162}}
\thanks{These two authors contributed equally to this work.}
 \altaffiliation[Present address: ]{Google Quantum AI, Mountain View, CA, 94043}
\affiliation{QC Ware Corporation, Palo Alto, CA, 94301, USA}
\author{Alicia R.~Welden~\orcidlink{0000-0002-2238-9825}}
\affiliation{QC Ware Corporation, Palo Alto, CA, 94301, USA}
\author{Robert M.~Parrish~\orcidlink{0000-0002-2406-4741}}
\email{rob.parrish@qcware.com}
\affiliation{QC Ware Corporation, Palo Alto, CA, 94301, USA}
\author{Thomas Fox~\orcidlink{0000-0002-1054-4701}}
\affiliation{Medicinal Chemistry, Boehringer Ingelheim Pharma GmbH \& Co. KG, Birkendorfer Stra{\ss}e 65, 88397 Biberach an der Ri\ss, Germany}
\author{Matthias Degroote~\orcidlink{0000-0002-8850-7708}}
\author{Elica ~Kyoseva~\orcidlink{0000-0002-9154-0293}}
\author{Nikolaj Moll~\orcidlink{0000-0001-5645-4667}}
\email{nikolaj.moll@boehringer-ingelheim.com}
\author{Raffaele Santagati~\orcidlink{0000-0001-9645-0580}}
\author{Michael Streif~\orcidlink{0000-0002-7509-4748}}
\affiliation{Quantum Lab, Boehringer Ingelheim, 55218 Ingelheim am Rhein, Germany}

\begin{abstract}
The computation of interaction energies on noisy intermediate-scale quantum (NISQ) computers appears to be challenging with straightforward application of existing quantum algorithms. 
For example, use of the standard supermolecular method with the variational quantum eigensolver (VQE) would require extremely precise resolution of the total energies of the fragments to provide for accurate subtraction to the interaction energy. 
Here we present a symmetry-adapted perturbation theory (SAPT) method that may provide interaction energies with high quantum resource efficiency. 
Of particular note, we present a quantum extended random-phase approximation (ERPA) treatment of the SAPT second-order induction and dispersion terms, including exchange counterparts. Together with previous work on first-order terms, this provides a recipe for complete SAPT(VQE) interaction energies up to second order. 
The SAPT interaction energy terms are computed as first-level observables with no subtraction of monomer energies invoked, and the only quantum observations needed are the the VQE one- and two-particle density matrices.
We find empirically that SAPT(VQE) can provide accurate interaction energies even with coarsely optimized, low circuit depth wavefunctions from the quantum computer, simulated through ideal statevectors. The errors on the total interaction energy are orders of magnitude lower than the corresponding VQE total energy errors of the monomer wavefunctions.
\begin{description}
\item[Keywords]
Quantum Computing, Intermolecular Interactions, Quantum Chemistry, Drug Discovery, SAPT
\end{description}
\end{abstract}

\maketitle

\section{Introduction}\label{sec:intro}

Quantum computing has emerged as a promising platform to approach classically challenging problems in chemistry~\cite{Cao2018,bauer2020quantum}. The most interesting near-term application is the simulation of strongly-correlated systems for which the electronic structure cannot be described with a single Slater determinant. For such systems Kohn-Sham density functional theory~\cite{hohenberg1964inhomogeneous,kohn1965self} (KS-DFT) may fail to describe the electronic structure correctly; popular examples are the Fe-S clusters or the FeMo-cofactor~\cite{Reiher2017,li2019electronic}. 

Classically, a proper treatment of these strongly correlated systems is achieved with multi-reference methods where the naïve combinatorial scaling of the wavefunction ansatz limits its applications. Note however that much progress has been made on classical heuristics for wavefunction methods that exhibit less than combinatorial scaling and that may be highly accurate for a broad range of problems~\cite{white1999ab,chan2011density,luchow2011quantum,garniron2018selected,levine2020casscf,holmes2016heat,sharma2017semistochastic}.
Alternatively, quantum algorithms~\cite{Aspuru-Guzik2005,McArdleRev2020,Cao2019,bauer2020quantum} might be used to solve the Schr\"odinger equation with a resource cost that scales polynomially with the number of qubits. Unfortunately, the currently available noisy intermediate-scale quantum (NISQ) hardware~\cite{Preskill2018} suffers from relatively poor gate fidelity and low qubit count~\cite{cohn2021quantum} which poses two key challenges. First, it is important for NISQ-tailored  quantum algorithms~\cite{Bharti2022} to minimize quantum resources. The most prominent NISQ methods are hybrid quantum-classical algorithms like the variational quantum eigensolver (VQE)~\cite{Peruzzo2014,McCleanVQE2016}, quantum Krylov methods~\cite{parrish2019quantum,Stair2020,huggins2020non,cohn2021quantum,klymko2022real} or the fermionic quantum Monte Carlo method~\cite{huggins2022unbiasing}. The second challenge is to find specific applications that could harness quantum computing~\cite{elfving2020will,Liu2022}.
Many application studies in chemistry use either reduced model systems or molecules with a simple electronic structure~\cite{kirsopp2021quantum,rice2021quantum,kim2022fault,greene2022modelling}. There are several promising application areas for quantum chemistry in computer aided drug design~\cite{HeifetzCADD2020} namely, exploring potential energy surfaces~\cite{o2021efficient}, simulating metalloenzymes~\cite{goings2022reliably} and computing protein-ligand interaction energies~\cite{malone2022towards}, the last of which we consider in this work.

The computation of non-covalent interaction energies is a routine task in classical quantum chemistry~\cite{rezac2016benchmark,kodrycka2019platinum} and the standard procedure is the supermolecular approach: the interaction energy is calculated as the difference between the ground state energies of the dimer and two monomers separated to infinity~\cite{BoysCP1970}. 
However, transferring this approach to a NISQ type quantum computer is difficult for several reasons: First, the VQE total energies (on the order of thousands of kcal/mol) need to be tightly converged to resolve interaction energies on the order a few kcal/mol with the supermolecular approach. This is a disadvantage on NISQ hardware because the total energy expectation value is obtained statistically and high-precision expectation values require a high number of measurements. Furthermore, converging the VQE total energy to high accuracy requires deep circuits associated with a large sets of parameters, where it becomes increasingly difficult to reach the global minimum on the parameter surface~\cite{McClean2018}.
Second, accounting for the basis set superposition error (BSSE) in the supermolecular approach~\cite{BoysCP1970} is commonly achieved by expanding the basis in the monomer calculations to the size of the dimer basis.  This unnecessarily increases the qubit count requirements for the individual monomers and can potentially lead to convergence issues for VQE ~\cite{Peruzzo2014,McCleanVQE2016}.

To this end, this work provides an alternative pathway towards interaction energies with high accuracy and low quantum resource requirements by using symmetry-adapted perturbation theory (SAPT)~\cite{Jeziorski1976,Jeziorski1994,patkowski2020recent} in combination with VQE monomer wave functions [SAPT(VQE)]. This approach directly computes the interaction energy as a sum of small expectation values; in contrast, the supermolecular approach computes the small interaction energy (several kcal/mol) as a difference of large total energies (thousands of kcal/mol). This work builds on our previous work~\cite{malone2022towards} where we presented the implementation of the first-order SAPT(VQE) terms of electrostatics and exchange. However, the first-order SAPT(VQE) terms alone are not capable of computing accurate interaction energies - standard levels of SAPT also require the computation of the second-order induction and dispersion terms \cite{Jeziorski1994}. Up to now, the absence of a SAPT(VQE) recipe for the complete second-order SAPT terms has been a major potential weakness of the approach - indeed, other authors\cite{kirsopp2021quantum} have noted that ``[first-order SAPT(VQE)] computed interaction energies did not reproduce ligand rankings yielded by more accurate 2nd order SAPT calculations,
due to the missing induction and dispersion components in
the 1st order approximation,'' and conclude that ``[the first-order SAPT(VQE)] workflow is limited
by truncation of the SAPT expansion at 1st order and it is
not clear how their method can be effectively extended to
higher orders.''  

In this work, we ameliorate this crucial deficiency by direct implementation of the second-order induction and dispersion terms, together with their exchange counterparts. Our approach follows the SAPT(FCI) approach of Korona~\cite{korona1997convergence} but with the naively-exponential-scaling FCI piece replaced by an active-space VQE wavefunction which is intended to be implemented on a forthcoming NISQ computer (in this work we use ideal statevector simulators for the numerical tests). To implement the second-order terms, we follow the extended random phase approximation (ERPA) formalism for SAPT(CASSCF) (complete active space self-consistent field (CASSCF)) developed by Hapka \textit{et al.}~\cite{hapka2021symmetry}, with VQE standing in for the FCI solver in CASSCF. In the approach, the response equations for the coupled polarization propagators are carried out in a truncated hole-particle basis reminiscent of the Casida expansion~\cite{casida1998molecular} in TD-DFT or the quantum subspace expansion (QSE) for VQE excited states~\cite{mcclean2017hybrid}. While this treatment necessarily does not include all Hamiltonian states even in the FCI limit, it does use a set of coupled hole-particle states that span the full range of energetic scales of the Hilbert space, and which are empirically known to provide highly-converged results for the induction and dispersion energies. Notably, the use of the ERPA formalism in SAPT(VQE) allows for the computation of the second-order induction and dispersion terms, together with exchange counterparts (here in the $S^2$ approximation) with the active-space one- and two-particle density matrices of the VQE ground state wavefunctions appearing as the only ``new'' quantum observables. These observables are polynomial scaling and typically readily available as a byproduct of the VQE optimization procedure. The subsequent ERPA equations, induction and dispersion contractions, and exchange counterparts are polynomial scaling classical operations. They are significantly more complicated than the first-order terms - roughly 200 equations are needed to describe the implementation (see supplemental material for full details), and the naive CuPy implementation of the equations implemented here is restricted to smaller systems than our previous paper due to classical postprocessing overhead (though optimizations such as hole/active/particle separations and density fitting might significantly reduce this overhead). Complexity notwithstanding, the approach presented provides a recipe for a SAPT(VQE) doppelganger of SAPT(FCI) complete through all second order terms.  SAPT methods complete through second order  are well known to produce interaction energies with high accuracy even with modest basis sets~\cite{korona2013coupled,parker2014levels}. 

Below, we first lay out the motivation and high-level theory for the ERPA treatment of the second-order SAPT(VQE) approach. The ERPA equations, contractions to induction and dispersion terms, and exchange counterparts are straightforward but extremely verbose, so much of their explicit presentation is deferred to the supplemental material. We then demonstrate the numerical performance of all four SAPT(VQE) terms and total interaction energies for several small multireference dimers and a model heme-nitrosyl hydrogen bonding complex, using classical ideal statevector simulators to emulate the VQE.
\section{Theory}\label{sec:theory}
The most direct ``supermolecular'' route to the interaction energy of two monomers $A$ and $B$ is to simply compute the total energy of the combined system $E_{AB}$ and subtract the total energies of its non-interacting constituents $E_{A} + E_{B}$, i.e., 
\begin{equation}
E_{\mathrm{int}} = E_{AB} - E_{A} - E_{B},
\end{equation}
ideally using the exact wave function for each system. 

An alternative approach to computing intermolecular interaction energies is symmetry adapted perturbation theory (SAPT), which obtains the interaction energy with a different approach.
In particular, we can write the Hamiltonian of the combined system as
\begin{equation}
\hat{H} = \hat{H}_A + \hat{H}_B + \hat{V},
\end{equation}
where we assume $\hat{H}_X\lvert\Psi_X\rangle = E_X \lvert \Psi_X \rangle$, where $\lvert\Psi_X\rangle$ is the ground state wavefunction of monomer $X$ and $\hat{V}$ contains only the Coulombic interactions between monomer $A$ and $B$. 
With this partitioning of the Hamiltonian we can build a perturbation theory for the intermolecular interaction energy directly, thus avoiding computing potentially very large total energies.
More explicitly we have
\begin{equation}
E_{\mathrm{int}} = \sum_n (E_{\mathrm{pol}}^{(n)} + E_\mathrm{exch}^{(n)}),
\end{equation}
where $E_{\mathrm{pol}}^{(n)}$ and $E_{\mathrm{exch}}^{(n)}$ are $n$th-order polarization and exchange energies respectively.
The combination of both first and second order terms yields the following expression for the interaction energy (see SI sections~1 to~3 for a detailed derivation):
\begin{equation}
E_{\mathrm{int}} \approx E_\mathrm{SAPT} = E_{\mathrm{elst}} + E_{\mathrm{exch}} + E_{\mathrm{ind,u}} +  E_{\mathrm{disp}}
\end{equation}
where $E_{\mathrm{elst}}$ denotes the electrostatic contribution to the total interaction energy energy, $E_{\mathrm{exch}}$ the exchange energy, $E_{\mathrm{ind,u}}$ the induction energy and  $E_{\mathrm{disp}}$ the dispersion energy.
If we use high quality monomer wavefunctions approaching full configuration interaction (FCI), we conceptually approch the SAPT(FCI) method of Korona and co-workers. In Korona's original work, the computation of the second-order induction and dispersion terms is performed by a direct response property treatment of the static (induction) and frequency dependent dynamic susceptibility tensors (dispersion), followed by contractions of these property tensors to form the polarization and exchange SAPT contributions.\cite{korona1997convergence,KoronaExchange2008,Korona2008,korona2008dispersion,korona2009exchdisp}
In this work, we instead employ the extended random phase approximation (ERPA)~\cite{ChatterjeeERPA2012} as pioneered within SAPT(CASSCF) by Hapka~\cite{hapka2018second,hapka2019second,hapka2021casscf,hapka2021symmetry} to avoid computing excited state properties on the quantum computer as they often require a significant measurement overhead on NISQ-era quantum computers ~\cite{parrish2019quantum,Parrish2019transitions,CaiMolecularResponse2020,OllitraultEOM2020}. We note that the use of ERPA as a proxy for explicit response properties of VQE resembles the quantum subspace expansion (QSE) method~\cite{mcclean2017hybrid}, wherein a basis of single and double excitations out of a VQE reference is used to provide an excited state ansatz that is truncated in character (but not in excitation energy).
Further details of SAPT and the ERPA procedure are provided in the supporting information (SI sections~1--3). We note that SAPT interaction energies using this second-order truncation of SAPT are typically highly accurate, even with Hartree-Fock wavefunctions (for the case of single-reference systems), and well established to produce accurate interaction energies in many common cases~\cite{parker2014levels}. Third- and higher-order extensions are likely possible along similar response property or ERPA lines as used here, but are typically found to not improve the SAPT interaction energy significantly past the second-order level.

The current NISQ-era hardware is limited to tens of qubits (spin-orbitals); therefore an active space formalism is necessary to describe realistic chemical systems. In the active space approach, we partition the one-electron orbital set into
core orbitals, active orbitals and virtual orbitals. Ideally, the active orbitals contain the orbitals required to describe the entangled electrons properly. The active space of the wave function is then is calculated on a quantum computer (see SI section 1.2 for more information).
In the SAPT(VQE) approach, one or both of the monomer active space wavefunctions are generated by VQE-type quantum circuits:
\begin{equation}
\vert \Psi_{\mathrm{VQE}}\rangle \equiv\hat U_{\mathrm{VQE}}\vert\Phi_I\rangle
\end{equation}

where $\vert\Phi_I\rangle$ is some initial state (typically the Hartree--Fock determinant). From these wavefunctions we obtain single-particle and two-particle reduced density matrices that go into the computation of the SAPT interaction energy.
In this work we use a modified version of the unitary cluster Jastrow wavefunction ansatz ~\citep{Matsuzawa2020} (VQE)  which takes the form
\begin{equation}
    \vert \Psi_{\mathrm{VQE}} \rangle
=
\prod_{k}
\exp(-\hat K^{(k)})
\exp(\hat T^{(k)})
\exp(+\hat K^{(k)})
\vert \Phi_{I} \rangle,
\label{eq:ansatz}
\end{equation}
where $\hat{K}^{(k)}$ and $\hat{T}^{(k)}$ are one- and two-body operators, and $k$ is a parameter that controls the depth of the circuit and as a result its variational freedom.
We use a slightly modified $k$-uCJ ansatz from Ref.~\citenum{Matsuzawa2020}, which we denote as $k$-muCJ for clarity, with the `m' standing for modified (see SI section~1.3 for more details). The SAPT(VQE) workflow is outlined at a high level in Fig.~\ref{fig:workflow}.
We note that the SAPT(VQE) method as formulated within the ERPA picture is independent of the quantum algorithm  used to determine the density matrices and thus can likely be readily adopted to any quantum algorithm of choice.

\section{Results}\label{sec:results}
In a first step, we test SAPT(VQE) with two classic intermolecular interaction motifs (water dimer and t-shaped benzene dimer). In a second step, we apply SAPT(VQE) to a heme-nitrosyl model complex; these systems are highly relevant in both biological~\cite{maia2014biology} and pharmaceutical~\cite{moncada1991nitric} chemistry. The SAPT(VQE) results presented in this section are the result of ideal statevector VQE simulations (see SI section~4 for for more details). 
In all examples, we benchmark the accuracy of the VQE/SAPT(VQE) results by comparing to classical SAPT(CAS-CI) energies using the same orbitals and active space (see Figs.~\ref{fig:water_results}--\ref{fig:mn_results}). The complete active space configuration interaction (CAS-CI) wavefunction represents the exact wavefunction within the active space approximation and thus, SAPT(CAS-CI) results represent the best possible interaction energy within the SAPT approximation but not the exact interaction energy (see Fig.~S6 for a comparison of VQE and CAS-CI wave functions). We note that this comparison is only possible for small active space sizes due to the combinatorial scaling of the CAS-CI wavefunction ansatz.

\subsection{Multi-Reference Benchmark Systems}
The chemistry of non-covalent interactions governs a wide range of  interaction motifs such as hydrogen bonds or dispersion bound systems. However, the electronic structure of these simple systems is often well described by classical single reference methods. Therefore, we modified two of the classic systems in our previous study~\cite{malone2022towards}, namely the water dimer and the t-shaped benzene dimer to make the electronic structure strongly correlated and thus challenging  to compute accurately for conventional single reference methods.

The first test case is a hydrogen bonding motif: the stretched water dimer complex, which is depicted in Fig.~\ref{fig:water_results}~(a). The two partially broken single bonds make this system strongly correlated and require a multi reference treatment to accurately describe  the electronic structure. We included all eight valence electrons of the stretched monomer and eight spatial orbitals (8e, 8o) in the active space (for a detailed procedure on how the active orbitals were selected for the CASSCF calculations see SI section~1). The CASSCF natural orbital occupation numbers (NOON, see Fig.~S2) exhibit deviations from integer values, which is an indicator of strong correlation. Consequently, the single reference RHF method fails to describe this system as apparent by the large deviation of the absolute energy of the monomer (see Fig.~\ref{fig:water_results}~(b) black dotted line). This system also provides a challenge for the quantum algorithm as the $k$-muCJ ansatz needs a repetition factor of $k > 10$ to converge the absolute energy to the stretched monomer below 1~kcal/mol.

In sharp contrast to this, the errors of the total interaction energy as well as the errors of each individual SAPT energy term are multiple orders of magnitude lower than the absolute VQE errors (see Fig.~\ref{fig:water_results}~(b)). In fact, the very shallow $k = 1$ circuit is accurate enough to provide interaction energies in comparison to the SAPT(CAS-CI) results. In a next step, we probed the bond dissociation of the water complex (along the intermolecular \ce{O-H} bond labeled $r$ in Fig.~\ref{fig:water_results}~(a)). We find a similar behavior: the errors in the interaction energies and each energy component are below 1~kcal/mol for all intermolecular distances and several orders of magnitude lower than the error in absolute energies. For $k = 4$, each interaction energy is below the 1~kcal/mol threshold; for $k = 1$ some errors are slightly larger at small intermolecular distances.

The second test case is a dispersion bound complex: the T-shaped benzene p-benzyne dimer, depicted in Fig.~\ref{fig:bz_results}~(a).
The p-benzyne monomer  has a biradical ground state, which is difficult to describe with classical single reference methods~\cite{crawford2001problematic}. The key findings are identical to the previous test case.  Thus, these findings hold for very different types of intermolecular interaction motifs, intermolecular distances, different active spaces and different type difficult electronic structures.

\subsection{Hydrogen Bonding to Heme-Nitrosyl Model Complexes}
As an application example, we study hydrogen bonding to a Manganese nitrosyl complex. Nitric oxide (NO) is a small molecule with important biological implications such as signal transduction~\cite{bruckdorfer2005basics, moncada1991nitric} or as a key intermediate in the nitrogen cycle~\cite{maia2014biology, lehnert2018reversing}. At the center of these processes are metalloporpyhrins~\cite{kadish2000porphyrin, lehnert2013structure, hunt2015heme}, where NO binds to the metal center as a nitrosyl ligand~\cite{santolini2011molecular, walker2005nitric}.

In order to understand and control these biological processes, the chemistry around the \ce{Metal-NO} bonds must be elucidated in terms of electronic structure and reactivity~\cite{hayton2002coordination} as illustrated by theoretical~\cite{radon2008binding}, experimental~\cite{lehnert2021biologically} and medicinal~\cite{tfouni2010tailoring,hickok2010nitric,serafim2012nitric} work. Unfortunately, the metal-NO bond in nitrosyl complexes poses a challenge for many quantum chemistry methods due to the redox active nature of NO ligand~\cite{ampssler2020not}. There are three possible oxidation states for the NO moiety: NO$^-$, NO$^\bullet$ and NO$^+$, which is illustrated for a generic \ce{M(II)-NO} complex in Figure~\ref{fig:mnno_scheme}. In many cases, the bond is best described in terms of a superposition of these states. This strong correlation makes this a challenging system for many single reference methods such as DFT~\cite{boguslawski2011can}. This results in a large variety of recommended functionals depending on the specific nitrosyl complex studied~\cite{radon2008binding,siegbahn2010significant,goodrich2013trans,lehnert2013structure}.

In this work, we study the hydrogen bonding to a heme-model manganese-nitrosyl complex. This model can serve as a proxy of how a metal-heme bound NO interacts with  a protein environment as depicted in Fig.~\ref{fig:mnno_scheme}. The metal-coordinating  cyano and ammonia ligands resemble a porphyrin coordination environment in terms of ligand field, total charge and $\pi$-acidity (see Fig.~\ref{fig:mnno_scheme}). As hydrogen bond donors we chose \ce{HF}, \ce{H2O}, \ce{NH3} and \ce{CH4}, which cover a wide range of donor strengths similarly to active sites in proteins. The resulting four hydrogen complexes are depicted in  Fig.~\ref{fig:mnno_scheme} and are abbreviated as \mnhf, \mnoh, \mnnh\ and \mnch. 

As a first step, we analyzed the electronic structure of  \ce{[Mn(CN)2(NH3)3NO]^0} and found the system to be strongly correlated (see SI section~5 for a more detailed discussion). We used a (6e, 6o) active space for subsequent CASSCF and VQE calculations to include all 6 electrons of the \ce{Mn-NO} bond. The key six active  orbitals are centered around the Mn-NO moiety and are very similar to the active orbitals in real heme nitrosyl complexes~\cite{radon2010electronic} (see Fig.~S2 and S5). The VQE as the $k$-uCJ ansatz required up to $k = 8$ layers to converge to the CASSCF energy within 1~kcal/mol (the variations are caused by the additional ghost-basis functions from the different hydrogen bonding donors). In contrast, the SAPT interaction energy is already significantly below that threshold even for $k = 1$. It is noteworthy that this finding holds true for the wide range of interaction energies in this series ($-$0.3 to $-$9.3~kcal/mol, see Fig.~\ref{fig:mn_results}~(a)). Thus, we confirm the core finding of this work for different interaction motifs, a wide range of interaction energies, different strongly correlated electronic structure systems and both equilibrium and non-equilibrium bond distances.

In addition to the interaction energy, SAPT provides a decomposition into physical meaningful terms helping to unravel the origin of the interaction. Fig.~\ref{fig:mn_dft_results}~(b) plots each component of the SAPT(VQE) ($k=4$) calculation of the series of hydrogen bonded complexes plus the stretched water dimer as a reference of a typical hydrogen bond (see see SI section 5.1, Table~S1 and Fig.~S7 for more details including bond distances). The main driving force for binding is the electrostatic term as expected for hydrogen bonds. 
The strongest contrast between the water dimer and the \ce{Mn-NO} hydrogen bonds is observed in the exchange energy and is the main driving force for the difference in interaction energies. This may be rationalized by difference in the diffuseness of the lone pairs: the bound NO becomes (partly) \ce{NO^+}, which makes the lone pair  more compact in space than the lone pair in the water dimer,  thus resulting in less exchange repulsion.

At last, we compare the SAPT(VQE)  interaction energy to DFT based supermolecular (BSSE corrected~\cite{boys1970calculation}) interaction energies, the standard approach on classical hardware. The exact comparison is difficult as there is no standard procedure to obtain accurate interaction energies for strongly correlated systems. In addition, we use a small basis set [due to technical limits in the current \textsc{CuPy} classical implementation of SAPT(VQE)]. However, nitrosyl complexes are an example of the non-universality problem of approximate density functionals as the hydrogen bonding moiety and the nitrosyl moiety prefer different approximate density functionals~\cite{radon2008binding, radon2010electronic, mardirossian2014omegab97x} and thus reliable predictions are only possible with careful system specific benchmarking when experimental data is available~\cite{lehnert2013structure}. In contrast, SAPT is expected to robustly give accurate results for hydrogen bonds given proper monomer wavefunctions, e.g.\ via a quantum algorithm in SAPT(VQE). To illustrate this point, Fig.~\ref{fig:mn_dft_results} plots the interaction energies of  SAPT(VQE) , SAPT(CAS-CI) and several popular DFT functionals (through the supermolecular approach). We included many popular functionals as well as several top performing functionals for non-covalent interactions~\cite{mardirossian2017thirty}. We see in Fig.~\ref{fig:mn_dft_results} that the SAPT(VQE) ($k=4$) is almost identical to the SAPT(CAS-CI) in all four cases. The DFT functionals exhibit a significant spread for each complex. The B97-D functional predicted the  smallest interaction energy in all four cases, but the highest interaction energy is predicted in each case by a different functional. Furthermore, we see the relative ordering of the functionals change for each system (color sequence in each plot). This illustrates the non-universality problem for approximate exchange correlation functionals even for very similar nitrosyl complexes (this also holds true for larger basis set as illustrated in Fig.~S8). Note that the SAPT(CAS-CI) results are the reference for the SAPT(VQE)  calculations and do \emph{not} represent the true interaction energy, thus, only the SAPT(VQE)  and not the DFT interaction energies should be compared against this reference.

To demonstrate that the erroneous behavior of DFT is related with the system studied here, we calculate nitric acid hydrogen complexes with \ce{HF}, \ce{H2O} (see Fig.~S9). These hydrogen complexes are the main group analogues of the nitrosyl complex where we replace the \ce{Mn-NO} with a \ce{H-NO} bond. 
This results in a much simpler electronic structure without strong correlation where we can generate reference energies using coupled cluster wave function methods. We find that many DFT functionals perform within 1~kcal/mol accuracy. Interestingly, we see that the relative ordering of the functionals changes notably from the MnNO to the HNO systems. Furthermore, we note that SAPT predicts the interaction energies with $<$0.5~kcal/mol error in both cases using the optimal basis set (see Fig.~S9~(a) and (b); see SI~S5.2 for details on the reference energies). We can expect SAPT(VQE) ans{\"a}tze to yield similarly accurate results for strongly correlated examples as presented above when the optimal basis is used. Therefore, the SAPT formalism, presented in this work, is able to provide accurate interaction energies both for simple and difficult electronic structures, while the accuracy of DFT deteriorates for the latter.

\section{Conclusion}\label{sec:conc}

With the developments of the present manuscript, we have what we believe represents a minimally complete path to accurate determination of intermolecular interaction energies on a NISQ-type computer. Our previous study~\cite{malone2022towards} established the theoretical framework of SAPT(VQE) but was limited to only first order terms of electrostatics and exchange for a proof-of-concept demonstration. This work obviates this limitation by including the second order terms of induction and dispersion, including their exchange counterparts, which results in a level of SAPT well-established to produce accurate interaction energies with chemical accuracy~\cite{parker2014levels}. In this hybrid quantum-classical procedure, we obtain the monomer wavefunctions on the quantum computer via the VQE algorithm and measure the one- and two-particle reduced density matrices of the monomers (simulated through ideal statevectors in the present work). On the classical computer we compute the first and second order SAPT contributions based on the reduced density matrices from VQE calculation. The direct computation of excited states for the second order terms is avoided via an extended random phase approximation (ERPA) formalism.

We find empirically that SAPT(VQE) can provide accurate interaction energies even with coarsely optimized, low circuit depth wavefunctions from the quantum computer. The resulting errors of first and second order contributions, in addition to the the total interaction energies, are orders of magnitude lower than the corresponding VQE total energies of the monomer wavefunctions.
Our empirical findings are based on the application of the SAPT(VQE) method to several systems with strongly correlated electronic structures: two classic intermolecular interaction motifs  and several hydrogen bonding complexes of a heme-nitrosyl model complex, a class of biological highly relevant metalloenzymes where classic quantum chemistry methods such as DFT struggle to obtain accurate interaction energies. Thus, this works paves the way to obtain accurate interaction energies on a NISQ-era quantum computer with few quantum resources. It is a first step to alleviate one of the major challenges in quantum chemistry where in-depth knowledge of both the method and system is required \textit{a priori} to reliably generate accurate interaction energies.

While a basic path to NISQ-type computations of interaction energies is now reasonably clear, much remains to improve the details of the operational concept. One basic direction that needs improvement is the classical acceleration of the ERPA and SAPT terms - our na\"ive \textsc{CuPy} code for this consideration was severely limited in system size due to a non-optimal treatment of core/active/virtual simpliciation and a lack of density fitting of the response functions.~\cite{korona2008dispersion} In another instance, it may be that the ERPA formalism for the second order terms could be improved by direct treatment of the response of the monomers to external perturbations, including coupled response of the quantum and classical monomer wavefunction parameters. It might also be the case that more-advanced treatment of the required observables along the lines of double factorization~\cite{motta2021low,huggins2021efficient} could significantly reduce the number of required measurements when the time comes to deploy this method in the presence of shot noise. Another highly interesting question is how the present approach might map (if at all) to the fault tolerant regime where statistical evaluation of expectation values is definitionally prohibitive. Finally, beyond the concept of interaction energies, this work represents our general hypothesis of the level of specialization needed to converge various important chemical observables - we believe that other properties such as gradients, polarizabilities, spectroscopies, etc may require similar quantum adaption of rather verbose classical methods like SAPT to provide good convergence of observables on NISQ devices.

\begin{figure*}
    \centering
    \includegraphics[width=1.0\textwidth]{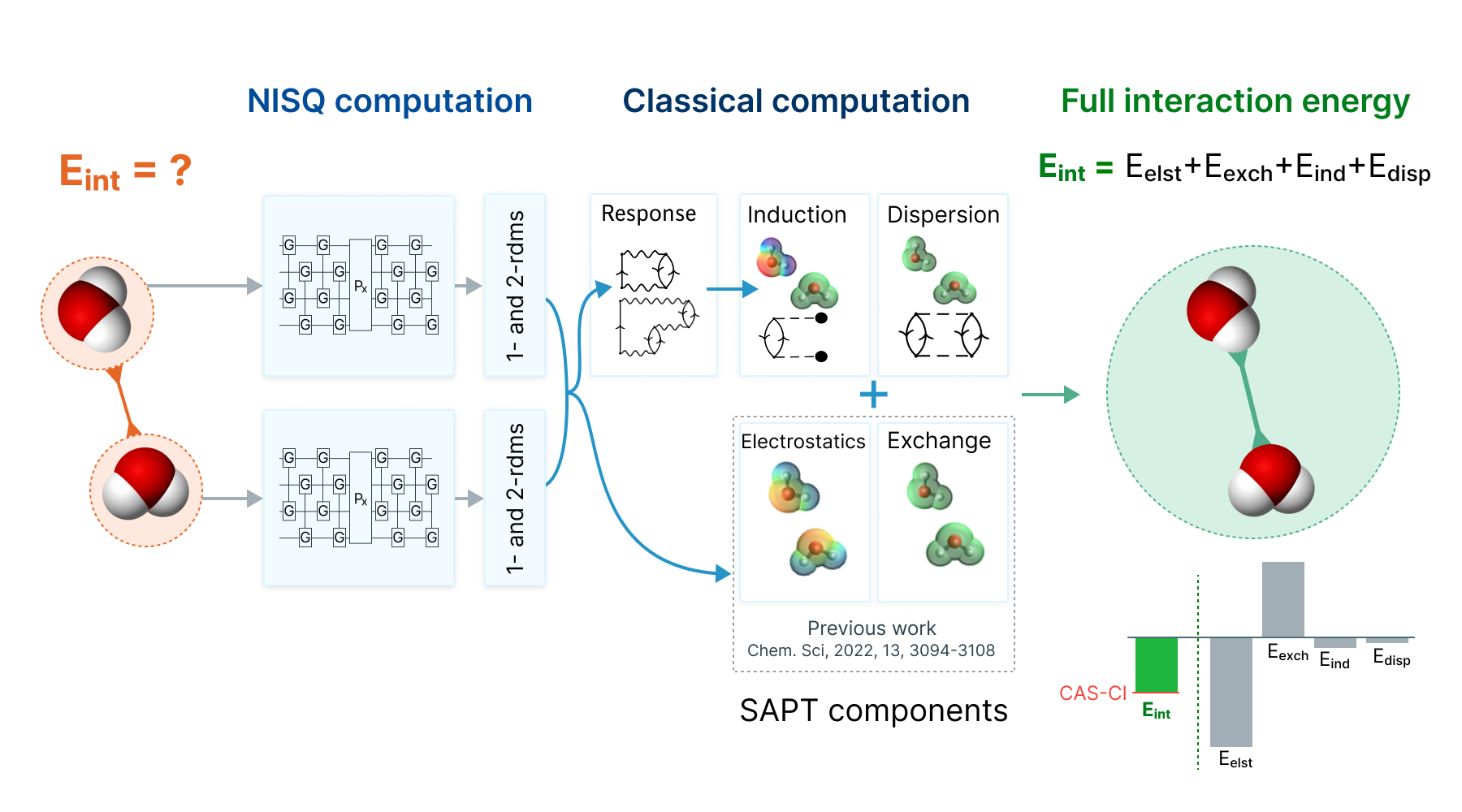}
    \caption{The workflow of SAPT(VQE) to obtain accurate interaction energies using a NISQ quantum algorithm:
    first, we define two monomers and the interaction of interest (exemplary here: two water molecules); second, we define the active space of the electrons which are simulated on the quantum computer; third, the quantum computer is used to compute the electronic structure via a quantum algorithm such as VQE. The converged wavefunction yields the reduced one- and two-particle density matrices (1- and 2-RDM); fourth, the classical computer computes the interaction energy as a post processing step using the RDMs via a sum of electrostatic, exchange, induction and dispersion energies. The former two terms were published by some of us previously~\cite{malone2022towards}, the latter two are presented in this study and emplpy an extended random phase approximation (ERPA) formalism. They require solving response equations and are necessary to obtain accurate interaction energies;   fifth, the SAPT energy components and interaction energies provide an in-depth understanding of intermolecular interaction of interest (note that in this work the quantum computing results were obtained on a simulator).}
    \label{fig:workflow}
\end{figure*}

\begin{figure*}[htbp]
    \centering
    \begin{subfigure}[b]{0.35\textwidth}
    \includegraphics[width=\textwidth]{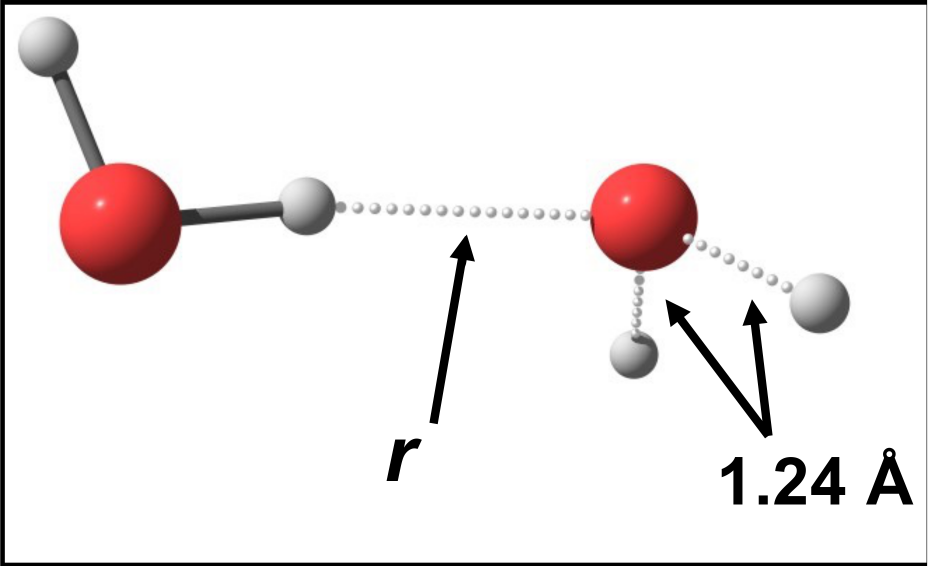}
    \caption{ }
 \end{subfigure}\\
     \begin{subfigure}[b]{0.45\textwidth}
    \includegraphics[width=\textwidth]{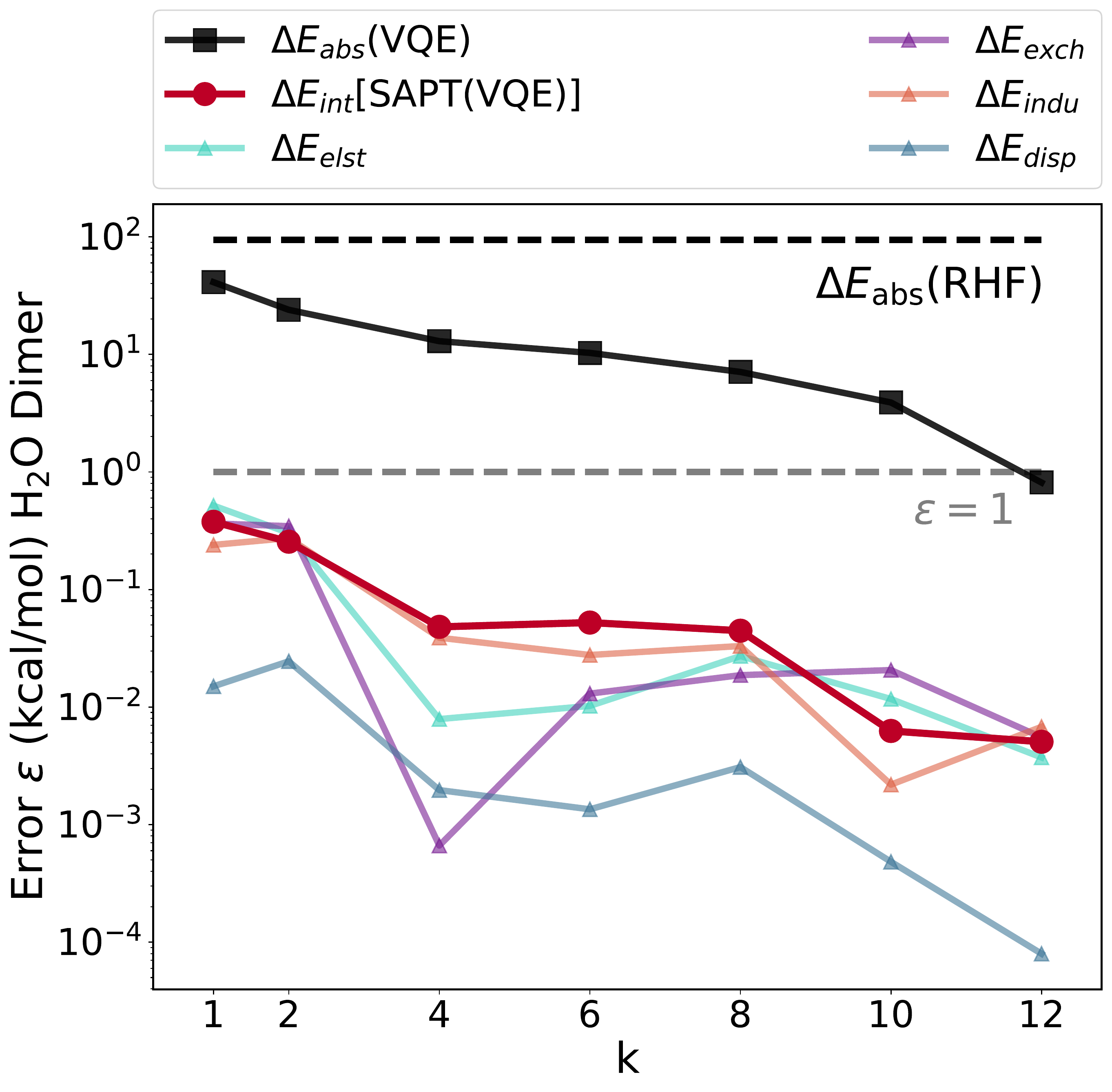}
    \caption{ }
 \end{subfigure}
    \begin{subfigure}[b]{0.45\textwidth}
    \includegraphics[width=\textwidth]{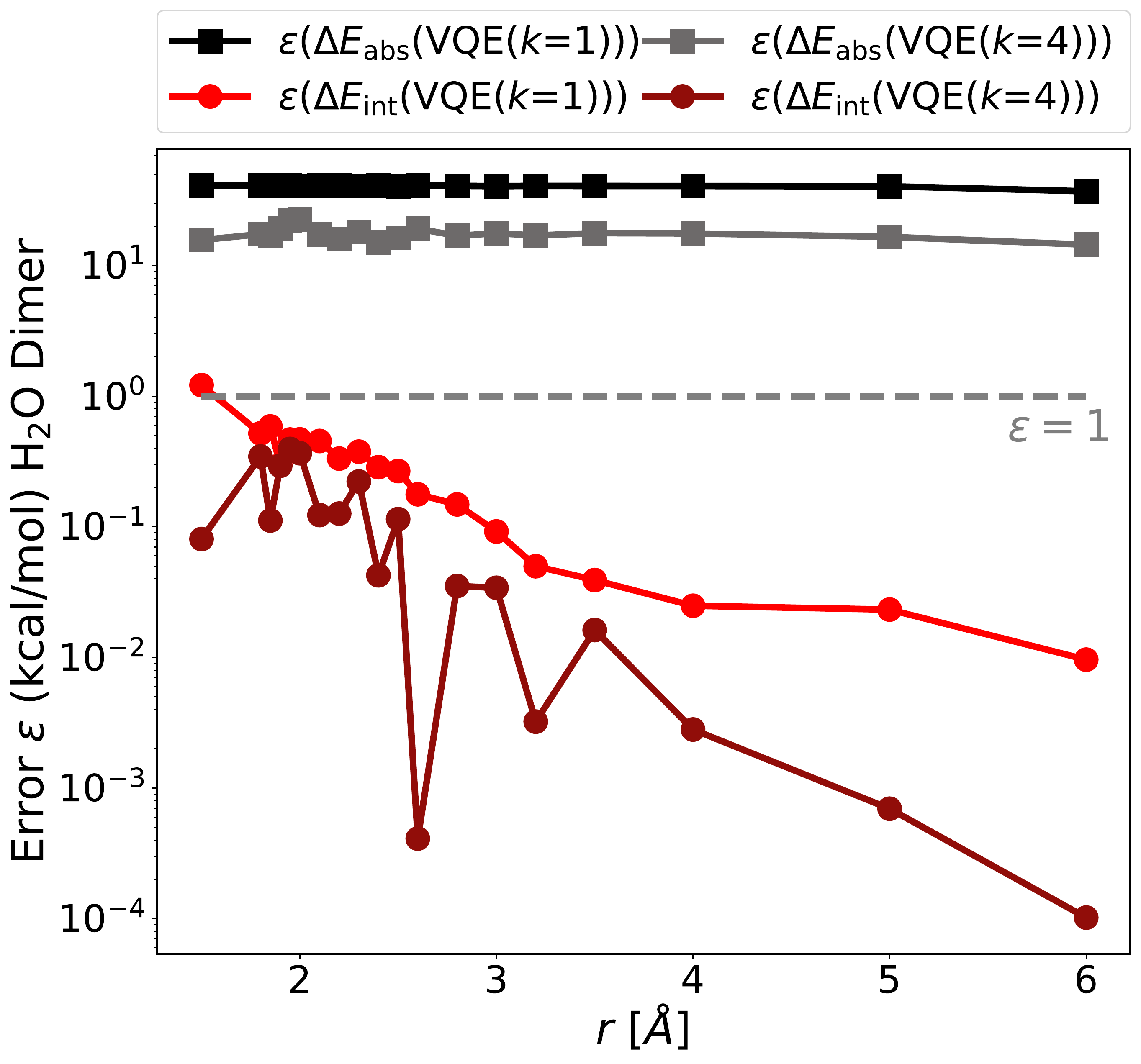}
    \caption{ }
 \end{subfigure}
    \caption{\small (a) Structure of the stretched water dimer; 
    (b) absolute errors of the VQE total energy (stretched monomer) and each SAPT(VQE)  energy term as a function of the repetition factor $k$ at $r$~=~2.0~\AA . This shows that accurate interaction energies can already be obtained with coarsely optimized, low circuit depth wavefunctions from the quantum computer (errors relative  to the CAS-CI and SAPT(CAS-CI) energies, the dotted gray line represents the chemical accuracy 1~kcal/mol threshold); (c) absolute error of  the VQE total energy (stretched water monomer) and each SAPT(VQE) ($k=1$) and SAPT(VQE) ($k=4$) energy term as a function of the intermolecular distance $r$. This shows that this empirical finding also holds for non-equilibrium bond distances.}
    \label{fig:water_results}
\end{figure*} 

\begin{figure*}[htbp]
\centering
    \begin{subfigure}[b]{0.25\textwidth}
    \includegraphics[width=\textwidth]{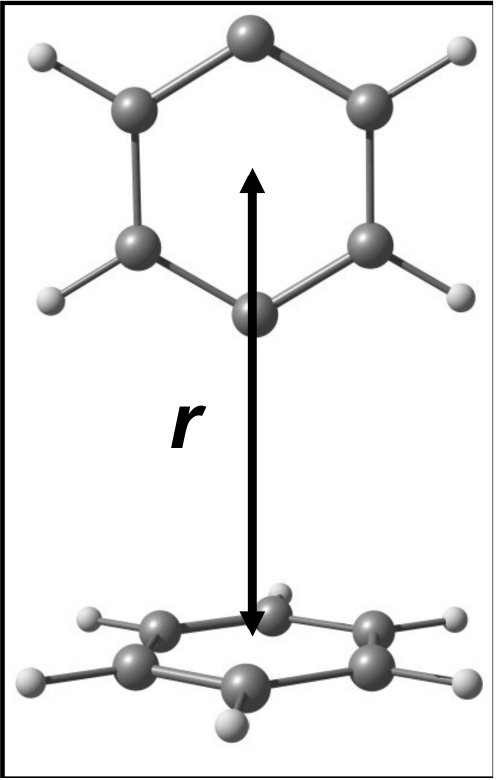}
    \caption{ }
 \end{subfigure}

         \begin{subfigure}[b]{0.49\textwidth}
    \includegraphics[width=\textwidth]{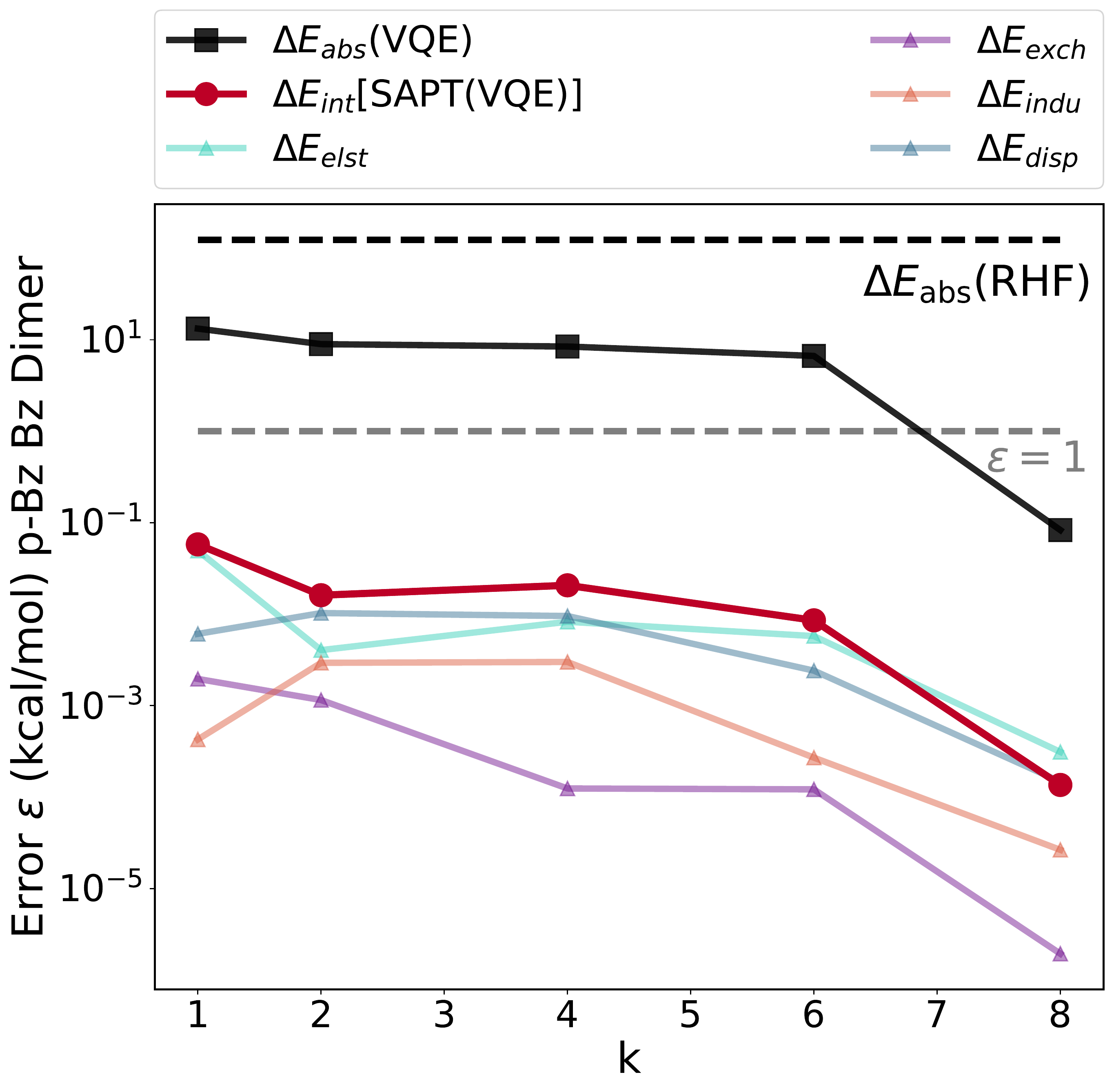}
    \caption{ }
 \end{subfigure}
     \begin{subfigure}[b]{0.49\textwidth}
    \includegraphics[width=\textwidth]{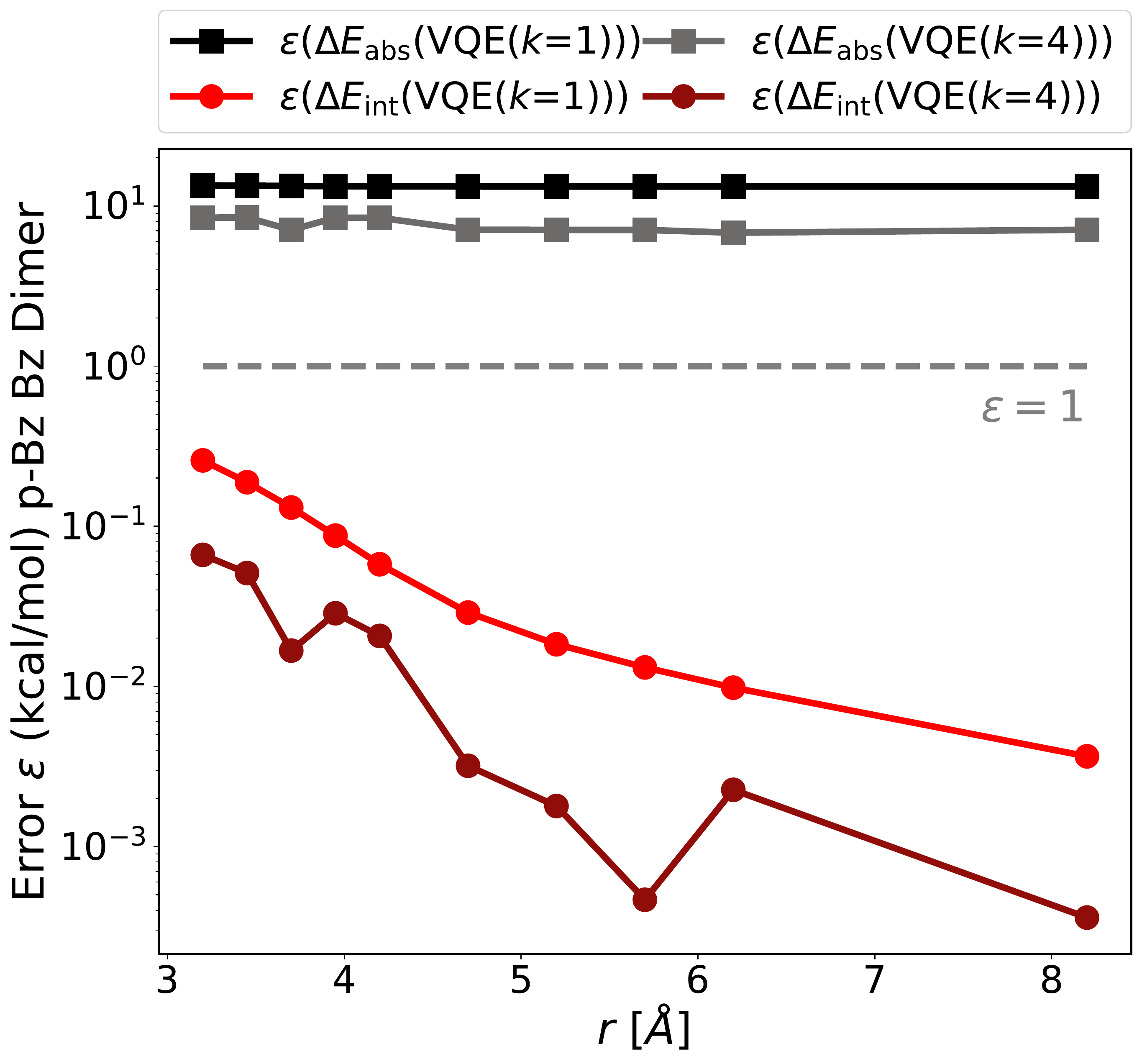}
    \caption{ }
 \end{subfigure}
      \caption{(a) Structure of the benzene p-benzyne dimer; 
       (b) absolute errors of the VQE total energy (p-benzyne monomer) and each SAPT(VQE)  energy term as a function of the repetition factor $k$ at r~=~3.9~\AA\ (error relative  to the CAS-CI and SAPT(CAS-CI) energies, the dotted gray line represents the chemical accuracy 1~kcal/mol threshold);
    (c) absolute error of of the VQE total energy (p-benzyne monomer) and each SAPT(VQE) ($k=1$) and SAPT(VQE) ($k=4$) energy term as a function of the intermolecular distance r. This shows that our empirical finding about accurate interaction energies also holds for different interaction motifs and different strongly correlated electronic structures.}
    \label{fig:bz_results2}
\end{figure*}

\begin{figure*}[htbp]
    \centering
    \includegraphics[width=1.0\textwidth]{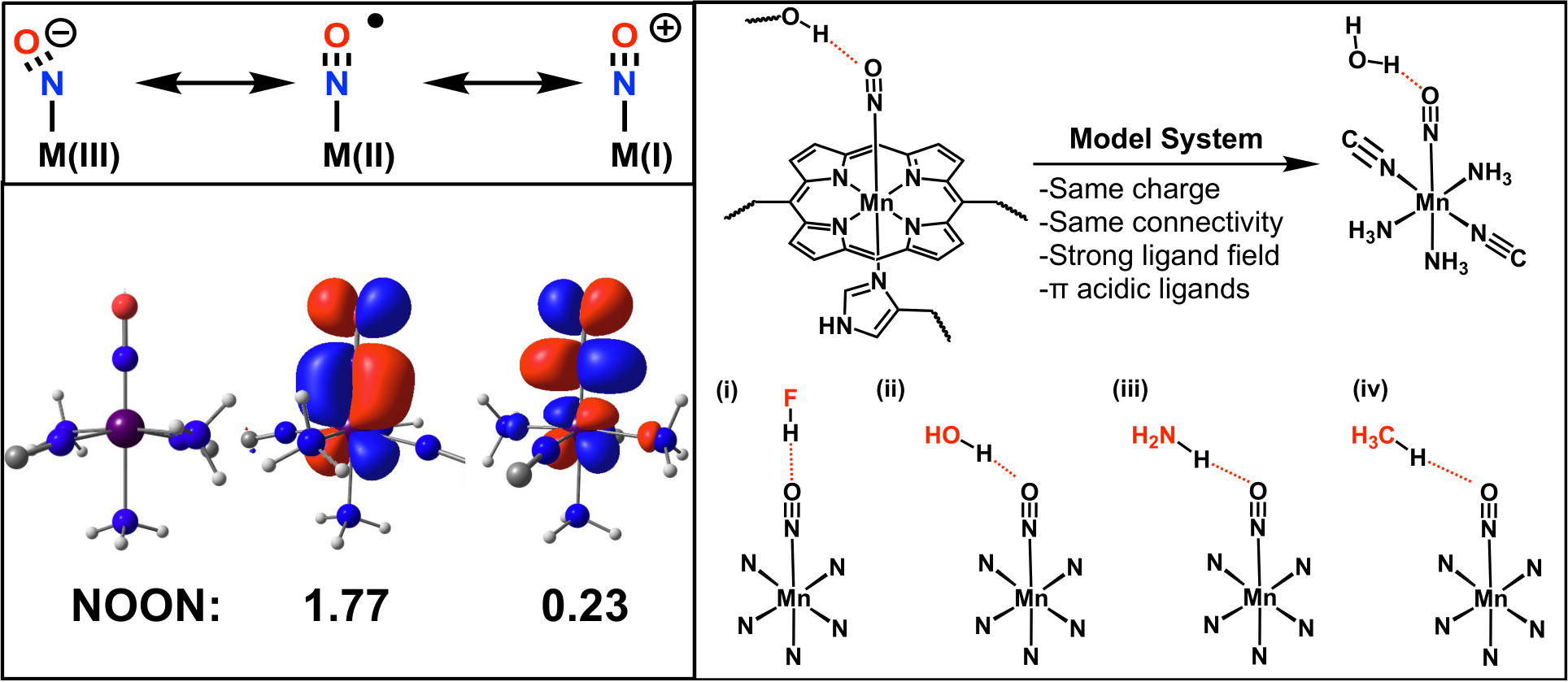}
    \caption{Upper left panel: different oxidation states of a transition metal bound NO; bottom left panel: structure of the \ce{[Mn(CN)2(NH3)3NO]^0} complex and the natural orbitals (NOON) with the largest deviation from integer value occupation ($\pi$ type metal to ligand backbonding); right panel: simplification of the heme ligand framework and schematic representation of the four manganese nitrosyl hydrogen bond complexes.}
    \label{fig:mnno_scheme}
\end{figure*}

\begin{figure*}[htbp]
    \centering
    \begin{subfigure}[b]{0.6\textwidth}
    \includegraphics[width=\textwidth]{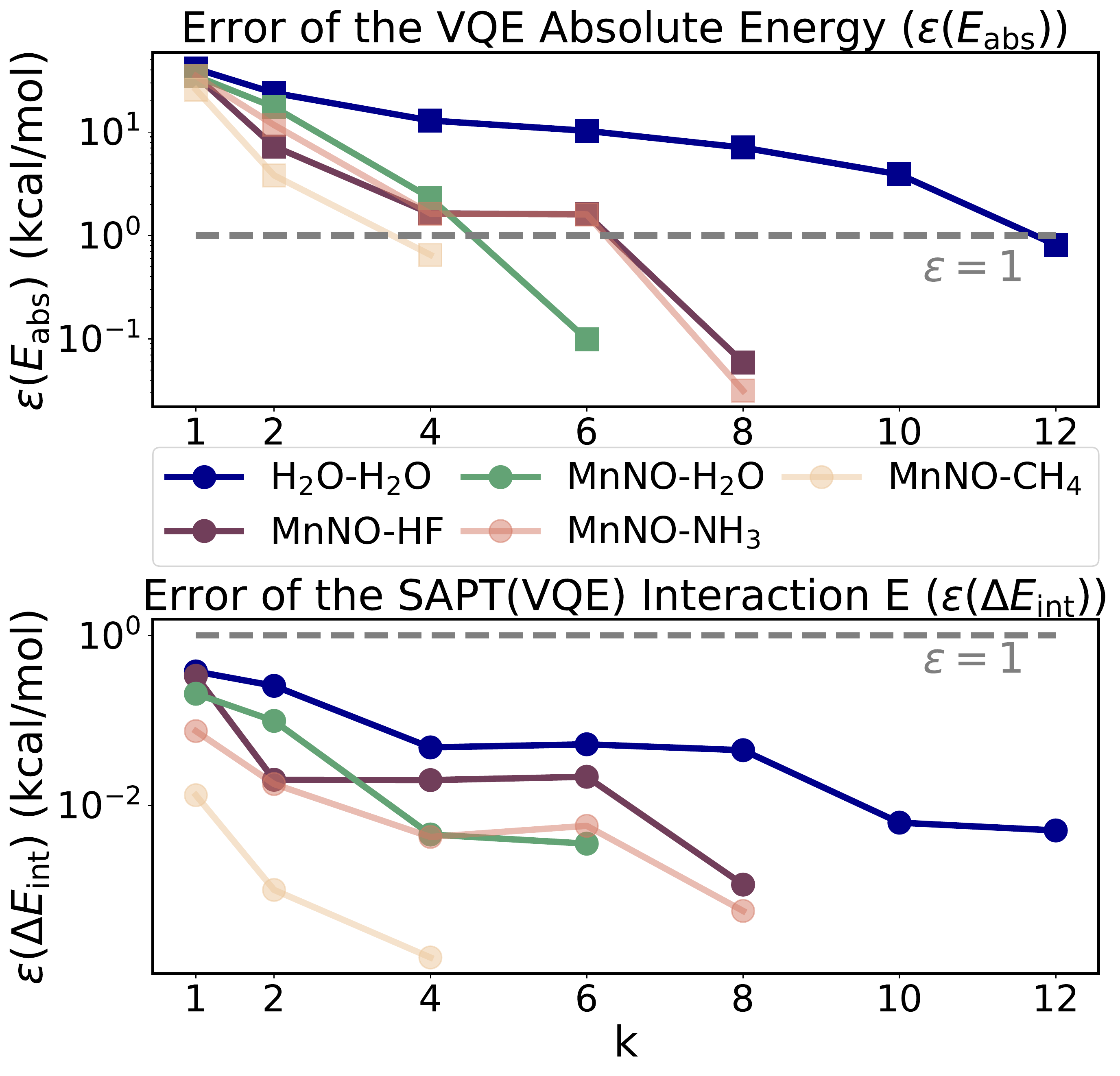}
    \caption{ }
 \end{subfigure}
 
     \begin{subfigure}[b]{0.6\textwidth}
    \includegraphics[width=\textwidth]{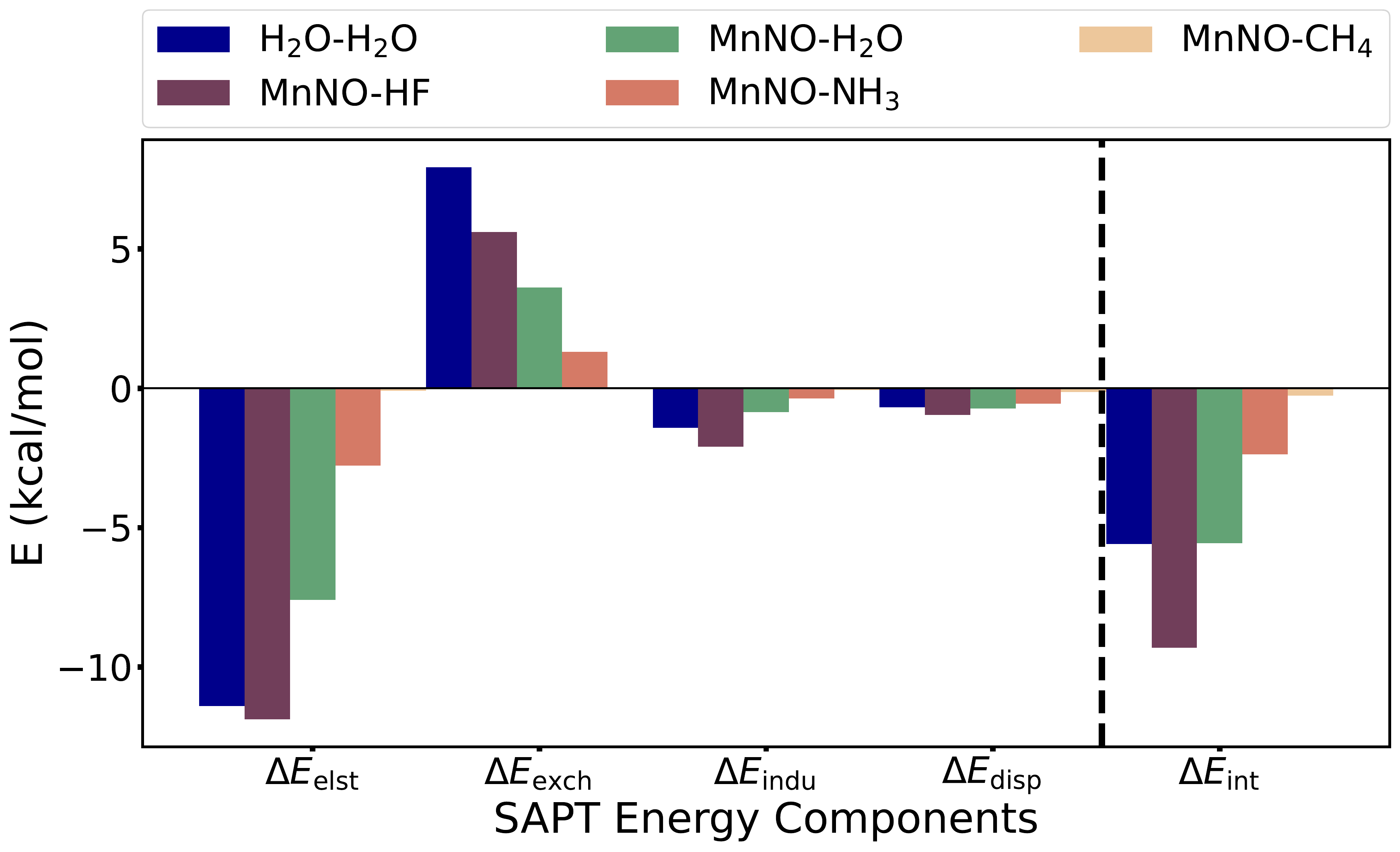}
    \caption{ }
 \end{subfigure}
      \caption{(a) Top figure: absolute errors of the VQE total energy for the \ce{[Mn(CN)_2(NH3)_3NO]^0} complex and the ``stretched'' water;  bottom figure: the SAPT(VQE)  interaction energy as a function of the repetition factor $k$ for the water dimer and each model heme-nitrosyl hydrogen bond complex showing that our empirical finding about accurate interaction energies also holds for a wide range of interaction energies (error relative  to the CAS-CI and SAPT(CAS-CI) energies, the dotted gray line represents the 1~kcal/mol  threshold);  (b) term-by-term decomposition of the SAPT(VQE) ($k=4$) interaction energies of each model heme-nitrosyl hydrogen complex (see Tab.~S1 for details).}
    \label{fig:mn_results}
\end{figure*}

\begin{figure*}[htbp]
    \centering

    \begin{subfigure}[b]{\textwidth}
    \includegraphics[width=\textwidth]{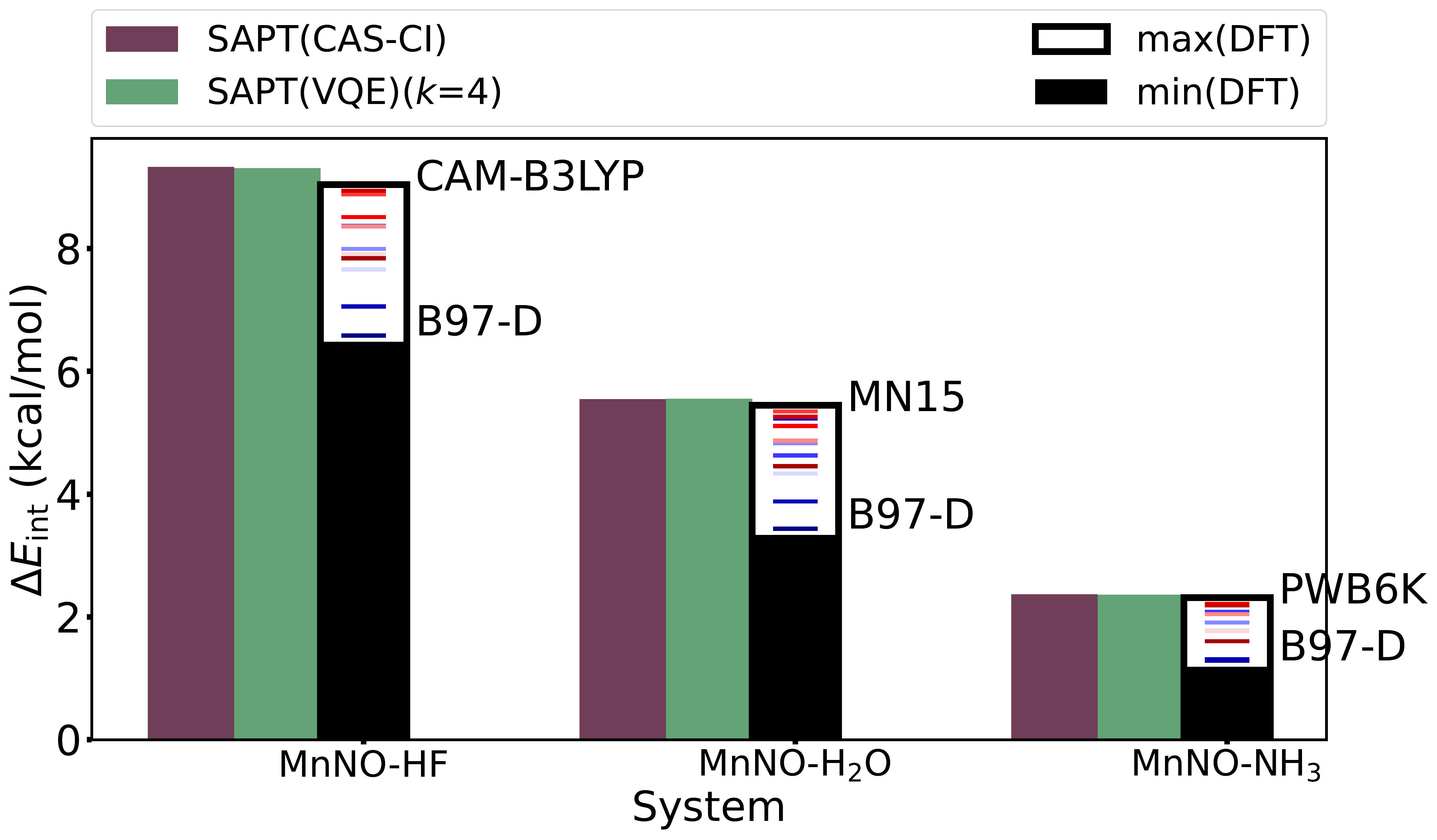}
 \end{subfigure}
 
     \begin{subfigure}[b]{\textwidth}
    \includegraphics[width=\textwidth]{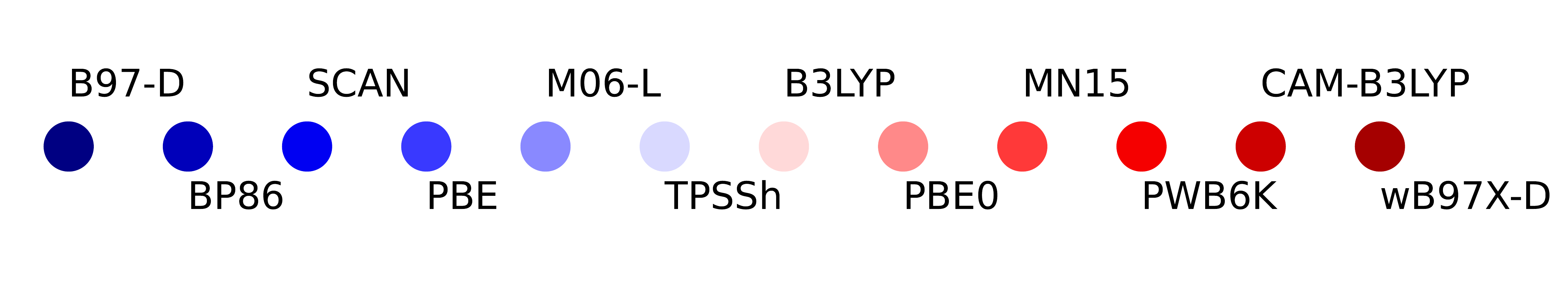}
 \end{subfigure}
      \caption{Interaction energies using several approximate DFT exchange correlation functionals (supramolecular approach) in comparison to the SAPT(CAS-CI) and SAPT(VQE) ($k=4$) interaction energies showing the sensitivity of the results for a given hydrogen bonding complex - exchange correlation functional pair (see Table~S3 for more details).}
    \label{fig:mn_dft_results}
\end{figure*}
\cleardoublepage
\section*{Data availability}
XYZ structures and NPZ files for the \mnhf\ example. We also include python code to obtain and save the active space Hamiltonian via \textsc{PySCF}.
\section*{Acknowledgments}
QC Ware Corp. acknowledges generous funding from Boehringer Ingelheim for this research project.
We thank Clemens Utschig-Utschig for insightful discussions and Daria Miroshko for the graphics design of Fig.~\ref{fig:workflow}.
\section*{Author contributions}
All authors discussed and designed the formalism of an ERPA-based extension of SAPT(VQE) for induction and dispersion. F. D. M., A. R. W., and R. M. P. carried out the initial derivation of the equations. F.D.M. performed extensive simplifications and spin-adaption of the equations. F. D. M. wrote the \textsc{CuPy} implementation of the method. M. L., T. F., M. D, and N. M. prepared the geometries and performed the classical reference computations. M.L. and F.D.M. performed the simulated SAPT(VQE) computations. All authors assisted with the production and analysis of the applications results and with the writing of the manuscript. 
\noindent\textbf{Correspondence} should be addressed to R.M.P. or N.M.

\section*{Competing interest}
M. L., A. R. W. and R. M. P. own stock/options in QC Ware Corp.

\appendix
\renewcommand\thefigure{S\arabic{figure}}  
\renewcommand\thetable{S\Roman{table}}    

\setcounter{figure}{0}    

\section{Theory \label{suppl:theory}}
\subsection{Indices and Labels}

We adopt the following notation for orbital sets used in this work:
\begin{itemize}
\item $A/B$ - monomer nuclear indices.
\item $\mu/\nu$ - nonorthogonal atomic spatial orbital basis indices (i.e.,
Gaussian basis indices).
\item $p/q$ - orthogonal molecular spatial orbital basis indices.
\item $i/j$ - orthogonal occupied spatial orbital basis indices.
\item $t/u$ - orthogonal active spatial orbital basis indices.
\item $a/b$ - orthogonal virtual spatial orbital basis indices.
\end{itemize}
Repeated indices within a monomer will be denoted with primes, e.g., $p, p',
p'', p'''$.  When dealing with spin-orbital quantities, we use the context
specific notation of an ``unbarred'' orbital index to denote $\alpha$ and a
``barred'' orbital index to denote $\beta$, i.e., $p^\dagger$ is an $\alpha$
spin-orbital creation operator on spatial orbital index $p$, while $\bar
p^\dagger$ is a $\beta$ spin-orbital creation operator on spatial orbital index
$p$.
\subsection{Symmetry Adapted Perturbation Theory}
The interaction energy between two monomers is defined as the the total energy of the combined system and subtract the total energies: 
\begin{equation}
E_{\mathrm{int}} = E_{AB} - E_{A} - E_B,
\end{equation}
where the total energies represent the full CI solution at the basis set limit.  In practice, approximate methods such as density functional theory or coupled cluster methods are used to compute accurate enough total energies to resolve the binding energy accurately.

An alternative approach to computing intermolecular interaction energies is symmetry adapted perturbation theory (SAPT) which is valid for non-covalent interactions.
Instead of computing total energies, SAPT assumes that the intermolecular interactions are weak and thus we can compute the interaction energy via perturbation theory.
In particular, we can write the Hamiltonian of the combined system as
\begin{equation}
\hat{H} = \hat{H}_A + \hat{H}_B + \hat{V},
\end{equation}
where we assume $\hat{H}_X\lvert\Psi_X\rangle = E_X \lvert \Psi_X \rangle$, where $\lvert\Psi_X\rangle$ is the ground state wavefunction of monomer $X$ and $\hat{V}$ contains only the Coulombic interactions between monomer $A$ and $B$. 
With this partitioning of the Hamiltonian we can build a perturbation theory for the intermolecular interaction energy directly thus avoiding computing potentially very large total energies.
More explicitly we have
\begin{equation}
E_{\mathrm{int}} = \sum_n (E_{\mathrm{pol}}^{(n)} + E_\mathrm{exch}^{(n)}),
\end{equation}
where $E_{\mathrm{pol}}^{(n)}$ and $E_{\mathrm{exch}}^{(n)}$ are $n$th-order polarization and exchange energies respectively.
An added benefit of SAPT is that we obtain an intuitive breakdown of the interaction energy components, into electrostatic, induction, dispersion and exchange components which can be used to provide chemical insight into the binding process.
At the lowest order this gives rise to SAPT0\cite{Jeziorski1994} yielding:
\begin{widetext}
\begin{equation} \label{eq:sapt}
E_{\mathrm{int}} \approx E_{SAP0} = E^{(1)}_{\mathrm{elst}} +  E^{(1)}_{\mathrm{exch}} + E^{(2)}_{\mathrm{ind,u}} + E^{(2)}_{\mathrm{disp}} + E^{(2)}_{\mathrm{exch-disp}} + E^{(2)}_{\mathrm{exch-ind,u}},
\end{equation}
\end{widetext}

where $ E^{(1)}_\mathrm{{elst}}$ corresponds to the first order electrostatics term, $ E^{(1)}_\mathrm{{exch}}$ to the first order exchange term,  $ E^{(1)}_\mathrm{{ind,u}}$ to the first order induction term,  $ E^{(2)}_\mathrm{{disp}}$ to the second order dispersion term, $ E^{(2)}_{\mathrm{exch-disp}}$ to the second order exchange-dispersion term, $ E^{(2)}_{\mathrm{exch-ind,u}}$ to the second order exchange-induction term (see below for a detailed derivation). In the main manuscript we combine both dispersion and inductions terms and drop the superscripts yielding:
\begin{equation}
E_{\mathrm{int}} \approx E_{\mathrm{elst}} + E_{\mathrm{exch}} + E_{\mathrm{ind,u}} +  E_{\mathrm{disp}}.
\end{equation}
Ref.~\citenum{parker2014levels} shows that this SAPT0 using Hartee-Fock wavefunctions and a medium size jun-cc-pvdz basis can yield highly accurate results for a broad range of non-covalent interactions.

As in-depth discussed in our previous work\cite{malone2022towards}, we use the density matrix formulation of SAPT \cite{Moszynski1994,KoronaExchange2008,Korona2008,Korona2009} as recently fully implemented for complete active space self consistent field (CASSCF) wavefunctions by Hapka and {\it et al.}~\cite{hapka2021casscf}
This formalism allows for the evaluation of the terms appearing in \cref{eq:sapt} using just the ground state one- and two-particle reduced density matrices of the monomers with additional response terms for the second order terms. Instead of Hartee-Fock density matrices, we use a quantum computer via a VQE ansatz to determine an accurate ground state wave function of the system (for more details on those algorithms see the next sections). The detailed derivation of the first order terms ($E^{(1)}_\mathrm{{elst}}$, $E^{(1)}_\mathrm{{exch}}$) be found in Ref.~\cite{malone2022towards}.

The second order SAPT terms for induction and dispersion energies (as well as their exchange counterparts) requires the calculation of excited state properties on a quantum computer (see below for a more detailed derivation).
Although several approaches have been suggested in the literature to compute excited state properties on NISQ-era quantum computers, they often require a significant measurement overhead~\cite{parrish2019quantum,Parrish2019transitions,CaiMolecularResponse2020,OllitraultEOM2020}.
To reduce this burden we employ the extended random phase approximation (ERPA)~\cite{ChatterjeeERPA2012} which requires only the one- and two-body reduced density matrices to be evaluated on the quantum computer~\cite{mcclean2017hybrid}.
This approximation has previously been shown by others to produce quite accurate interaction energies when employed in SAPT based on CASSCF wavefunctions and we will show that this carries over for VQE wavefunctions.

Given that NISQ-era devices are currently limited to tens of qubits (spin-orbitals) we will use an active space approach analogous to CAS-CI methods. In the active space approach we partition the one-electron orbital set into
$N_c$ core orbitals, $N_a$ active orbitals and $N_i$ virtual orbitals.
This partitioning gives rise to modified monomer Hamiltonian given by (for example for monomer $A$)
\begin{equation}
    \begin{split}
    \hat{H_{A}}' =
    &\sum_{tt'\sigma} \tilde{h}_{tt'} a^\dagger_{t\sigma}a_{t'\sigma}+\\
    &\frac{1}{2}\sum_{\sigma\sigma'}\sum_{tt't''t'''} (tt''|t't''') a^\dagger_{t\sigma}a^\dagger_{t'\sigma'} a_{t'''\sigma'} a_{t''\sigma}
    \end{split}\label{eq:fzc},
\end{equation}
where the modified one-electron integrals $\tilde{h}_{tt'}$ now include core-active space interactions
\begin{equation}
    \tilde{h}_{tt'} = h_{tt'} + \sum_{ii'} \left[(tt'|ii') - \frac{1}{2} (ti'|it')\right]\gamma_{ii'}\label{eq:core_ham}.
\end{equation}
The key approximation in the active space approach is that a  (small) set of ``active'' orbitals and electrons are defined \textit{a priori} and the FCI expansion is constrained to that subset of electrons and orbitals. The quality of the CAS-CI and VQE results depends strongly on the selected active space. \cite{veryazov2011how,sun2014exact,sayfutyarova2017automated}.

\subsection{Variational Quantum Eigensolver\label{sec:VQE}}
This section briefly summarizes the VQE ansatz used in this work, it is identical to the approach taken in our previous work~\cite{malone2022towards} and discussed in more detail. The SAPT post processing step only relies on the reduced density matrices which can in principle be generated with an quantum algorithm such as other VQE flavors\cite{anselmetti2021local} or Quantum Krylov methods\cite{parrish2019quantum}.

We generate the active space wave function of the strongly correlated monomer using the VQE ansatz described below (it is also possible that both monomer wave function are evaluated on the quantum computer but in practice usually only one monomer will exhibit strong correlation and thus require a VQE treatment):

\begin{equation}
|\Psi_{\mathrm{VQE}}\rangle
\equiv
\hat U_{\mathrm{VQE}}
|\Phi_I\rangle
\end{equation}
where $|\Phi_I\rangle$ guess state (typically the Hartree--Fock).

Throught this study we only use real active space
wavefunction $|\Psi_\mathrm{VQE}\rangle$ and they will be a definite
eigenfunction of the $\hat N_{\alpha}$, $\hat N_{\beta}$, and $\hat S^2$
operators.

The paper is using the following Jordan-Wigner representation:
\begin{equation}
p^{\pm}
=
\bigotimes_{p' = 0}^{p' = p-1}
\hat Z_{p'}
(\hat X_{p} \mp i\hat Y_{p}) / 2,
\end{equation}
\begin{equation}
\bar p^{\pm}
=
\bigotimes_{p' = 0}^{p' = N_\alpha - 1}
\hat Z_{p'}
\bigotimes_{\bar p' = 0}^{\bar p' = \bar p-1}
\hat Z_{\bar p'}
(\hat X_{\bar p} \mp i\hat Y_{\bar p}) / 2,
\end{equation}
where $p^+ = p^\dagger$ and $p^-=p$ and we order the Jordan-Wigner strings in
$\alpha$-then-$\beta$ order and $\hat{Z}, \hat{Y}$ and $\hat{X}$ are the usual Pauli operators.

In this work we use a modified version of the unitary cluster Jastrow wavefunction
\cite{Matsuzawa2020} ($k$-uCJ) which takes the form
\begin{equation}
    | \Psi_{\mathrm{VQE}} \rangle
=
\prod_{k}
\exp(-\hat K^{(k)})
\exp(\hat T^{(k)})
\exp(+\hat K^{(k)})
| \Phi_{I} \rangle,
\label{eq:ansatzsi}
\end{equation}
where $\hat{K}^{(k)}$ and $\hat{T}^{(k)}$ are one- and two-body operators, and $k$ is a parameter that controls the depth of the circuit and as a result its variational freedom.
The key difference in our $k$-uCJ ansatz from Ref.~\citenum{Matsuzawa2020} is that the two-body operator and the restriction  to real anti-symmetric matrices.

The one-body  orbital transformations (spin restricted) are definad as
\begin{equation}
    \hat K^{(k)}
\equiv
\sum_{pp'}
\kappa_{pp'}^{(k)}
\left [
(p^\dagger p' - p'^\dagger p)
+
(\bar p^\dagger \bar p' - \bar p'^\dagger \bar p)
\right ]
\end{equation}
where $\kappa_{pp'}^{(k)} = -\kappa_{p'p}^{(k)}$ is a real, antisymmetric $N_{a} \times N_{a}$
matrix of orbital rotation generators, which equivalent to the one-particle  spin-restricted orbital transformation:
\begin{equation}
    U_{pp'}^{(k)}
\equiv
\left [
    \exp (\kappa^{(k)})
\right ]_{pp'}
\end{equation}
This spin-restricted orbital rotation
is expressed as quantum circuits using a fabric of Givens
rotations \cite{KivlichanGivens2018}.

\begin{figure*}
\begin{center}
\Qcircuit @C=1em @R=.7em {
& \gate{\mathbf{G}} \qwx[2]  & \qw  &  \qw & \gate{\mathbf{G}} \qwx[2]  & \qw  & \multigate{3}{\mathbf{P_X}} & \gate{\mathbf{G}} \qwx[2]  & \qw  &  \qw & \gate{\mathbf{G}} \qwx[2]  & \qw\\
 & \qw & \gate{\mathbf{G}} \qwx[2]  & \qw  & \qw & \gate{\mathbf{G}} \qwx[2]  &  \ghost{\mathbf{P_X}}  & \qw & \gate{\mathbf{G}} \qwx[2]  & \qw  & \qw & \gate{\mathbf{G}} \qwx[2]\\
& \gate{\mathbf{G}}  & \qw &  \qw & \gate{\mathbf{G}}  & \qw &  \ghost{\mathbf{P_X}} & \gate{\mathbf{G}}  & \qw &  \qw & \gate{\mathbf{G}}  & \qw \\
 & \qw & \gate{\mathbf{G}}  & \qw &  \qw & \gate{\mathbf{G}}  & \ghost{\mathbf{P_X}} & \qw & \gate{\mathbf{G}}  & \qw &  \qw & \gate{\mathbf{G}} \\
}
\end{center}

\caption{Quantum circuit of a single layer ($k=1$) $k$-muCJ VQE entangler circuit for $M=2$ spatial orbitals or $N=4$ qubits. Even (odd) qubits represent $\alpha$ ($\beta$) spin-orbitals. The quantum circuit starts with two-quibit Givens rotation among $\alpha$ and $\beta$ orbitals. The next steps are  a double substitution operator (four qubit exchange gate) and another layer of Givens rotations.\label{fig:entangler}}
\end{figure*}
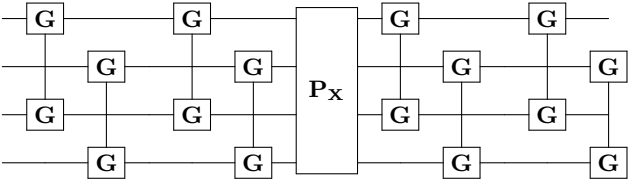

The two-particle operator is defined as 
\begin{widetext}
\begin{equation}
\begin{split}
\hat T^{(k)}
\equiv
\sum_{p=0}^{M-1}
\sum_{\substack{p'=p\ \mathrm{mod} \ 2 \\p+=2}}^{M-2}
\tau_{pp'}^{(k)}
[
&
(p'+1)^{\dagger}
\overline{(p'+1)}^{\dagger}
p'
\bar p'
\\
&
-
p'^{\dagger}
\bar p'^{\dagger}
{(p'+1)}
\overline{(p'+1)}
]
\end{split}
\label{eq:diag_doub}
\end{equation}
\end{widetext}
The uCJ implementation is similar but not exact to Refs.~\cite{anselmetti2021local} and \cite{OGormanSWAP2019}. This it is denoted as  $k$-muCJ for clarity, with the `m' standing for modified.
It is important to point out that  the choice of VQE ansatz is largely irrelevant from a SAPT perspective and is not a major point in this paper. An example of one layer of the muCJ circuit ansatz is given in \cref{fig:entangler}.

Using the ansatz discussed above, the VQE objective function is defined as
\begin{align}
E_{\mathrm{VQE}}
(\kappa_{pq}^{k}, \tau_{pq}^{k})
&\equiv
\langle \Psi_{\mathrm{VQE}} 
(\kappa_{pq}^{k}, \tau_{pq}^{k})
|
\hat H
|
\Psi_{\mathrm{VQE}}
(\kappa_{pq}^{k}, \tau_{pq}^{k})
\rangle
\\
&=
\langle \Phi_{\mathrm{I}} 
|
\hat U^{\dagger}
(\kappa_{pq}^{k}, \tau_{pq}^{k})
\hat H
\hat U 
(\kappa_{pq}^{k}, \tau_{pq}^{k})
|
\Phi_{\mathrm{I}}
\rangle.
\label{eq:obj}
\end{align}

The number of measurements scales $\mathcal{O}(N_a^4)$; however, we point out that efficient estimation of the density matrices are currently intensively investigated by several reaseach groups, e.g. Ref.~\citenum{AIQuantum2020} or Ref.~\citenum{TillyCASSCF2021}.

\section{Derivation of the Second Order SAPT Terms \label{supl:sapt_theory}}
It is helpful to first recall standard Rayleigh-Schr\"odinger perturbation theory for the intermolecular interaction energy.
To begin we can write
we write the Hamiltonian of the combined system as
\begin{equation}
\hat{H} = \hat{H}_A + \hat{H}_B + \hat{V},
\end{equation}
where we assume $\hat{H}_X\lvert\Psi_X\rangle = E_X \lvert \Psi_X \rangle$, where $\lvert\Psi_X\rangle$ is the ground state wavefunction of monomer $X$ and $\hat{V}$ contains only the Coulombic interactions between monomer $A$ and $B$: 
\begin{widetext}
\begin{equation}
\hat{V} = \sum_{i=1}^{N_A} \sum_{j=1}^{N_B}\frac{1}{|\mathbf{r}_i-\mathbf{r}_j|} - \sum_{i=1}^{N_A} \sum_{\beta}\frac{Z_\beta}{|\mathbf{R}_\beta-\mathbf{r}_i|} - \sum_j^{N_B} \sum_{\alpha} \frac{Z_\alpha}{|\mathbf{R}_\alpha - \mathbf{r}_j|} + \sum_{\alpha}\sum_{\beta}\frac{Z_\alpha Z_\beta}{|\mathbf{R}_\alpha-\mathbf{R}_\beta|},
\end{equation}
\end{widetext}

where it is understood that $i$ and $j$ are distinct indices and the sum over $\alpha /  \beta$ runs over the number of atoms in monomer $A/B$ respectively.
Under the assumption that the potential $\hat{V}$ is a small perturbation (valid for weak intermolecular interactions) we can in principle now use perturbation theory.
Following Ref. \citenum{Jeziorski1994} one typically starts SAPT by first constructing a perturbation series for the so-called polarization energy 
\begin{equation}
    |\Psi_{\mathrm{pol}}^{(n)}\rangle =-\hat{R}_0 \hat{V} |\Psi^{(n)}_\mathrm{pol}\rangle + \sum_{k}^{n-1} E_\mathrm{pol}^{(k)}\hat{R}_0|\Psi_\mathrm{pol}^{(n-k)}\rangle,
\end{equation}
where
\begin{equation}
    \hat{R}_0 = (\hat{H}_0 - E_0 + \hat{P}_0)^{-1} \hat{Q}_0,
\end{equation}
is the resolvent operator, $\hat{H}_0 = \hat{H}_A + \hat{H}_B$, $\hat{P} = |\Psi_0\rangle\langle\Psi_0|$, $\hat{Q} = 1-\hat{P}$ and $|\Psi_\mathrm{pol}^0\rangle = |\Psi_0\rangle = |\Psi_A^0\Psi_B^0\rangle$.
Here we have that $\hat{H}=\hat{H}_A + \hat{H}_B + \hat{V}$ and we assume $
\hat{H}_0|\Psi_0\rangle = (E^0_A + E^0_B)|\Psi_0\rangle$.
The first order polarization energy (usually  called the electrostatic energy  $E_{\mathrm{elst}}$) is then given by
\begin{align}
E_\mathrm{pol}^{(1)} &= \langle \Psi_0 | \hat{V}|\Psi_0\rangle \\ 
                     &= \bar{\gamma}_{pp'}v_{pq}^{p'q'} \bar{\gamma}_{qq'},\label{eq:elst}
\end{align}
where we used the definition $E_{\mathrm{pol}}^{(n)} = \langle \Psi_0 | \hat{V} | \Psi_\mathrm{pol}^{(n)}\rangle$

For the second order polarisation energy it is conventional in SAPT to split into the induction and dispersion energy, i.e., $E_{\mathrm{pol}}^{(2)} = E_\mathrm{ind} + E_\mathrm{disp}$.
The distinction arises by considering that the induction term accounts for contributions from terms where one monomer is in an excited state and the dispersion term considers contributions from terms where both monomers are in excited states.
For the induction energy we have $|\Psi_A^{(\mathrm{ind})}\rangle \equiv \hat{R}_0^A \hat{\Omega}^B |\Psi_A^0\rangle$, where $\hat{R}_0^A = (\hat{H}_A - E_0^A + \hat{P}_A)^{-1} \hat{Q}_A$ and
\begin{equation}
\hat{\Omega}^\mathrm{B} = \sum_i^{N_A} \left(\hat{V}_B(\mathbf{r}_i) + \int d\mathbf{r}_j \frac{\hat{\rho}_B(\mathbf{r}_j)}{|\mathbf{r}_i-\mathbf{r}_j|}\right)
\end{equation}
is the effective electrostatic field of monomer $B$.
Using the sum over states formula for the resolvent we have then that
\begin{align}
    |\Psi_A^{(\mathrm{ind})}\rangle &\equiv \hat{R}_0^A \hat{\Omega}^B |\Psi_A^0\rangle \\
    &=
    -\sum_\mu \frac{|\Psi^\mu_A\rangle\langle\Psi^\mu_A|\hat{\Omega}^B|\Psi_A^0\rangle}{E_A^\mu-E_A^0}\label{eq:psi_ind},
\end{align}
from which it follows
\begin{equation}
E_{\mathrm{ind}}(A\leftarrow B) = -
\sum_\mu \frac{|\langle\Psi^\mu_A|\hat{\Omega}^B|\Psi_A^0\rangle|^2}{E_A^\mu-E_A^0},\label{eq:ind2}
\end{equation}
with a similar expression for $E_
\mathrm{ind}(B\leftarrow A)$.
By introducing the (spin-summed) transition one-particle reduced density matrix 
\begin{equation}
    \bar{\gamma}_{pp'}^\mu = \langle \Psi^0 | p^{\dagger} p' | \Psi^\mu\rangle + \langle \Psi^0 | \bar{p}^{\dagger} \bar{p}' | \Psi^\mu\rangle \label{eq:opdm}
\end{equation}
we can write \cref{eq:ind2} as
\begin{equation}
E_{\mathrm{ind}}(A\leftarrow B) = -\sum_\mu
\frac{(\bar{\gamma}_{pp'}^\mu\Omega_{pp'})^2}{E_A^\mu-E_A^0}.\label{eq:ind}
\end{equation}
Similar expressions exist for monomer $B$.

For the dispersion contribution we first define 
\begin{equation}
    |\Psi^{\mathrm{disp}}_{AB}\rangle = \frac{|\Psi_A^\mu\Psi_B^\nu\rangle\langle\Psi_A^\mu\Psi_B^\nu|\hat{V}_{ee}|\Psi_A^0\Psi_B^0\rangle}{E_A^{\mu} + E_B^\nu},\label{eq:psi_disp}
\end{equation}
so that
\begin{align}
E_{\mathrm{disp}}^{(2)} &= \sum_{\mu\nu}\frac{|\langle \Psi_A^0\Psi^0_B | \hat{V}_{ee} |\Psi_A^\mu\Psi_B^\nu\rangle|^2}{E_A^{\mu} + E_B^\nu}\\
                        &= \sum_{\mu\nu} \frac{\left(\bar{\gamma}^\mu_{pp'} v_{pq}^{p'q'} \bar{\gamma}^\nu_{qq'}\right)^2}{E_A^{\mu} + E_B^\nu}\label{eq:disp}
\end{align}

\Cref{eq:elst,eq:ind,eq:disp} provide all polarization energy contributions up to second order in the intermolecular interaction energy.
Unfortunately, without further modification one finds the RS perturbation theory does not converge for many-electron systems\citep{Jeziorski1994}.
Symmetry adapted perturbation theory (SAPT) attempts to fix this issue by explicitly accounting for fermionic anti-symmetry in the wavefunction when electrons undergo exchange processes between monomers.

The SAPT expression for the intermolecular interaction energy is given by\citep{Jeziorski1994}
\begin{equation}
    E_{\mathrm{SAPT}}^{(n)} = \frac{\langle\Psi_0|\hat{V}|\mathcal{A}\Psi_{\mathrm{pol}}^{(n-1)}\rangle - \sum_{k=1}^{n-1} E_{\mathrm{SAPT}}^{(n)} \langle \Psi_0|\mathcal{A}\Psi_{\mathrm{pol}}^{(n-k)}\rangle}
    {\langle \Psi_0| \mathcal{A}\Psi_0\rangle},
\end{equation}
where $|\Psi_\mathrm{pol}^{(n)}\rangle$ are the $n$th-order polarization wavefunctions given previously and $\mathcal{A}$ is the antisymmetrizer operator. With this definition the SAPT interaction takes the following form 
\begin{equation}
E_{\mathrm{SAPT}}^{(n)} = E_\mathrm{pol}^{(n)} + E_{\mathrm{exch}}^{(n)}.
\end{equation}
If one neglects all electron exchange processes other than those that exchange a single electron between monomer $A$ and $B$ one arrives at the so-called $S^2$ approximation to the exchange energies (through second order)\citep{korona2008second,korona2009exchdisp,hapka2019second,hapka2021casscf}:
\begin{align}
    E_{\mathrm{exch}}^{(1)} &= \langle \Psi_0 | (\hat{V}-\bar{V})(\hat{P}-\bar{P}) | \Psi_0\rangle \label{eq:exch1}\\
    E_{\mathrm{exch-ind}}^{(2)}(B\rightarrow A)
    &= \langle \Psi_A^0 \Psi_B^0|(\hat{V}-\bar{V})(\hat{P}-\bar{P})|\Psi_A^{\mathrm{ind}}\Psi_B^0\rangle,\label{eq:exch_ind} \\
 E_{\mathrm{exch-ind}}^{(2)} &= \langle \Psi_A^0 \Psi_B^0|(\hat{V}-\bar{V})(\hat{P}-\bar{P})|\Psi^{\mathrm{disp}}_{AB}\rangle,\label{eq:exch_disp}
\end{align}
where $|\Psi^{\mathrm{ind}}\rangle$ and $|\Psi^{\mathrm{disp}}_{AB}\rangle$ have been defined previously and $\hat{P}$ is following electron exchange operator
\begin{equation}
    \hat{P} = \sum_i^{N_A} \sum_j^{N_B} \hat{P}_{ij}
\end{equation}
where $\hat{P}_{ij}$ exchanges an electron from monomer $A$ to monomer $B$.
Note a similar expression exists for \mbox{$E_{\mathrm{exch-ind}}(A\rightarrow B)$}.

It is helpful to note that \cref{eq:exch1,eq:exch_ind,eq:exch_disp} all share a similar structure.
Let us first example \cref{eq:exch1}:
\begin{equation}
    E_{\mathrm{exch}} = \langle \Psi_A^0 \Psi_B^0|(\hat{V}-\bar{V})\hat{P}
|\Psi^0_A\Psi_B^0\rangle
\label{eq:exch_t1_second_order}
\end{equation}
which can be written in the density matrix formalism of SAPT as
\begin{equation}
E_{\mathrm{exch}}^{(1)}(S^2)
=
\int
\gamma_{\mathrm{int}}(\bx_i,\bx_j)
\left(
    \tilde{v}(\br_i,\br_j)
    -
    \frac{
        E_{\mathrm{pol}}^{(1)}
    }
    {
        N_A N_B
    }
\right)
d\bx_i
d\bx_j
,
\label{eq:exch}
\end{equation}
where $\bx_i$ and $\bx_j$ denote both spin and spatial coordinates of electrons in monomer $A$ and $B$ respectively and the `interaction' density matrix is defined as
\begin{equation}
\begin{split}
&\gamma_{\mathrm{int}}(\bx_i,\bx_j)
=\\
&-\gamma_A(\bx_i,\bx_j)\gamma_B(\bx_j,\bx_i)\\
&- \int \gamma_A(\bx_i,\bx_{j'}) \Gamma_B(\bx_j,\bx_{j'},\bx_j,\bx_i) d\bx_{j'}\\
&- \int \Gamma_A(\bx_i,\bx_{i'},\bx_i,\bx_j) \gamma_B(\bx_j,\bx_{i'}) d\bx_{i'}\\
&- \int \int \Gamma_A(\bx_i,\bx_{i'},\bx_i,\bx_{j'}) \Gamma_B(\bx_j,\bx_{j'},\bx_j,\bx_{i'}) d\bx_{i'}d\bx_{j'}
\label{eq:intdm}.
\end{split}
\end{equation}
By inspection \cref{eq:exch_ind} is \emph{identical} to \cref{eq:exch_t1_second_order} up to the replacement of $|\Psi_A^0\rangle$ by the excited state wavefunction $|\Psi_A^\mu\rangle$.
Thus we have
\begin{widetext}
\begin{equation}
\langle \Psi_A^0 \Psi_B^0|(\hat{V}-\bar{V})\hat{P}
|\Psi_\mu^A\Psi_B^0\rangle = 
\int
\gamma^\mu_{\mathrm{int}}(\bx_i,\bx_j)
\left(
    \tilde{v}(\br_i,\br_j)
    -
    \frac{
        E_{\mathrm{pol}}^{(1)}
    }
    {
        N_A N_B
    }
\right)
d\bx_i
d\bx_j,
\end{equation}
\end{widetext}
where
\begin{equation}
\begin{split}
&\gamma^\mu_{\mathrm{int}}(\bx_i,\bx_j)
=\\
&-\gamma^\mu_A(\bx_i,\bx_j)\gamma_B(\bx_j,\bx_i)\\
&- \int \gamma^\mu_A(\bx_i,\bx_{j'}) \Gamma_B(\bx_j,\bx_{j'},\bx_j,\bx_i) d\bx_{j'}\\
&- \int \Gamma^\mu_A(\bx_i,\bx_{i'},\bx_i,\bx_j) \gamma_B(\bx_j,\bx_{i'}) d\bx_{i'}\\
&- \int \int \Gamma^\mu_A(\bx_i,\bx_{i'},\bx_i,\bx_{j'}) \Gamma_B(\bx_j,\bx_{j'},\bx_j,\bx_{i'}) d\bx_{i'}d\bx_{j'}
\label{eq:intdm_indu}.
\end{split}
\end{equation}
where $\gamma^\mu$ and $\Gamma^\mu$ are transition one- and two-particle density matrices.
Expanding the transition density matrix in orbital space and performing spin summations we find
\begin{equation}
\begin{split}
\langle \Psi_A^0 \Psi_B^0|(\hat{V}-\bar{V})\hat{P}
|\Psi^\mu_A\Psi_B^0\rangle
&=
-\frac{1}{2}(\bar{\gamma}^\mu_{pp'} \bar{\gamma}_{qq'} \tilde{v}_{pq}^{q'p'}\\ 
&+\bar{\gamma}^\mu_{pp'}
\bar{\Gamma}_{q''q'''}^{qq'}
S_{p'q'}
\tilde{v}_{pq}^{q''' q''}\\
&+\bar{\gamma}_{qq'}
[\bar{\Gamma}^\mu]_{p''p'''}^{pp'}
S_{q'p'}
\tilde{v}_{pq}^{p'' p'''}\\
&+[\bar{\Gamma}^\mu]_{p''p'''}^{pp'}
\bar{\Gamma}_{q''q'''}^{qq'}
S_{p'q'''}
S_{q'p'''}
\tilde{v}_{pq}^{p'' q''}\\
&-E_{\mathrm{pol}}^{(1)}S_{pq'} \bar{\gamma}_{qq'} S_{qp'} \bar{\gamma}^\mu_{pp'}
).
\end{split}
\label{eq:exch_dm_indu}
\end{equation}

It is helpful to define the exchange functional
\begin{equation}
\begin{split}
F^{\mu\nu}\equiv F[\bar{\gamma}^\mu, \bar{\Gamma}^\mu, \bar{\gamma}^\nu, \bar{\Gamma}^\nu]
&=
-\frac{1}{2}([\bar{\gamma}^\mu]_{pp'} [\bar{\gamma}^\nu]_{qq'} \tilde{v}_{pq}^{q'p'} \\
&+[\bar{\gamma}^\mu]_{pp'}
[\bar{\Gamma}^\nu]_{q''q'''}^{qq'}
S_{p'q'}
\tilde{v}_{pq}^{q''' q''}\\
&+[\bar{\gamma}^\nu]_{qq'}
[\bar{\Gamma}^\mu]_{p''p'''}^{pp'}
S_{q'p'}
\tilde{v}_{pq}^{p'' p'''}\\
&+[\bar{\Gamma}^\mu]_{p''p'''}^{pp'}
[\bar{\Gamma}^\nu]_{q''q'''}^{qq'}
S_{p'q'''}
S_{q'p'''}
\tilde{v}_{pq}^{p'' q''}\\
&-E_{\mathrm{pol}}^{(1)}S_{pq'} [\bar{\gamma}^\nu]_{qq'} S_{qp'} [\bar{\gamma}^\mu]^{pp'}
),
\end{split}
\label{eq:exch_dm_func}
\end{equation}
with the understanding that $\gamma^0 = \gamma$ etc.,
so that for the exchange-dispersion contribution we have 
\begin{equation}
    \langle \Psi_A^0 \Psi_B^0|(\hat{V}-\bar{V})\hat{P}|\Psi_A^\mu\Psi_B^\nu\rangle = F^{\mu\nu} = F[\bar{\gamma}_A^\mu,\bar{\Gamma}_A^\mu,\bar{\gamma}_B^\nu,\bar{\Gamma}^\nu_B],
\end{equation}
which follows by inspection.

With these notational conventions we can evaluate exchange-induction term we have (in the $S^2$ approximation)
\begin{equation}
    E_{\mathrm{exch-ind}}(B\rightarrow A)
    = \langle \Psi_A^0 \Psi_B^0|(\hat{V}-\bar{V})(\hat{P}-\bar{P})|\Psi_A^{(\mathrm{ind})}\Psi_B^0\rangle.\label{eq:exch_ind2}
\end{equation}

Inserting \cref{eq:psi_ind} into \cref{eq:exch_ind2} and breaking the expectation value into two terms we find (dropping $B\rightarrow A$ notation)
\begin{widetext}
\begin{equation}
E_{\mathrm{exch-ind}}
= -\sum_{\mu}
\frac{\langle \Psi_A^0 \Psi_B^0|(\hat{V}-\bar{V})(\hat{P}-\bar{P})
|\Psi^\mu_A\Psi_B^0\rangle\langle\Psi^\mu_A\Psi_B^0|\hat{\Omega}^B|\Psi_A^0\Psi_B^0\rangle}
{E_A^\mu-E_A^0}.
\label{eq:exch_ind_sos}
\end{equation}
\end{widetext}

Let us first look at the term in the numerator, $\langle \Psi_A^0 \Psi_B^0|(\hat{V}-\bar{V})(\hat{P}-\bar{P})
|\Psi^\mu_A\Psi_B^0\rangle$ which we will break down into two pieces:
\begin{equation}
\langle \Psi_A^0 \Psi_B^0|(\hat{V}-\bar{V})\hat{P}
|\Psi^\mu_A\Psi_B^0\rangle \label{eq:exch_ind_exch_t1}
\end{equation}
and
\begin{equation}
    \bar{P}\langle \Psi_A^0 \Psi_B^0|(\hat{V}-\bar{V})
|\Psi^\mu_A\Psi_B^0\rangle \label{eq:exch_ind_exch_t2}.
\end{equation}

\cref{eq:exch_ind_exch_t1} is given by $F^{\mu}$ as discussed previously
while \cref{eq:exch_ind_exch_t2}
is simply (note $\langle\Psi_A^0|\Psi_A^\mu\rangle = 0$ so the $\bar{P}\bar{V}$ contribution is zero)
\begin{align}
    \bar{P}\langle \Psi_A^0 \Psi_B^0|\hat{V} |\Psi^\mu_A\Psi_B^0\rangle &=
    \bar{P}\langle \Psi_A^0 \Psi_B^0|\hat{\Omega}|\Psi^\mu_A\Psi_B^0\rangle\\
    = \bar{P} \bar{\gamma}_{pp'}^\mu \Omega_{pp'}^B
\end{align}
and by inspection with \cref{eq:exch}
\begin{align}
    \bar{P} \equiv
    \langle \Psi_A^0 \Psi_B^0|\hat{P} |\Psi_A^0\Psi_B^0\rangle 
    &=
    \frac{1}{N_A N_B} \int d\mathrm{x}\ d\mathrm{x}' \ \gamma_{\mathrm{int}}(\mathbf{x},\mathbf{x}')\\
    &= -\frac{1}{2} S_{pq'} \bar{\gamma}_{qq'} S_{qp'} \bar{\gamma}_{pp'}\label{eq:pbar}.
\end{align}
So, putting all of this together we have
\begin{align}
    E_{\mathrm{exch-ind}}(B\rightarrow A)
    =
    -\left(\frac{F^\mu t^\mu}{\omega_A^\mu}
    + E_{\mathrm{ind}}(B\rightarrow A) \bar{P} \right),
\end{align}
where
\begin{equation}
    t^\mu = \gamma_{pp'}^\mu \Omega^B_{pp'},
\end{equation}
and we have used the fact that
\begin{equation}
    \frac{(\bar{\gamma}^\mu_{pp'}\Omega^B_{pp'})^2}{\omega^\mu_A} \equiv -E_\mathrm{ind}(B\rightarrow A).
\end{equation}
The full expression is then
\begin{widetext}
\begin{equation}
    E_{\mathrm{exch-ind}} = E_{\mathrm{exch-ind}}(B\rightarrow A) +  E_{\mathrm{exch-ind}}(A\rightarrow B) = 
    -\left(\frac{F^\mu t^\mu}{\omega_A^\mu}
    +\frac{F^\nu t^\nu}{\omega_B^\nu}
+ E_{\mathrm{ind}}\bar{P} \right).\label{eq:exch_indu_full}
\end{equation}
\end{widetext}

The second-order exchange-dispersion energy can be found in a similar fashion. 
We can start from the definition in the $S^2$ approximation
\begin{equation}
    E_{\mathrm{exch-disp}}(S^2)
    = \langle \Psi_A^0 \Psi_B^0|(\hat{V}-\bar{V})(\hat{P}-\bar{P})|\Psi^{\mathrm{disp}}_{AB}\rangle,\label{eq:exch_disp2}
\end{equation}
In contrast to the induction term, the dispersion interaction includes terms where both monomers are in an excited state.
Much like the exchange-induction interaction we can split \cref{eq:exch_disp2} into two terms.
Inserting Eq.~\cref{eq:psi_disp} into \cref{eq:exch_disp2} we first have to evaluate
\begin{equation}
    \langle \Psi_A^0 \Psi_B^0|(\hat{V}-\bar{V})\hat{P}|\Psi_A^\mu\Psi_B^\nu\rangle = F^{\mu\nu} = F[\bar{\gamma}_A^\mu,\bar{\Gamma}_A^\mu,\bar{\gamma}_B^\nu,\bar{\Gamma}^\nu_B],
\end{equation}
which follows by inspection.
We also identify the usual induction numerator term
\begin{equation}
    \langle \Psi_A^0 \Psi_B^0|V_{ee}|\Psi_A^\mu\Psi_B^\nu\rangle = v_{pq}^{p'q'} \gamma_{pp'}^\mu \gamma_{qq'}^\nu = t^{\mu\nu}.
\end{equation}
Noting that
\begin{equation}
    E^{(2)}_\mathrm{disp} = - \sum_{\mu\nu} \frac{(t^{\mu\nu})^2}{\omega_A^\mu+\omega_B^\nu}
\end{equation}
and combining all of these with \cref{eq:pbar} we find
\begin{equation}
E_{\mathrm{exch-disp}}
=
-\left(\frac{F^{\mu\nu}t^{\mu\nu}}{\omega_A^\mu+\omega_B^\nu} + E_{\mathrm{disp}}\bar{P} \right).
\end{equation}
\section{Extended Random Phase Approximation \label{supl:erpa}}

Evaluating the SAPT expressions given in \cref{supl:sapt_theory} requires a knowledge of the one- and two-particle transition density matrices at a given level of theory.
There are two challenges to achieving this given current NISQ hardware.
First is that we can only simulate a small fragment of the problem in a quantum computer, which is achieved here using an active space approach.
Second is that in principle we would need to compute all eigenvalues and excited states to compute the transition density matrices which is significantly more challenging than just computing the ground state energy.

To overcome these issues, we instead include only the subset of particle-hole excitations. In particular, we follow Hapka {\it et al.}~\citep{hapka2018second,hapka2019second,hapka2021casscf} and use the extended random phase approximation~\citep{ChatterjeeERPA2012,Vanaggelen2013,pernal2014accurate} (ERPA) to approximately determined excited state properties from CASSCF quality wavefunctions.
For notational clarity in this section we will not distinguish between monomers, and orbitals are general spatial orbitals.

Following Pernal~\citep{ChatterjeeERPA2012,pernal2014accurate}, in the ERPA solves the following eigenvalue problem 
\begin{equation}
\begin{bmatrix}
A & B\\
B & A
\end{bmatrix}
\begin{bmatrix}
  X_{\nu}\\
  Y_{\nu}
  \end{bmatrix}
=\omega_{\nu}
\begin{bmatrix}
  -N & 0 \\
  0 & N
  \end{bmatrix}
\begin{bmatrix}
  X_{\nu}\\
  Y_{\nu}
\end{bmatrix}\label{eq:erpa}.
\end{equation}
where
\begin{align}
A_{pq,rs} &= \langle \Psi_0 |[p^\dagger q, [\hat{H}, s^\dagger r] |\Psi_0 \rangle \\
B_{pq,rs} &= \langle \Psi_0| [p^\dagger q, [\hat{H}, r^\dagger s]|\Psi_0 \rangle \\
N_{pq,rs} &= \langle \Psi_0 | [p^\dagger q, s^\dagger r] | \Psi_0\rangle = \delta_{ps} \delta_{qr}(n_q - n_p),
\end{align}
where the above equations are valid in the natural spin-orbital basis corresponding to $|\Psi_0\rangle$ and we take $p>q$ and $r>s$. 
Computing the Hessian matrices in~\cref{eq:erpa} is somewhat tedious but expressions are available in the literature~\citep{ChatterjeeERPA2012,pernal2014accurate}.

Note that using excited states from the ERPA introduces two approximations in the quality of the resulting SAPT interaction energies. 
The first, is that even with an exact FCI wavefunction the ERPA will not yield all the excited states of the Hamiltonian as only particle-hole-like excitations are included.
The second issue is that the CASSCF wavefunction is not an eigenstate of the zeroth order Hamiltonian in the SAPT perturbation theory.
In principle, this issue can be addressed within the context of the adiabatic-connection formalism based upon the ERPA~\citep{pernal2018electron,pastorczak2018correlation,hapka2018second,hapka2019second,hapka2021casscf} which yields the so-called coupled approximation used in this work. This amounts to using the full Hamiltonian in~\cref{eq:erpa} with the CASSCF (or VQE) wavefunction as $|\Psi_0\rangle$. The adiabatic-connection approach assumes the one- and two-particle reduced density matrices remain constant across the adiabatic-connection pathway.
This approximation appears to perform well in the sense that SAPT(CASSCF) compares favourably to FCI interaction energies~\citep{hapka2018second,hapka2019second,hapka2021casscf}.

\subsection{Spin-Summation}
In this work we only target singlet excited states and thus $X_{\beta\alpha} = X_{\alpha\beta} = 0$, thus we explicitly form the spin-restricted ERPA equations.  
In particular we form
\begin{align}
A_{pq,rs} &= \langle \Psi_0 |\hat{E}_{pq}, [\hat{H}, \hat{E}_{sr}] |\Psi_0 \rangle \label{eq:amat_erpa}\\
B_{pq,rs} &= \langle \Psi_0| [\hat{E}_{pq}, [\hat{H}, \hat{E}_{rs}]|\Psi_0 \rangle \label{eq:bmat_erpa} \\
N_{pq,rs} &= \langle \Psi_0 | [\hat{E}_{pq}, \hat{E}_{sr}] | \Psi_0\rangle = \delta_{ps} \delta_{qr}(n_q - n_p),
\end{align}
where $\hat{E}_{rs} = (r^{\dagger}s + \bar{r}^\dagger \bar{s})$.
It is helpful to note the following identities:
\begin{align}
    [A,BC]  &= A\{B,C\} - \{A,C\}B\\
    [AB,CD] &= [AB, C] D + C[AB, D]
\end{align}
So, for example,
\begin{equation}
\begin{split}
[\hat{E}_{rs},\hat{E}_{sr}]
&=
\sum_{pq'}
[r^\dagger s + \bar{r}^\dagger \bar s, q^\dagger p + \bar{q}^\dagger \bar{p}]
\\
&= 
\sum_{pq'}
[r^\dagger s , q^\dagger p]
+
[r^\dagger s , \bar q^\dagger \bar p]
+
[\bar r^\dagger \bar s , q^\dagger p]
+
[\bar r^\dagger \bar s , \bar q^\dagger \bar p],
\end{split}
\end{equation}
and
\begin{align}
[r^\dagger s , q^\dagger p]
&=
[r^\dagger  s, q^\dagger] p + q^\dagger [r^\dagger s, p]
\\
&=
(r^\dagger \{s,q^\dagger\} - \{r^\dagger, q^\dagger\}s)p
+ q^\dagger (r^\dagger \{s,p\} - \{r^\dagger, p\}s)\\
&=
r^\dagger p\delta_{sq}
-q^\dagger s \delta_{rp}\\
[\bar r^\dagger \bar s , \bar q^\dagger \bar p]
&=
\bar r^\dagger \bar p\delta_{\bar s \bar q}
-\bar q^\dagger \bar s \delta_{\bar r \bar p}\\
    [\bar r^\dagger \bar s , q^\dagger p] &= 
[r^\dagger s , \bar q^\dagger \bar p] = 
0
\end{align}

Similarly\citep{helgaker2014molecular},
\begin{align}
[\hat{E}_{mn}, \hat{E}_{pq}]
&=
\hat{E}_{mq}\delta_{pn} - \hat{E}_{pn}\delta_{mq} \\
[\hat{E}_{rs}, [\hat{E}_{ij}, \hat{E}_{qp}]]
&=
[\hat{E}_{rs}, \hat{E}_{ip}]\delta_{qj} - [\hat{E}_{rs}, \hat{E}_{qj}]\delta_{ip}\\
&=
\hat{E}_{rp}\delta_{is}\delta_{qj} - \hat{E}_{is}\delta_{rp}\delta_{qj}
-
\hat{E}_{rj}\delta_{qs}\delta_{ip} + \hat{E}_{qs}\delta_{rj}\delta_{ip}
\\
[\hat{E}_{mn}, \hat{e}_{pqrs}]
&=
\delta_{pn}\hat{e}_{mqrs} - \delta_{mq} \hat{e}_{pnrs} + \delta_{rn} \hat{e}_{pqms} - \delta_{ms} \hat{e}_{pqrn}\label{eq:gamma_commutator},
\end{align}
where $\hat{e}_{pqrs} = p^\dagger r^\dagger s q + p^\dagger \bar{r}^\dagger \bar{s} q + \bar{p}^\dagger r^\dagger s \bar{q} + \bar{p}^\dagger \bar{r}^\dagger \bar{s} \bar{q}$.
We can now evaluate the commutator on the left hand side of  \cref{eq:amat_erpa} where we take 
\begin{equation}
\hat{H} = \sum_{pq} h_{pq} (p^\dagger q + \bar{p}^\dagger \bar{q})+ \frac{1}{2}\sum_{pqrs} (pq|rs) \hat{e}_{pqrs}.
\end{equation}
Let us begin with the one-body term (dropping hats on operators for brevity, and here all orbital labels refer to general spatial orbitals, no distinction is made between active, core or inactive)
\begin{widetext}
\begin{align}
\sum_{ij} h_{ij} [E_{rs}, [E_{ij}, E_{qp}]]
&=
\sum_{ij} h_{ij}(
\hat{E}_{rp}\delta_{is}\delta_{qj} - \hat{E}_{is}\delta_{rp}\delta_{qj}
-
\hat{E}_{rj}\delta_{qs}\delta_{ip} + \hat{E}_{qs}\delta_{rj}\delta_{ip}
)
\\
&=
h_{sq}\hat{E}_{rp}
-
\sum_i h_{iq}\hat{E}_{is}\delta_{rp}
-\sum_j h_{pj} \hat{E}_{rj} \delta_{qs}
+
h_{pr} \hat{E}_{qs}
\end{align}
\end{widetext}

If we work in the natural orbital basis $\langle \Psi_0 | \hat{E}_{pq} | \Psi_0\rangle = \bar{\gamma}_{pq} = n_p \delta_{pq}$, with $0 \le n_p \le 2 $\footnote{Note we assume $n_p < n_q$ for $p > q$ (i.e. the basis is ordered in descending order of natural orbital occupancy) which is the opposite ordering that is usually taken in the ERPA literature~\citep{ChatterjeeERPA2012,pernal2014accurate}}, then we have
\begin{widetext}

\begin{align}
\langle \sum_{ij} h_{ij} [E_{rs}, [E_{ij}, E_{qp}]] \rangle &=
h_{sq} \delta_{rp} (n_r - n_s) + h_{pr} \delta_{qs} (n_q-n_r).
\end{align}

For the two-body part we have to evaluate expressions like (see \cref{eq:bmat_erpa})
\begin{align}
\sum_{ijkl} (ij|kl) [E_{rs}, [e_{ijkl}, E_{pq}]] &
\end{align}
Let us look at
\begin{align}
[E_{rs}, [e_{ijkl}, E_{pq}]]
&=-[E_{rs}, [E_{pq}, e_{ijkl}]]
=
-[E_{rs},
\delta_{iq}\hat{e}_{pjkl} - \delta_{pj} \hat{e}_{iqkl} + \delta_{kq} \hat{e}_{ijpl} - \delta_{pl} \hat{e}_{ijkq}]\\
&=
-[E_{rs},
\delta_{iq}\hat{e}_{pjkl} - \delta_{pj} \hat{e}_{iqkl} + \delta_{kq} \hat{e}_{ijpl} - \delta_{pl} \hat{e}_{ijkq}]
\end{align}
\begin{align}
[\hat{E}_{rs}, \delta_{iq} \hat{e}_{pjkl}] &= \delta_{iq}(
\delta_{ps}\hat{e}_{rjkl}
-\delta_{rj}\hat{e}_{pskl}
+\delta_{ks}\hat{e}_{pjrl}
-\delta_{rl}\hat{e}_{pjks}
)
\\
[\hat{E}_{rs}, \delta_{pj} \hat{e}_{iqkl}] &= \delta_{pj}(
\delta_{is}\hat{e}_{rqkl}
-\delta_{rq}\hat{e}_{iskl}
+\delta_{ks}\hat{e}_{iqrl}
-\delta_{rl}\hat{e}_{iqks}
)
\\
[\hat{E}_{rs}, \delta_{kq} \hat{e}_{ijpl}] &= \delta_{kq}(
\delta_{is}\hat{e}_{rjpl}
-\delta_{rj}\hat{e}_{ispl}
+\delta_{ps}\hat{e}_{ijrl}
-\delta_{rl}\hat{e}_{ijps}
)
\\
[\hat{E}_{rs}, \delta_{pl} \hat{e}_{ijkq}] &= \delta_{pl}(
\delta_{is}\hat{e}_{rjkq}
-\delta_{rj}\hat{e}_{iskq}
+\delta_{ks}\hat{e}_{ijrq}
-\delta_{rq}\hat{e}_{ijks}
)
\end{align}
so,
\begin{equation}
\begin{split}
\langle \sum_{ijkl} (ij|kl) [E_{rs}, [e_{ijkl}, E_{pq}]] \rangle
=
-\sum_{ijkl} (ij|kl)
&\bigg[
\delta_{iq}(
\delta_{ps}\bar{\Gamma}_{rjkl}
-\delta_{rj}\bar{\Gamma}_{pskl}
+\delta_{ks}\bar{\Gamma}_{pjrl}
-\delta_{rl}\bar{\Gamma}_{pjks}
)
\\
&
-
\delta_{pj}(
\delta_{is}\bar{\Gamma}_{rqkl}
-\delta_{rq}\bar{\Gamma}_{iskl}
+\delta_{ks}\bar{\Gamma}_{iqrl}
-\delta_{rl}\bar{\Gamma}_{iqks}
)\\
&
+
\delta_{kq}(
\delta_{is}\bar{\Gamma}_{rjpl}
-\delta_{rj}\bar{\Gamma}_{ispl}
+\delta_{ps}\bar{\Gamma}_{ijrl}
-\delta_{rl}\bar{\Gamma}_{ijps}
)\\
&
-
\delta_{pl}(
\delta_{is}\bar{\Gamma}_{rjkq}
-\delta_{rj}\bar{\Gamma}_{iskq}
+\delta_{ks}\bar{\Gamma}_{ijrq}
-\delta_{rq}\bar{\Gamma}_{ijks}
)
\bigg],
\end{split}
\end{equation}
where the spin-summed two-particle reduced density matrix is
\begin{equation}
\bar{\Gamma}_{pqrs} = \langle \Psi_0 | \hat{e}_{pqrs} | \Psi_0\rangle.
\end{equation}

Removing the Kroenecker deltas we find
\begin{equation}
\begin{split}
\frac{1}{2}\langle \sum_{ijkl} (ij|kl) & [E_{rs}, [e_{ijkl}, E_{pq}]] \rangle
= \\
&-
\frac{1}{2}
\bigg[
\sum_{jkl} (qj|kl)
\delta_{ps}\bar{\Gamma}_{rjkl}
-\sum_{kl} (qr|kl)
\bar{\Gamma}_{pskl}
+\sum_{jl} (qj|sl)
\bar{\Gamma}_{pjrl}
-\sum_{jk} (qj|kr)
\bar{\Gamma}_{pjks}
\\
&
-
\sum_{kl} (sp|kl)
\bar{\Gamma}_{rqkl}
+
\sum_{ikl} (ip|kl)
\delta_{rq}\bar{\Gamma}_{iskl}
-
\sum_{il} (ip|sl)
\bar{\Gamma}_{iqrl}
+
\sum_{ik} (ip|kr)
\bar{\Gamma}_{iqks}
\\
&
+
\sum_{jl} (sj|ql)
\bar{\Gamma}_{rjpl}
-
\sum_{il} (ir|ql)
\bar{\Gamma}_{ispl}
+
\sum_{ijl} (ij|ql)
\delta_{ps}\bar{\Gamma}_{ijrl}
-
\sum_{ij} (ij|qr)
\bar{\Gamma}_{ijps}
\\
&
-
\sum_{jk} (sj|kp)
\bar{\Gamma}_{rjkq}
+
\sum_{ik} (ir|kp)
\bar{\Gamma}_{iskq}
-
\sum_{ij} (ij|sp)
\bar{\Gamma}_{ijrq}
+
\sum_{ijk} (ij|kp)
\delta_{rq}\bar{\Gamma}_{ijks}
\bigg]
\end{split}
\label{eq:rhs_erpa}
\end{equation}
\end{widetext}

which can be simplified a using symmetries in the ERIs and TPDMs.
Note \cref{eq:rhs_erpa} reduces to the expression given in Ref.\citenum{ChatterjeeERPA2012} if working in the natural spin orbital basis.

\subsection{Numerical Solution }
With the expressions for the Hessians at hand we can now solve \cref{eq:erpa}.
First, it is helpful to bring this expression into a symmetric eigenvalue problem of half the dimension.
This has two advantages, namely the computational cost will be reduced by a factor of eight and we can use a symmetric eigenvalue solver.

To proceed it is important to note some features \cref{eq:erpa}.
First we choose the normalization:
\begin{align}
    Y_\nu^{\dagger} N Y_\nu
    -
    X_\nu^{\dagger} N X_\nu
    &= 1,\\
    (Y-X)^{\dagger} N (Y+X) &= 1.\label{eq:erpa_norm_2}
\end{align}
Next, we note that
\begin{align}
    (A+B)(X+Y) &= \omega_n N(Y-X)\\
    (A-B)(Y-X) &= \omega_n N(X+Y),
\end{align}
which implies
\begin{align}
    (A+B)(X+Y) = \omega_n^2 N (A-B)^{-1} N (X+Y)
\end{align}
or
\begin{widetext}
\begin{align}
    (A+B)N^{-1/2}N^{1/2}(X+Y) &= \omega_n^2 N^{1/2}N^{1/2} (A-B)^{-1} N^{1/2} N^{1/2} (X+Y)\\
    A_+ T_n &= \omega_n^2 A_-^{-1} T_n\\
    A_- A_+ T_n &= \omega_n^2 T_n
\end{align}
\end{widetext}
if $(A-B)$ is positive definite.

If $(A+B)$ is also positive definite we can write
\begin{align}
    A_+^{1/2} A_+^{1/2} A_- A_+^{1/2} A_+^{1/2} T_n &= \omega_n^2 A_+ T_n \\
    A_+^{1/2} A_- A_+^{1/2} A_+^{1/2} T_n &= \omega_n^2 A_+^{1/2} T_n \\
    A_+^{1/2} A_- A_+^{1/2} Q_n &= \omega_n^2 Q_n \\
\end{align}
where
\begin{equation}
    Q_n = A_+^{1/2} T_n = A_+^{1/2} N^{1/2} (X+Y),
\end{equation}
$Q^{\dagger}Q = 1$, and
\begin{equation}
    X + Y = N^{-1/2} A_+^{-1/2} Q_n \label{eq:erpa_xy_def}.
\end{equation}
This allows us to write combinations of the $X$ and $Y$ matrices necessary for building transition density matrices in terms of the solutions of the Hermitian eigenvalue problem.
The factors of $N^{-1/2}$ can be determined using canonical orthogonalization with a threshold to discard small eigenvalues.
\subsection{Normalization}
A useful sanity check at this point is that we have
\begin{align}
    (A+B)(X+Y) &= \omega_n N(Y-X) \label{eq:xy_norm_sub}.
\end{align}
Multiplying on the right by $(X+Y)^{\dagger}$ we have and using \cref{eq:erpa_norm_2}
\begin{align}
    (X+Y)_n(A+B)(X+Y)_n &= \omega_n (X+Y)N(Y-X) \\
    (X+Y)_n(A+B)(X+Y)_n &= \omega_n \label{eq:eig_check}
\end{align}
Inserting \cref{eq:erpa_xy_def} in \cref{eq:eig_check} we find
\begin{widetext}
\begin{align}
    (X+Y)_n(A+B)(X+Y)_n &= Q_n A_+^{-1/2} N^{-1/2} (A+B) N^{-1/2} A_+^{-1/2} Q_n =  Q_n A_+^{-1/2} A_+ A_+^{-1/2} Q_n = 1 \ne \omega!
\end{align}
\end{widetext}

A more consistent normalization of \cref{eq:erpa_xy_def} can be found by
\begin{equation}
    X + Y \equiv T_+ = \sqrt{\omega_n} N^{-1/2} A_+^{-1/2} Q_n \label{eq:erpa_xy_def2}
\end{equation}
and by \cref{eq:erpa_norm_2} we can deduce
\begin{equation}
    Y - X \equiv T_- = \frac{1}{\sqrt{\omega_n}} N^{-1/2} A_+^{1/2} Q_n \label{eq:erpa_xy_def3}
\end{equation}
from which it follows that
\begin{align}
    X &= \frac{1}{2}(T_+-T_-) \\
    Y &= \frac{1}{2}(T_++T_-).
\end{align}
To compute the symmetrized (spin summed) transition density matrix we follow Pernal \citep{ChatterjeeERPA2012,pernal2014intergeminal} and use the identity
\begin{align}
    \bar{\gamma}^{\nu}_{pq} &= \langle \Psi_0|\hat{E}_{pq}|\Psi_\nu\rangle = \langle 0 | [ \hat{E}_{pq}, \hat{O}^\dagger_\nu ] | 0 \rangle,
\end{align}
then
\begin{align}
    \gamma_{pq}^\nu + \gamma_{qp}^\nu &= [N(Y-X)]_{pq} \qquad \forall_{p>q}.
\end{align}
For the transition two-particle reduced density matrix we have
\begin{align}
    \bar{\Gamma}^\nu_{pqrs} =  \langle\Psi_0 | [\hat{e}_{pqrs}, \hat{O}_\nu^\dagger]|\Psi_0\rangle.
\end{align}
Using \cref{eq:gamma_commutator} and looking at the $X$ part of $\hat{O}_\nu^\dagger$ first we have
\begin{widetext}
\begin{align}
    \bar{\Gamma}^{\nu (X)}_{pqrs}
    &= -\sum_{m > n} X_{mn}^\nu \langle [\hat{E}_{mn}, \hat{e}_{pqrs}] \rangle \\
    &= -\sum_{m > n} X_{mn}^\nu \left(\delta_{pn}\bar{\Gamma}_{mqrs} - \delta_{mq} \bar{\Gamma}_{pnrs}+\delta_{rn}\bar{\Gamma}_{pqms} - \delta_{ms} \bar{\Gamma}_{pqrn}\right)
\end{align}
and similarly we have
\begin{align}
    \bar{\Gamma}^{\nu (Y)}_{pqrs}
    &= -\sum_{m > n} Y_{mn}^\nu \langle [\hat{E}_{nm}, \hat{e}_{pqrs}] \rangle \\
    &= -\sum_{m > n} Y_{mn}^\nu \left(\delta_{pm}\bar{\Gamma}_{nqrs} - \delta_{nq} \bar{\Gamma}_{pmrs}+\delta_{rm}\bar{\Gamma}_{pqns} - \delta_{ns} \bar{\Gamma}_{pqrm}\right)
\end{align}
and naturally $\bar{\Gamma}^{\nu }_{pqrs}= \bar{\Gamma}^{\nu (X)}_{pqrs} + \bar{\Gamma}^{\nu (Y)}_{pqrs}$.

It is helpful remove the Kroenecker deltas so that 
\begin{equation}
\begin{split}
    \bar{\Gamma}_{pqrs}^\nu
    =
    &-\left(\sum_{m=p+1}^{M} X^\nu_{mp} \bar{\Gamma}_{mqrs} + \sum_{m=0}^{p} Y^\nu_{pm} \bar{\Gamma}_{mqrs}\right)\\
    &+ \left(\sum_{m=0}^{q} X^\nu_{qm} \bar{\Gamma}_{pmrs} + \sum_{m=q+1}^{M} Y^\nu_{mq} \bar{\Gamma}_{pmrs}\right)\\
    &-\left(\sum_{m=r+1}^{M} X^\nu_{mr} \bar{\Gamma}_{pqms} + \sum_{m=0}^{r} Y^\nu_{rm} \bar{\Gamma}_{pqms}\right)\\
    &+\left(\sum_{m=0}^{s} X^\nu_{sm} \bar{\Gamma}_{pqrm} + \sum_{m=s+1}^{M} Y^\nu_{sm} \bar{\Gamma}_{pqrm}\right),
\end{split}
\end{equation}
\end{widetext}

where $M$ is the number of (natural) spatial orbitals.
If we define
\begin{align}
    Q_{pq}^\nu &= X_{pq}^\nu  \ \forall \ p > q \\
    Q_{pq}^\nu &= Y_{qp}^\nu \ \ \forall \ q > p \\
    R_{pq}^\nu &= Y_{pq}^\nu \ \ \forall \ p > q \\
    R_{pq}^\nu &= X_{qp}^\nu \ \forall \ q > p
\end{align}
then we have
\begin{equation}
    \bar{\Gamma}_{pqrs}^\nu = \sum_m \left(-Q^\nu_{mp} \bar{\Gamma}_{mqrs} + R^\nu_{mq} \bar{\Gamma}_{pmrs} - Q^\nu_{mr} \bar{\Gamma}_{pqms} + R^\nu_{ms} \bar{\Gamma}_{pqrm} \right).\label{eq:trans_tpdm_intermediates}
\end{equation}
A helpful reference point to benchmark is the Hartree--Fock limit $|\Psi_0\rangle \rightarrow |\mathrm{RHF}\rangle$
then
\begin{align}
    \bar{\Gamma}_{pqrs}
    &=
    \bar{\gamma}_{pq}\bar{\gamma}_{rs} - \frac{1}{2} \bar{\gamma}_{ps}\bar{\gamma}_{rq} \\
    &=
    n_{q}\delta_{pq} n_s \delta_{rs} - \frac{1}{2} n_s\delta_{ps} n_{q}\delta_{rq},\label{eq:trans_tpdm_hf}
\end{align}
where $n_i = 2$ and $n_a=0$ for $i$ and occupied MO and $a$ a virtual MO.
Let us first insert the Coulomb like terms from \cref{eq:trans_tpdm_hf} into the first two terms of \cref{eq:trans_tpdm_intermediates}.
We have
\begin{widetext}
\begin{align}
    \sum_m -Q_{mp} \bar{\Gamma}_{mqrs} + R_{mq} \bar{\Gamma}_{pmrs}
    &\rightarrow
    \sum_m (R_{mq} n_p \delta_{mp} n_s - Q_{mp} n_q \delta_{mq} ) n_s \delta_{rs} \\
    &= (R_{pq} n_p - Q_{qp} n_q) n_s \delta_{rs}.
\end{align}
\end{widetext}

Now for $p>q$ we have
\begin{align}
    \sum_m -Q_{mp} \bar{\Gamma}_{mqrs} + R_{mq} \bar{\Gamma}_{pmrs}
    &\rightarrow (Y_{pq}^\nu n_p - Y_{pq} n_q) n_s \delta_{rs}\\
    &= (n_p-n_q)Y_{pq}n_s\delta_{rs}
\end{align}
\begin{widetext}
and for $p<q$ we have
\begin{align}
    \sum_m -Q_{mp} \bar{\Gamma}_{mqrs} + R_{mq} \bar{\Gamma}_{pmrs}
    &\rightarrow (X_{qp}^\nu n_p - X_{qp} n_q) n_s \delta_{rs}\\
    &= (n_p-n_q)X_{qp}n_s\delta_{rs} \\
    &= -(n_p-n_q)X_{pq}n_s\delta_{rs} \ \forall \ p > q
\end{align}
\end{widetext}
where we relabelled $p$ and $q$ in the last line.
Thus we have
\begin{equation}
    \sum_m -Q_{mp} \bar{\Gamma}_{mqrs} + R_{mq} \bar{\Gamma}_{pmrs}
    \xrightarrow[]{\text{Coulomb-like}}
    \gamma_{pq}^\nu n_s \delta_{rs},
\end{equation}
and by inspection
\begin{equation}
    \sum_m -Q_{mr} \bar{\Gamma}_{pqms} + R_{ms} \bar{\Gamma}_{pqrm}
    \xrightarrow[]{\text{Coulomb-like}}
    \gamma_{rs}^\nu n_q \delta_{pq}.
\end{equation}
For the exchange terms we just need to set $q\rightarrow s$ so putting it together we find
\begin{equation}
    \bar{\Gamma}^\nu_{pqrs}
    =
    \gamma_{pq}^\nu n_{s} \delta_{rs}
    + \gamma_{rs}^\nu n_p \delta_{pq}
    - \frac{1}{2}\gamma_{ps}^\nu n_q \delta_{rq}
    - \frac{1}{2}\gamma_{rq}^\nu n_s \delta_{ps}.
\end{equation}
This expression is useful for checking that exchange-induction and exchange-dispersion terms reduce to their appropriate SAPT0 expressions. 

\subsection{Numerical Implementation of SAPT Expressions Within the ERPA}\label{sec:numsapt}
In principle one could proceed to evaluate the SAPT expressions in \cref{supl:sapt_theory} by building the transition one- and two-particle density matrices.
Indeed, this is possible for the first-order SAPT terms that do not contain any response term, and the second-order polarization energies which are functions of the transition one-particle reduced density matrix only.
However, forming the transition two-particle density matrix, which is required for exchange-induction and exchange-induction is not practical in general so some work needs to be done. We follow Refs.~\citenum{hapka2019second,hapka2021casscf} and try to optimize the contraction order of the intermediate tensors.
Let us first look at the exchange-induction term.
The first two terms only contain the one-particle (transition) density matrix; we start with the derivation of the third term:
\begin{align}
    F^\mu_{3a} &= -\sum_{m>p} \tilde{Q}^
    \mu_{mp}
    \bar{\Gamma}^{mp'}_{p''p'''} \bar{\gamma}_{qq'} S_{q'p'}\tilde{v}^{p''p'''}_{pq}.
\end{align}
Let
\begin{equation}
    T^{mp'}_{pq} = \bar{\Gamma}^{mp'}_{p''p'''} \tilde{v}_{pq}^{p''p'''}
\end{equation}
and
\begin{equation}
    N_{p'q} = \bar{\gamma}_{qq'}S_{q'p'}
\end{equation}
then
\begin{equation}
    W_{pm} = T_{pq}^{mp'} N_{p'q}
\end{equation}
so
\begin{equation}
    F_{3a}^\mu = -\sum_{m>p} \tilde{Q}^
    \mu_{mp} W_{pm}.
\end{equation}
\begin{align}
    F^\mu_{3b} &= \sum_{m>p''} \tilde{R}^
    \mu_{mp''}
    \bar{\Gamma}^{pp'}_{mp'''} \bar{\gamma}_{qq'} S_{q'p'}\tilde{v}^{p''p'''}_{pq}\\
               &= \sum_{m>p''} \tilde{R}^\mu_{mp''} \left(\bar{\Gamma}_{mp'''}^{pp'}\tilde{v}_{pq}^{p''p'''}\right)\left(\bar{\gamma}_{qq'}S_{q'p'}\right) \\
               &= \sum_{m>p''} \tilde{R}^\mu_{mp''} T_{mq}^{p''p'} N_{p'q} \\
               &= \sum_{m>p''} \tilde{R}^\mu_{mp''} W_{p''m}.
\end{align}
Next we have
\begin{align}
    F^\mu_{3c} &= -\sum_{m>p'} \tilde{Q}^
    \mu_{mp'}
    \bar{\Gamma}^{pm}_{p''p'''} \bar{\gamma}_{qq'} S_{q'p'}\tilde{v}^{p''p'''}_{pq}\\
               &= -\sum_{m>p'} \tilde{Q}^\mu_{mp'} \left(\bar{\Gamma}_{p''p'''}^{pm}\tilde{v}_{pq}^{p''p'''}\right)\left(\bar{\gamma}_{qq'}S_{q'p'}\right) \\
               &= -\sum_{m>p'} \tilde{Q}^\mu_{mp'} T_{mq} N_{p'q} \\
               &= -\sum_{m>p'} \tilde{Q}^\mu_{mp'} W_{p'm}
\end{align}
and finally
\begin{align}
    F^\mu_{3d} &= \sum_{m>p'''} \tilde{R}^\mu_{mp'''}
    \bar{\Gamma}^{pp'}_{p'm} \bar{\gamma}_{qq'} S_{q'p'}\tilde{v}^{p''p'''}_{pq}\\
               &= \sum_{m>p'''} \tilde{R}^\mu_{mp'''} \left(\bar{\Gamma}_{p''m}^{pp'}\tilde{v}_{pq}^{p''p'''}\right)\left(\bar{\gamma}_{qq'}S_{q'p'}\right) \\
               &= \sum_{m>p'''} \tilde{R}^\mu_{mp'''} T_{mq}^{p'p'''} N_{p'q} \\
               &= \sum_{m>p'''} \tilde{R}^\mu_{mp'''} W_{p''m}.
\end{align}

For the fourth term we have to evaluate something like
\begin{equation}
F^\mu_{4a} = \left(\sum_{m > p} X_{mp}^\nu + \sum_{p > m} Y_{pm}^\nu\right)
\bar{\Gamma}^{mp'}_{p''p'''} \bar{\Gamma}_{q''q'''}^{qq'}S_{p'q'''}S_{q'p'''}\tilde{v}^{p''q''}_{pq}
\end{equation}
we can define
\begin{equation}
    U_{pp'}^{p''q'} = \bar{\Gamma}_{q''q'''}^{qq'}\tilde{v}_{pq}^{p''q''} S_{p'q'''}
\end{equation}
and
\begin{equation}
    M_{p''q'}^{mp'} = \bar{\Gamma}_{p''p'''}^{mp'} S_{q'p'''}
\end{equation}
then
\begin{equation}
    Z_{pm} = U_{pp'}^{p''q'} M_{p''q'}^{mp'}
\end{equation}
\begin{equation}
    F^\mu_{4a} = \left(\sum_{m > p} X_{mp}^\nu + \sum_{p > m} Y_{pm}^\nu\right) Z_{pm}
\end{equation}

The exchange-dispersion part is much more verbose and let us look at the most complex part which contains contributions like
\begin{widetext}
\begin{equation}
    F^{\mu\nu}_4 = \left(\sum_{m > p} X_{mp}^\mu + \sum_{p > m} Y_{pm}^\mu\right) \left(\sum_{n > q} X_{nq}^\nu + \sum_{p > n} Y_{pn}^\nu\right)
\bar{\Gamma}^{mp'}_{p''p'''} \bar{\Gamma}_{q''q'''}^{nq'}S_{p'q'''}S_{q'p'''}\tilde{v}^{p''q''}_{pq}
\end{equation}
\end{widetext}
of which there are sixteen different terms arising from all combinations of the $m$ and $n$ indices in the TPDMs.
Let us go about this systematically, and build some common intermediates (inspired by Hapka\citep{hapka2019second}). For convenience let us use upper case indices for monomer B and lower case indices for monomer A and lower all indices. Given the finite number of letters in the alphabet, intermediates will often have reuse symbols. For safety we will have to assume they are only defined locally (i.e. only for one of the 16 terms),
although by inspection many of them are identical (up to permutation of indices).
Let us first at least define the protected intermediates (similar to Hapka\citep{hapka2019second}) which can be used in practice:
\begin{align}
    N_{mqrR}^A &= \Gamma_{mqrs} S_{sR} \\
    N_{MQRr}^B &= \Gamma_{MQRS} S_{rS} \\
    Z_{mpMP}^{AB} &= N_{mqrR} N_{MQRr} \tilde{v}_{pP}^{qQ} \\
    U_{rsPQ} &= \Gamma_{pqrs} \tilde{v}_{pP}^{qQ} \\
    W_{rsRS}^{AB} &= \Gamma_{pqrs} \Gamma_{PQRS} \tilde{v}_{pP}^{qQ} \\
                  &= U_{rsPQ} \Gamma_{PQRS},
\end{align}
with obvious analogues if monomer $B$'s indices are contracted over.
\begin{align}
    \Gamma_{mqrs} \Gamma_{MQRS} S_{rS} S_{sR} \tilde{v}_{pqPQ}
     &= 
     N^A_{mqrR} N^B_{MQRr} \tilde{v}_{pqPQ} \\
     &= T^{AB}_{mqMQ} \tilde{v}_{pqPQ} \\
     &= V^{AB}_{mpMP}
\end{align}
then
\begin{align}
    F^{\mu\nu}_{4(1)}
    &= Q^\mu_{mp} Q^\nu_{MP} V^{AB}_{mpMP} \\
    &= G^{\mu}_{MP} Q^{\nu}_{MP}.
\end{align}
Now there is some pattern in this, namely we will have contributions where the contracted $Q/R$ matrix index ($m,M$) will be in either half of the 2PDM.
Let us proceed systematically along $\mu$ first then $\nu$.
For the second term we have:
\begin{align}
    \Gamma_{pmrs} \Gamma_{MQRS} S_{rS} S_{sR} \tilde{v}_{pqPQ}
     &=
     N^A_{pmrR} N^B_{MQRr} \tilde{v}_{pqPQ} \\
     &= T^{AB}_{pmMQ} \tilde{v}_{pqPQ} \\
     &= V^{AB}_{mqMP}
\end{align}
\begin{align}
    F^{\mu\nu}_{4(2)}
    &= R^\mu_{mq} Q^\nu_{MP} V^{AB}_{mqMP} \\
    &= G^{\mu}_{MP} Q^{\nu}_{MP}.
\end{align}
For the third term we have:
\begin{align}
    \Gamma_{pqms} \Gamma_{MQRS} S_{rS} S_{sR} \tilde{v}_{pqPQ}
     &=
     N^A_{pqmR} N^B_{MQRr} \tilde{v}_{pqPQ} \\
     &= T^{AB}_{mRPQ} N^B_{MQRr} \\
     &= V^{AB}_{mrMP} \ \ \text{or} \\
     &= U_{rsPQ} (S_{sR} N^B_{MQRr}) \\
     &= V^{AB}_{mrMP}
\end{align}
\begin{align}
    F^{\mu\nu}_{4(3)}
    &= Q^\mu_{mr} Q^\nu_{MP} V_{mrMP} \\
    &= G^{\mu}_{MP} Q^{\nu}_{MP}.
\end{align}
For the fourth term we have:
\begin{align}
    \Gamma_{pqrm} \Gamma_{MQRS} S_{rS} S_{sR} \tilde{v}_{pqPQ}
     &=
     N^A_{pqSm} N^B_{MQsS} \tilde{v}_{pqPQ} \\
     &= T^{AB}_{mSPQ} N^B_{MQsS} \\
     &= V^{AB}_{msMP}
\end{align}
\begin{align}
    F^{\mu\nu}_{4(4)}
    &= R^\mu_{ms} Q^\nu_{MP} V^{AB}_{msMP} \\
    &= G^{\mu}_{MP} Q^{\nu}_{MP}.
\end{align}
For the fifth term we have:
\begin{align}
    \Gamma_{mqrs} \Gamma_{PMRS} S_{rS} S_{sR} \tilde{v}_{pqPQ}
     &= 
     N^A_{mqrR} N^B_{PMRr} \tilde{v}_{pqPQ} \\
     &= T^{AB}_{mqMP} \tilde{v}_{pqPQ} \\
     &= V^{AB}_{mpMQ}
\end{align}
\begin{align}
    F^{\mu\nu}_{4(5)}
    &= Q^\mu_{mp} R^\nu_{MQ} V^{AB}_{mpMQ} \\
    &= G^{\mu}_{MQ} R^{\nu}_{MQ}.
\end{align}
For the sixth term we have:
\begin{align}
    \Gamma_{pmrs} \Gamma_{PMRS} S_{rS} S_{sR} \tilde{v}_{pqPQ}
     &=
     N^A_{pmrR} N^B_{PMRr} \tilde{v}_{pqPQ} \\
     &= T^{AB}_{pmPM} \tilde{v}_{pqPQ} \\
     &= V^{AB}_{mqMQ}
\end{align}
\begin{align}
    F^{\mu\nu}_{4(6)}
    &= R^\mu_{mq} R^\nu_{MQ} V^{AB}_{mqMQ} \\
    &= G^{\mu}_{MQ} R^{\nu}_{MQ}.
\end{align}
For the seventh term we have:
\begin{align}
    \Gamma_{pqms} \Gamma_{PMRS} S_{rS} S_{sR} \tilde{v}_{pqPQ}
     &=
     N^A_{pqmR} N^B_{PMRr} \tilde{v}_{pqPQ} \\
     &= T^{AB}_{mRPQ} N^B_{PMRr} \\
     &= V^{AB}_{mrMQ}
\end{align}
\begin{align}
    F^{\mu\nu}_{4(7)}
    &= Q^\mu_{mr} R^\nu_{MQ} V_{mrMQ} \\
    &= G^{\mu}_{MQ} R^{\nu}_{MQ}.
\end{align}
For the eight term we have:
\begin{align}
    \Gamma_{pqrm} \Gamma_{PMRS} S_{rS} S_{sR} \tilde{v}_{pqPQ}
     &=
     N^A_{pqSm} N^B_{PMsS} \tilde{v}_{pqPQ} \\
     &= T^{AB}_{mSPQ} N^B_{PMsS} \\
     &= V^{AB}_{msMQ}
\end{align}
\begin{align}
    F^{\mu\nu}_{4(8)}
    &= R^\mu_{ms} R^\nu_{MQ} V^{AB}_{msMQ} \\
    &= G^{\mu}_{MQ} Q^{\nu}_{MQ}.
\end{align}
For the ninth term we have:
\begin{align}
    \Gamma_{mqrs} \Gamma_{PQMS} S_{rS} S_{sR} \tilde{v}_{pqPQ}
     &=
     N^A_{mqrR} N^B_{PQMr} \tilde{v}_{pqPQ} \\
     &= N^A_{mqrR} T^{AB}_{Mrpq} \\
     &= V^{AB}_{mpMR}
\end{align}
\begin{align}
    F^{\mu\nu}_{4(9)}
    &= Q^\mu_{mp} Q^\nu_{MR} V^{AB}_{mpMR} \\
    &= G^{\mu}_{MR} Q^{\nu}_{MR}.
\end{align}
For the tenth term we have:
\begin{align}
    \Gamma_{pmrs} \Gamma_{PQMS} S_{rS} S_{sR} \tilde{v}_{pqPQ}
     &=
     N^A_{pmrR} N^B_{PQMr} \tilde{v}_{pqPQ} \\
     &= N^A_{pmrR} T^{AB}_{Mrpq} \\
     &= V^{AB}_{mqMR}
\end{align}
\begin{align}
    F^{\mu\nu}_{4(10)}
    &= R^\mu_{mq} Q^\nu_{MR} V^{AB}_{mqMR} \\
    &= G^{\mu}_{MR} Q^{\nu}_{MR}.
\end{align}
For the eleventh term we have:
\begin{align}
    \Gamma_{pqms} \Gamma_{PQMS} S_{rS} S_{sR} \tilde{v}_{pqPQ}
     &= T^{AB}_{msPQ} N^B_{PQMr} S_{sR} \\
     &= V^{AB}_{mrMR}
\end{align}
\begin{align}
    F^{\mu\nu}_{4(11)}
    &= Q^\mu_{mr} Q^\nu_{MR} V^{AB}_{mrMR} \\
    &= G^{\mu}_{MR} Q^{\nu}_{MR}.
\end{align}
For the twelfth term we have:
\begin{align}
    \Gamma_{pqrm} \Gamma_{PQMS} S_{rS} S_{sR} \tilde{v}_{pqPQ}
     &= T^{AB}_{mrPQ} N^B_{PQMr} S_{sR} \\
     &= V^{AB}_{msMR}
\end{align}
\begin{align}
    F^{\mu\nu}_{4(12)}
    &= R^\mu_{ms} Q^\nu_{MR} V^{AB}_{msMR} \\
    &= G^{\mu}_{MR} Q^{\nu}_{MR}.
\end{align}
For the thirteenth term we have:
\begin{align}
    \Gamma_{mqrs} \Gamma_{PQRM} S_{rS} S_{sR} \tilde{v}_{pqPQ}
     &=
     N^A_{mqSR} T^{AB}_{pqRM} \\
     &= V^{AB}_{mpMS}
\end{align}
\begin{align}
    F^{\mu\nu}_{4(13)}
    &= Q^\mu_{mp} R^\nu_{MS} V^{AB}_{mpMS} \\
    &= G^{\mu}_{MS} Q^{\nu}_{MS}.
\end{align}
For the fourteenth term we have:
\begin{align}
    \Gamma_{pmrs} \Gamma_{PQRM} S_{rS} S_{sR} \tilde{v}_{pqPQ}
     &=
     N^A_{mqSR} T^{AB}_{pqRM} \\
     &= V^{AB}_{mqMS}
\end{align}
\begin{align}
    F^{\mu\nu}_{4(14)}
    &= R^\mu_{mq} R^\nu_{MS} V^{AB}_{mqMS} \\
    &= G^{\mu}_{MS} Q^{\nu}_{MS}.
\end{align}
For the fifteenth term we have:
\begin{align}
    \Gamma_{pqms} \Gamma_{PQRM} S_{rS} S_{sR} \tilde{v}_{pqPQ}
     &= \Gamma_{pqms} S_{rS} S_{sR} T^{AB}_{pqRM} \\
     &= U^{AB}_{msRM} S_{rS} S_{sR} \\
     &= V^{AB}_{mrMS}
\end{align}
\begin{align}
    F^{\mu\nu}_{4(15)}
    &= Q^\mu_{mr} R^\nu_{MS} V^{AB}_{mrMS} \\
    &= G^{\mu}_{MS} Q^{\nu}_{MS}.
\end{align}
For the sixteenth term we have:
\begin{align}
    \Gamma_{pqrm} \Gamma_{PQRM} S_{rS} S_{sR} \tilde{v}_{pqPQ}
     &= \Gamma_{pqrm} S_{rS} S_{sR} T^{AB}_{pqRM} \\
     &= U^{AB}_{rmRM} S_{rS} S_{sR} \\
     &= V^{AB}_{msMS}
\end{align}
\begin{align}
    F^{\mu\nu}_{4(16)}
    &= R^\mu_{ms} R^\nu_{MS} V^{AB}_{msMS} \\
    &= G^{\mu}_{MS} Q^{\nu}_{MS}.
\end{align}
Naturally,
\begin{equation}
    F_4^{\mu\nu} = \sum_{i} F_{4(i)}^{\mu\nu}.
\end{equation}

\cleardoublepage
\section{Methods (Computational Details)}
All density functional theory calculations were performed with the pyscf software package\cite{sun2020recent} and used a Lebedev grid used for the numerical integration of the exchange correlation functionals with 70, 105, 140  radial grid points and 590 770, 770 angular grid points, for atoms of the first, second and fourth period respectively (grid level 5). The geometry optimizations were performed with pyscf and the geomeTRIC\cite{wang2016geometry} package. 
All CASSCF calculation were performed with  pyscf software package\cite{sun2020recent} and selected CI (semistochastic heat-bath configuration interaction (SHCI)) calculations with the DICE\cite{holmes2016heat,sharma2017semistochastic} program package via the pyscf plugin. We used a loose screening parameter of 0.001 (sweep $\epsilon$) for computational efficiency and in order to afford large active spaces. We used the 6-31G\cite{hehre1972a,rassolov1998a} basis for the ``stretched water'' dimer and a mixed basis (C: 6-31G; H: STO-3G\cite{hehre1969self,pietro1983molecular}) for the benzene p-benzyne dimer. We also employed a mixed basis for the \ce{[Mn(NH3)_3(CN)_2NO]^0}$\cdots$X (X~= HF,\ce{H2O},\ce{NH3} and \ce{CH4}), where the crucial moieties were described with the 6-31G basis (Mn, NO and X) and the ligand framework (CN and \ce{NH3}) STO-3G was used.
The mixed basis was chosen due to the memory constraints of the GPU for the subsequent ERPA calculations. This is because our current implementation is limited in system size due to a non-optimal treatment of core/active/virtual simpliciation and a lack of density fitting of the response functions. 
Note that all VQE calculations were performed on ideal statevector simulators, which were carried out on an GPU accelerated in-house QC Ware package (quasar/vulcan). In addition, double factorization was used for evaluating the total energy.\cite{motta2021low,huggins2021efficient} All SAPT and VQE calculations were performed on the newest NVIDIA A100 GPUs provided by Amazon Web Services.

The geometries of the water dimer and benzene were taken from the SI of our previous study.\cite{malone2022towards}. The \ce{[Mn(NH3)_3(CN)_2NO]^0} monomer was optimized without constraints and the geometry was frozen for subsequent geometry optimization of the hydrogen bonded dimer complexes. This ensures that the active space orbitals remain as similar as possible in each complex.

\begin{figure*}[htbp]
    \centering
    \includegraphics[width=0.6\textwidth]{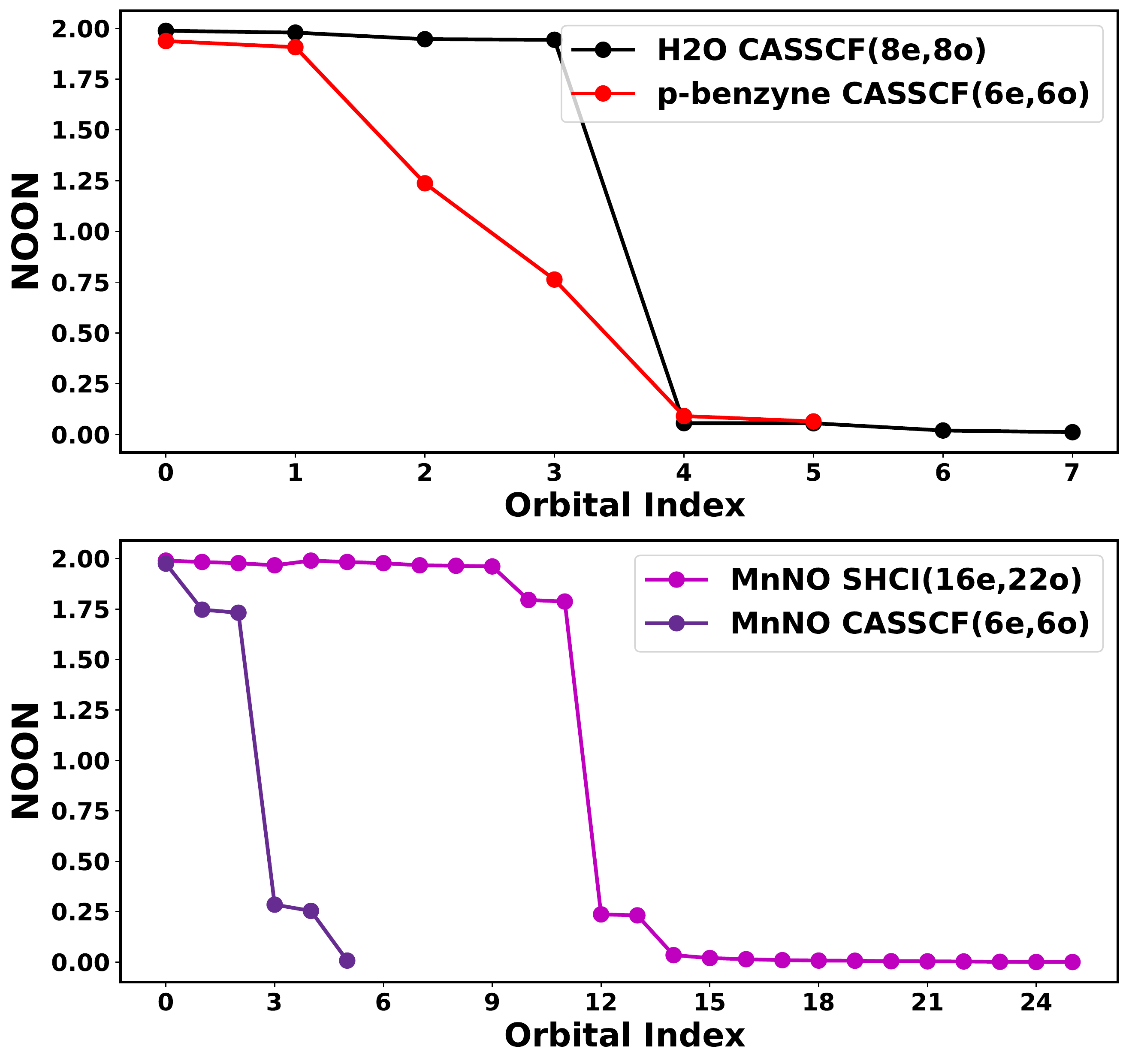}
    \caption{Natural occupation number of the SHCI/CASSCF natural orbitals for the strongly correlated monomer of each test case: upper panel ``stretched'' \ce{H2O} and p-benzyne; lower panel: \ce{[Mn(CN)2(NH3)3NO]^0}. }
    \label{fig:NOON}
\end{figure*}

The active spaces for the ``stretched'' water dimer and benzene p-benzyne dimer, were selected based on MP2 natural orbitals followed by a CASSCF calculation with (8e, 8o) and (6e, 6o), respectively.  The active space of \ce{[Mn(NH3)_3(CN)_2NO]^0} was selected via the automated construction of molecular active spaces from atomic valence orbitals (AVAS)\cite{sayfutyarova2017automated}, where we included all Mn 3-d and the p-orbitals of the NO ligand. Additionally, we systematically added orbitals to the active space based on SHCI resulting in a (16e, 22o) active space. The natural occupation numbers are plotted in \cref{fig:NOON} and show several orbitals  with strong deviation from integer values confirming the multi-reference character of the system. The subsequent CASSCF calculation used the SHCI natural orbitals as a starting guess and included (6e, 6o) in the active space. The corresponding natural orbitals are depicted in \cref{fig:MnNO_NOON}.  The strongest deviation are seen in \cref{fig:MnNO_NOON}~(c)--(f); the orbitals show both bonding and antibonding $\pi$ type orbitals of Mn-\ce{d_{xz}} and Mn-\ce{d_{yz}} with two NO $\pi^*$ orbitals. These orbitals correspond to  metal-to-ligand $\pi$ backbonding in the Dewar–Chatt–Duncanson picture\cite{chatt1953586,chatt1955directing}. The strong static correlation of these four electrons indicates that the electronic structure of this complex is a superposition of  the two configurations [Mn(II)-NO$^\bullet$] and [Mn(II)-NO$^+$]. The corresponding NOONs are depicted in \cref{fig:NOON} and  the values only slightly change in comparison to the  (approximate) larger  active space SHCI results.

In order to converge the $k$-mUCJ ansatz with larger repetition factors a read-in algorithm inspired by Ref.~\citenum{huggins2020non} was implemented.
For a given circuit repetition number $N$, the optimal VQE parameters from the $N-1$\ce{^{th}} step are used as a starting guess and the remaining extra parameters are populated with random parameters from a  normal Gaussian distribution ($\mathcal{N}$) with mean $\mu=0.0$ and variance $\sigma^2=0.001$ ($\mathcal{N}$(0.0, 0.001)). We increased the repetition factor in the following order: $N = 1,2,4,\dots ,m$. All calculations were performed in the dimer-centered basis (ghost atoms on the other monomer) and a gradient threshold of $1\times10^{-6}$ for the L-BFGS-B solver in scipy or with a maximum iteration of 1500 whichever occurred first. In cases when the maximum iteration was reached first we found the norm of the gradient to be $\approx 1\times10^{-5}$.

For the subsequent SAPT calculations we constructed the one- and two-particle reduced density matrices in the full set of natural orbitals. In the case of RHF these correspond to the canonical Hartree--Fock orbitals. For SAPT(CASSCF) and SAPT(VQE) we used the CASSCF natural orbitals. These density matrices were then used to construct the ERPA and SAPT equations given in \cref{eq:erpa} and \cref{sec:numsapt}. We used a threshold of 1 $\times 10^{-5}$ for the canonical orthogonalization step when solving the ERPA generalized (symmetric) eigenvalue problem. The ERPA and SAPT expressions given in in \cref{eq:erpa} and \cref{sec:numsapt} were implemented in an in-house python code. NumPy\citep{van2011numpy,harris2020array} was used for tensor manipulation and linear algebra operations with certain steps accelerated using CuPy\citep{nishino2017cupy}.
This prototyping code does not exploit core-active-virtual partitioning and is thus limited to systems containing roughly 130 orbitals.

\clearpage

\section{Additional Information for the Results Section}
\begin{sidewaystable}[htbp]
    \centering
    \caption{Detailed SAPT energy contributions for all system studied in this manuscript from Fig. 2, 3 and 5 from the main manuscript (energies in kcal/mol, intermolecular distance $r$ in \AA )}
    \label{tab:saptdata}
    \begin{ruledtabular}
    \begin{tabular}{ll|rrrrrrrrr}
      System &Method  & $E^{(1)}_{\mathrm{elst}}$ &  $E^{(1)}_{\mathrm{exch}}$ & $E^{(2)}_{\mathrm{disp}}$ & $E^{(2)}_\mathrm{{exch-disp}}$ &  $E^{(2)}_{\mathrm{ind,u}}$ & $E^{(2)}_{\mathrm{exch-ind,u}}$ & $E_{\mathrm{int}}$ & $r$\\
    \colrule
    \ce{H2O\bond{...}H2O}  & SAPT(CAS-CI) & -11.41 & 7.89 & -0.88 & 0.20 & -3.57 & 2.15 & -5.63 & 2.0 \\
    \ce{H2O\bond{...}H2O}   & SAPT(VQE) ($k$=1) & -10.89 & 7.75 & -0.90 & 0.19 & -3.33 & 1.93 & -5.25 & 2.0 \\
    \ce{H2O\bond{...}H2O}   & SAPT(VQE) ($k$=1) & -11.40 & 7.93 & -0.88 & 0.20 & -3.53 & 2.11 & -5.58 & 2.0 \\
    \rule{0pt}{1ex}    \\
    \ce{p-Bz\bond{...}Bz} & SAPT(CAS-CI) & 0.13 & 0.26 & -0.61 & 0.03 & -0.10 & 0.05 & -0.23 & 3.9 \\
    \ce{p-Bz\bond{...}Bz} & SAPT(VQE) ($k$=1) & 0.08 & 0.26 & -0.61 & 0.03 & -0.10 & 0.05 & -0.29 & 3.9 \\
    \ce{p-Bz\bond{...}Bz} & SAPT(VQE) ($k$=1) & 0.12 & 0.26 & -0.61 & 0.03 & -0.10 & 0.05 & -0.25 & 3.9\\
    \rule{0pt}{1ex}    \\
    \mnhf & SAPT(CAS-CI) & -11.88 & 5.6 & -1.09 & 0.13 & -3.21 & 1.12 & -9.33 & 1.8 \\
    \mnhf & SAPT(VQE) ($k$=1) & -12.19 & 5.58 & -1.08 & 0.13 & -3.23 & 1.12 & -9.67 & 1.8 \\
    \mnhf & SAPT(VQE) ($k$=1) & -11.87 & 5.60 & -1.09 & 0.13 & -3.21 & 1.12 & -9.31 & 1.8\\
    \rule{0pt}{1ex}    \\
    \mnoh & SAPT(CAS-CI) & -7.59 & 3.62 & -0.86 & 0.14 & -1.60 & 0.74 & -5.55 & 2.1 \\
    \mnoh & SAPT(VQE) ($k$=1) & -7.78 & 3.60 & -0.85 & 0.14 & -1.61 & 0.74 & -5.76 & 2.1 \\
    \mnoh & SAPT(VQE) ($k$=1) & -7.6 & 3.62 & -0.85 & 0.14 & -1.6 & 0.74 & -5.56 & 2.1 \\
    \rule{0pt}{1ex}    \\
    \mnnh & SAPT(CAS-CI) & -2.77 & 1.31 & -0.62 & 0.06 & -0.56 & 0.20 & -2.37 & 2.3 \\
    \mnnh & SAPT(VQE) ($k$=1) & -2.83 & 1.31 & -0.61 & 0.06 & -0.57 & 0.19 & -2.45 & 2.3 \\
    \mnnh & SAPT(VQE) ($k$=1) & -2.77 & 1.31 & -0.62 & 0.06 & -0.56 & 0.20 & -2.37 & 2.3 \\
    \rule{0pt}{1ex}    \\
    \mnch & SAPT(CAS-CI) & -0.10 & 0.02 & -0.13 & 0.00 & -0.07 & 0.00 & -0.27 & 3.5 \\
    \mnch & SAPT(VQE) ($k$=1) & -0.10 & 0.02 & -0.13 & 0.00 & -0.07 & 0.00 & -0.28 & 3.5\\
    \mnch & SAPT(VQE) ($k$=1) & -0.10 & 0.02 & -0.13 & 0.00 & -0.07 & 0.00 & -0.27& 3.5 \\

        \end{tabular}
    \end{ruledtabular}

\end{sidewaystable}
\cleardoublepage
\subsection{Detailed SAPT Analysis of the MnNO Complexes}

The detailed analysis of the  energy term decomposition helps to unravel the origin of the interaction. Fig.~5~(b) plots each component of the SAPT(VQE) (k~=~4) calculation of the series of hydrogen bonded complexes plus the ``stretched'' water dimer as a reference of a typical hydrogen bond. We see that the electrostatic term dominates the interaction as expected for hydrogen bonds. The \mnhf\ complex has the largest electrostatic contribution ($-$11.9~kcal/mol) which is stronger than a water hydrogen bond ($-$11.4~kcal/mol). The electrostatic interaction of \mnoh\ is significantly weaker despite the fact that the dipole moments of both \ce{H-F} and \ce{H2O} are around 1.8~D. When comparing the exchange terms, we see that the terms rapidly decrease from \mnhf\ to \mnch\ which can be rationalized by the change in bond distances and  by the exponential decay of this term. Interestingly, the water dimer and \mnhf\  have similar bond distances but \mnhf\ exhibits a significantly less repulsive term. This difference is the main driving force for the difference in interaction energies and can be rationalized by the diffuseness of the lone pairs. The bound NO becomes (partly) \ce{NO^+}, which makes the lone pair  more compact in space than the lone pair in the water dimer; thus, resulting in less exchange repulsion.

\subsection{Additional Comments on the Comparison Between DFT and SAPT(VQE)}
We compare the SAPT(VQE) interaction energies to DFT based supermolecular (BSSE corrected\cite{boys1970calculation}) interaction energies. We note that exact comparisons are difficult for two reasons: first, the static correlation makes it difficult to generate reliable reference energies as the ``gold standard'' CCSD(T) for non-covalent interactions is not reliable anymore\cite{bulik2015can}; second, SAPT0 produces most accurate energies with the jun-cc-pVDZ basis set\cite{parker2014levels} which cannot be employed due to technical limits in the current implementation. However,  nitrosyl complexes are an example of non-universality problem of approximate density functionals as the hydrogen bonding moiety and the nitrosyl moiety prefer different approximate density functionals\cite{radon2008binding,radon2010electronic,mardirossian2014omegab97x} and thus reliable prediction are only possible with careful system specific benchmarking when experimental data is available.\cite{lehnert2013structure} In contrast, SAPT is expected to give accurate results for hydrogen bonds given that proper monomer wavefunctions are used. 
The calculation of accurate interaction energies poses a challenge for DFT as different types of approximate density functionals are recommended for the different molecular moieties: non-hybrid functionals are recommended for an accurate description of the metal-NO moiety\cite{radon2008binding,radon2010electronic}, while for hydrogen bonds  range-separated hybrids are recommended\cite{mardirossian2017thirty}. These two competing requirements make these systems very sensitive to the choice of specific approximate exchange correlation functional illustrating the non-universality problem of approximate density functionals. When using DFT, careful benchmark against experimental data is necessary.\cite{lehnert2013structure}  Our findings indicate that the $k$-uCJ VQE ansatz is able to accurately describe the difficult electronic structure of the nitrosyl monomer in the hydrogen bonding complex (see~\cref{fig:pdm_error})
To illustrate this point, Fig.~6 (main text) plots the interaction energies of  SAPT(VQE), SAPT(CAS-CI) and several popular DFT functionals (supramolecular). We included many popular functionals as well as several top performing functionals for non-covalent interactions\cite{mardirossian2017thirty}. We see in Fig.~6 (main text) that the SAPT(VQE) (k~=~4) is almost identical to the SAPT(CAS-CI) in all four cases. We also note that the DFT functionals exhibit a significant spread for each complex. The B97-D functional predicted the  smallest binding energy in all four cases, but the highest interaction energy is predicted by a different functional. Furthermore, we see the relative ordering of the functionals change for each system (color sequence in each plot). This illustrates the non-universality problem for approximate exchange correlation functionals even for very similar nitrosyl complexes (this also holds true for larger basis set as illustrated in \cref{fig:DFT_large_basis}). We note that the SAPT(CAS-CI) results are the reference for the SAPT(VQE) calculations and do \emph{not} represent the true interaction energy, thus, only the SAPT(VQE)  not the DFT interaction energies should be compared against this reference.
B97-D\cite{grimme2006semiempirical}, BP86\cite{perdew1986accurate,becke1988density}, SCAN\cite{sun2015strongly}, PBE\cite{perdew1996generalized}, M06-L\cite{zhao2006new}, TPSSh\cite{staroverov2003comparative}, B3LYP\cite{becke1988density,Lee1988,Becke1993}, PBE0\cite{adamo1999toward}, MN15\cite{haoyu2016mn15}, PWB6K\cite{zhao2005design}, CAM-B3LYP\cite{yanai2004new}, $\omega$B97X-D\cite{chai2008long}. 
This diverse list includes local DFT functionals, hybrid functionals with a wide range of exact exchange (10\%--46\%) and range separated hybrids. 

\begin{figure*}[htbp]
    \centering
    \begin{subfigure}[b]{0.45\textwidth}
    \includegraphics[width=\textwidth]{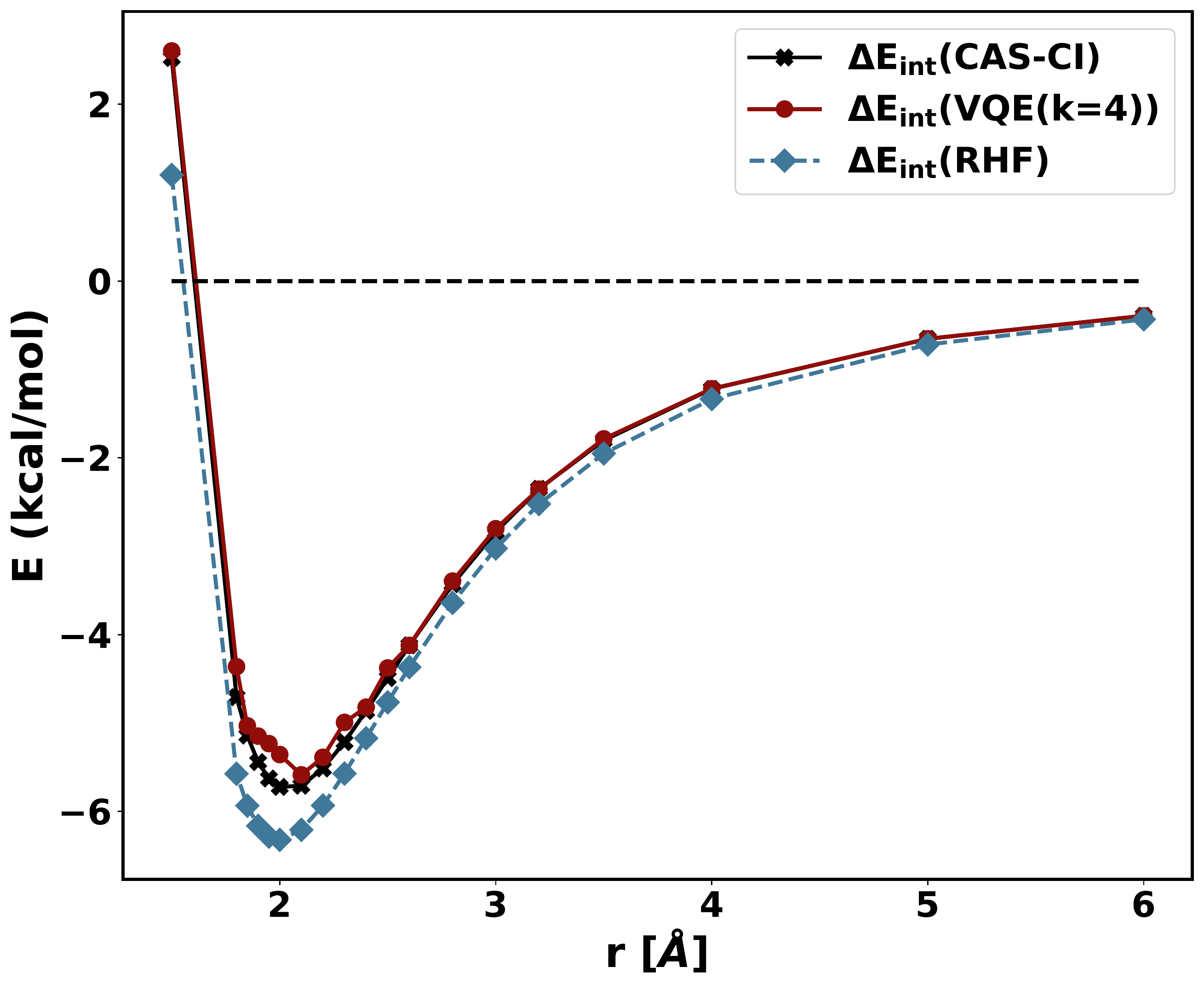}
    \caption{ }
 \end{subfigure}
     \begin{subfigure}[b]{0.45\textwidth}
    \includegraphics[width=\textwidth]{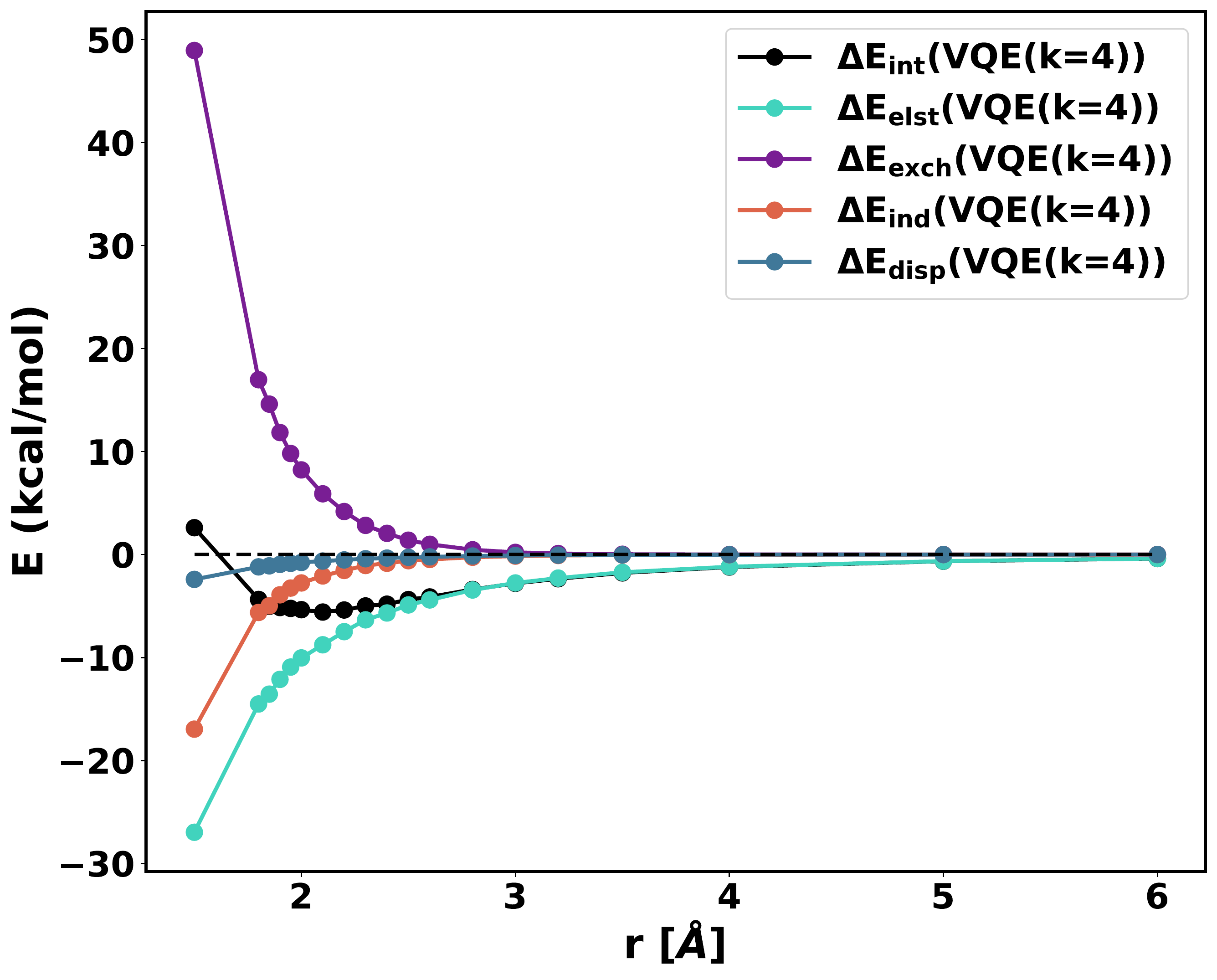}
    \caption{ }
 \end{subfigure}
    \caption{(a) potential energy scan along $r$ using SAPT(CAS-CI), SAPT(RHF) and SAPT(4-uCJ);  (b) distance dependence of each SAPT(4-uCJ) energy term with respect to the intermolecular distance r for the ``stretched'' water dimer.}
    \label{fig:water_results2}
\end{figure*} 
\begin{figure*}[htbp]
\centering
\includegraphics[width=0.6\textwidth]{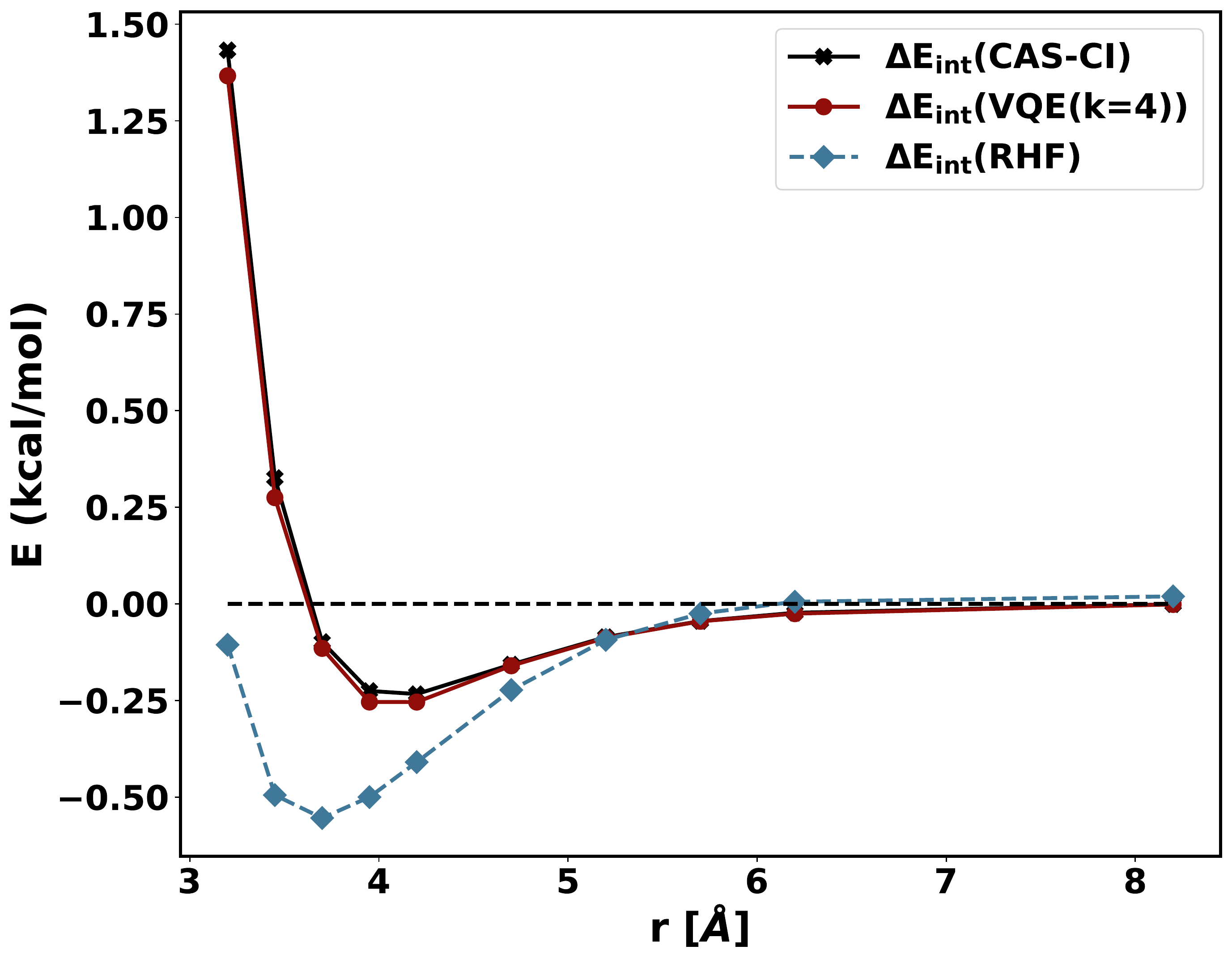}
      \caption{Potential energy scan along $r$ using SAPT(CAS-CI), SAPT(RHF) and SAPT(VQE) ($k$~=~4) for the p-benzyne-benzene dimer.}
    \label{fig:bz_results}
\end{figure*}

\begin{figure*}[htbp]
    \centering
    \begin{subfigure}[b]{0.2\textwidth}
    \includegraphics[width=\textwidth]{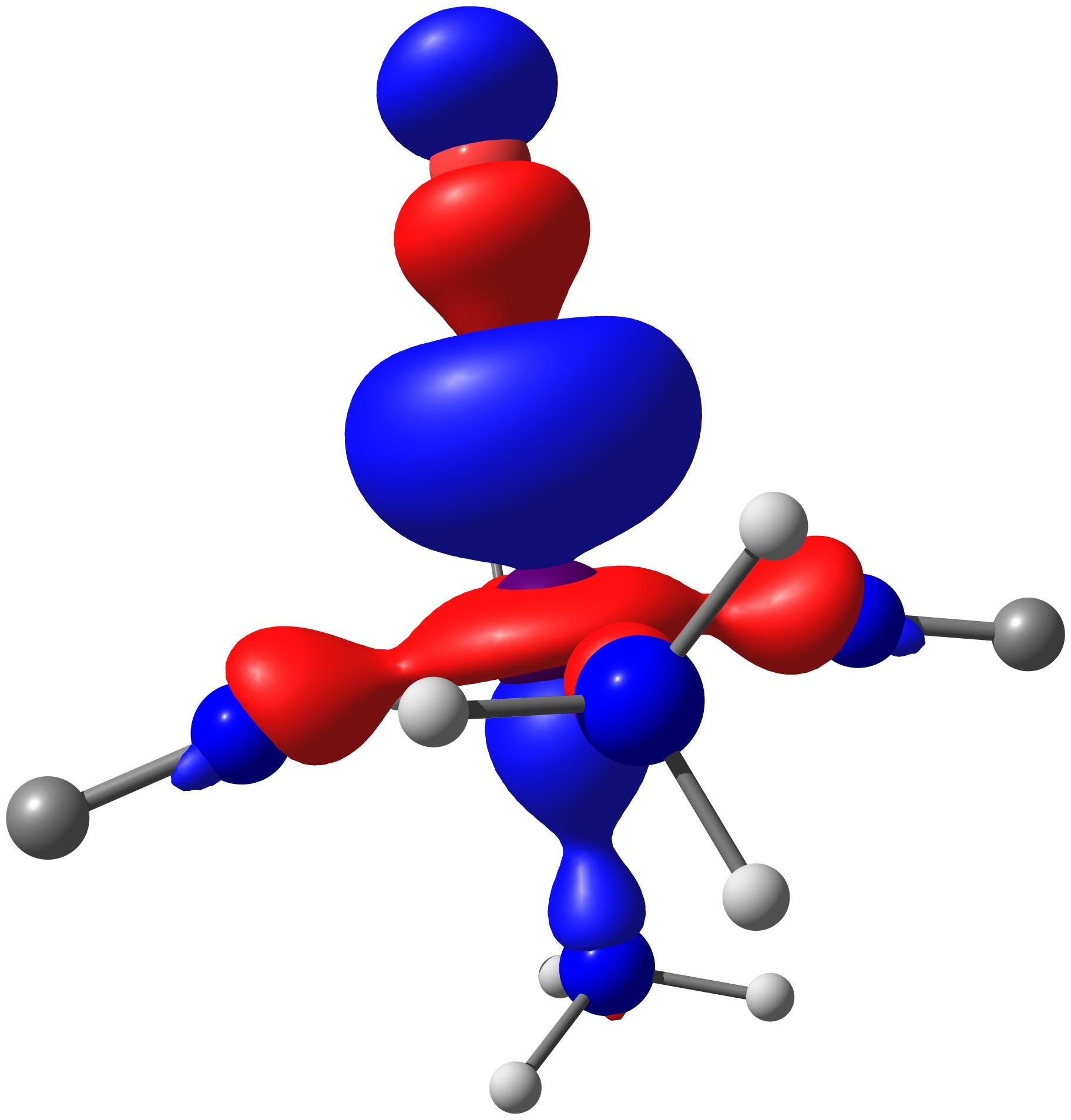}
    \caption{ }
 \end{subfigure}
     \begin{subfigure}[b]{0.2\textwidth}
    \includegraphics[width=\textwidth]{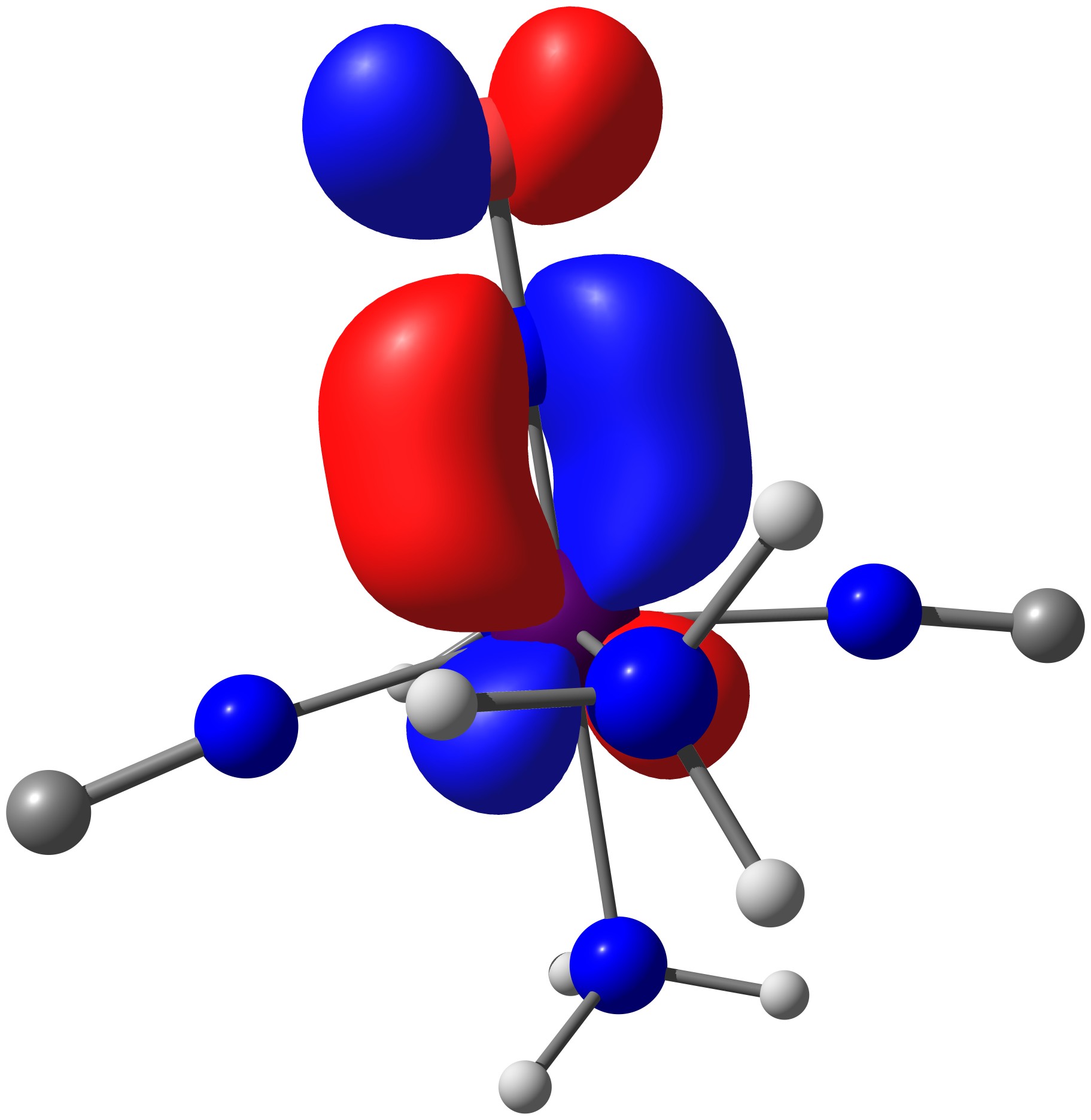}
    \caption{ }
 \end{subfigure}
    \begin{subfigure}[b]{0.2\textwidth}
    \includegraphics[width=\textwidth]{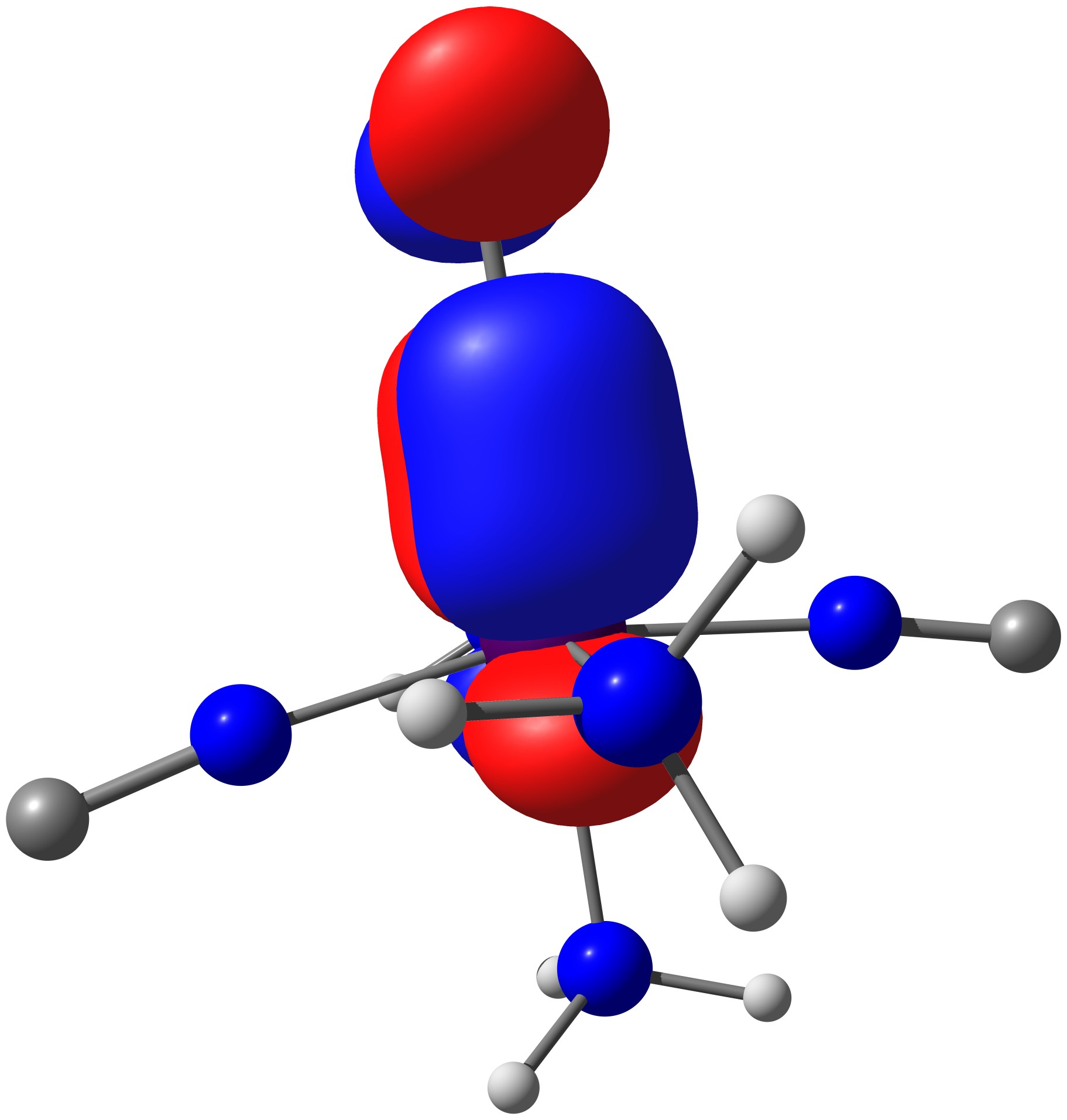}
    \caption{ }
 \end{subfigure}
 
     \begin{subfigure}[b]{0.2\textwidth}
    \includegraphics[width=\textwidth]{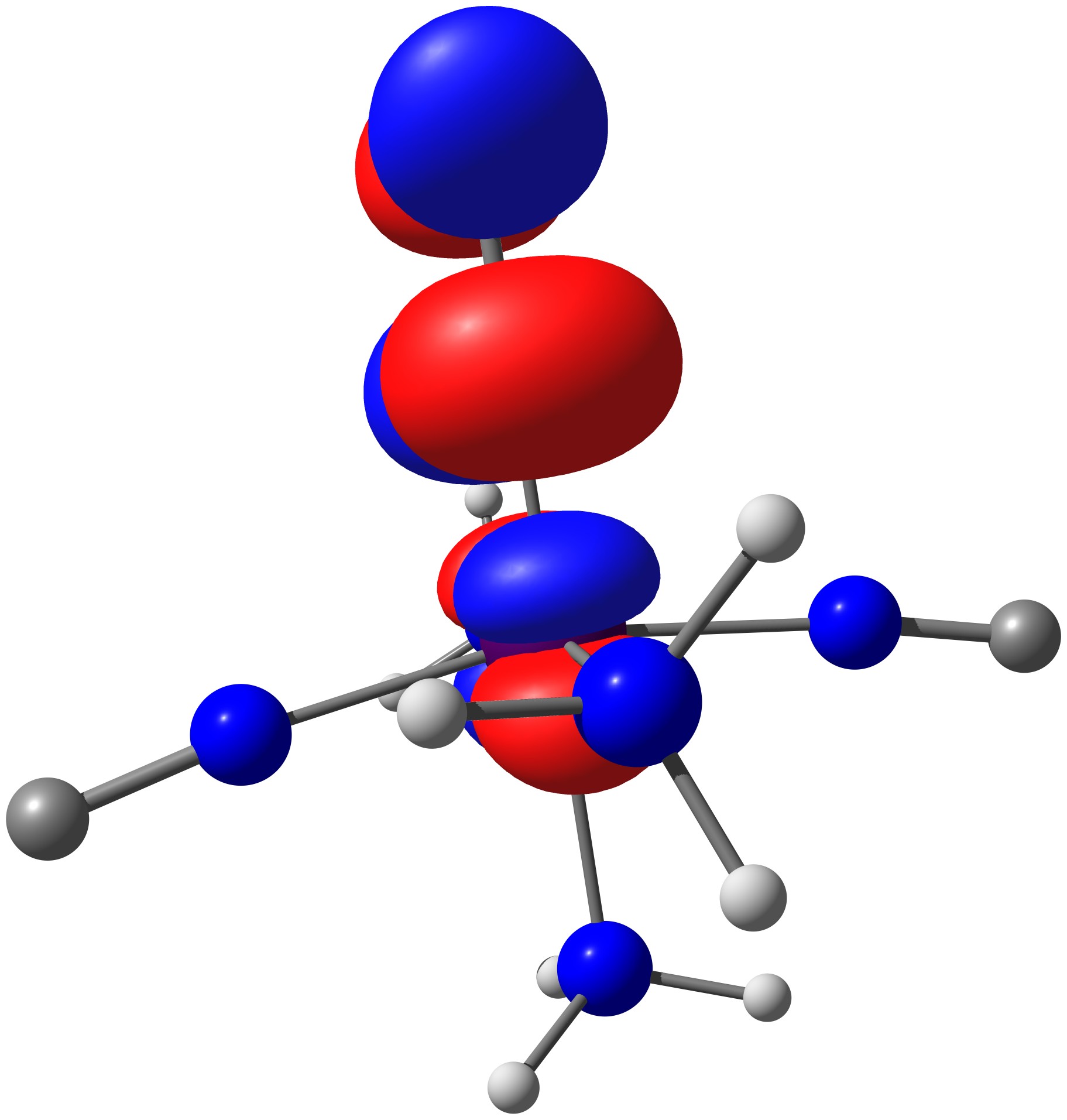}
    \caption{ }
 \end{subfigure}
     \begin{subfigure}[b]{0.2\textwidth}
    \includegraphics[width=\textwidth]{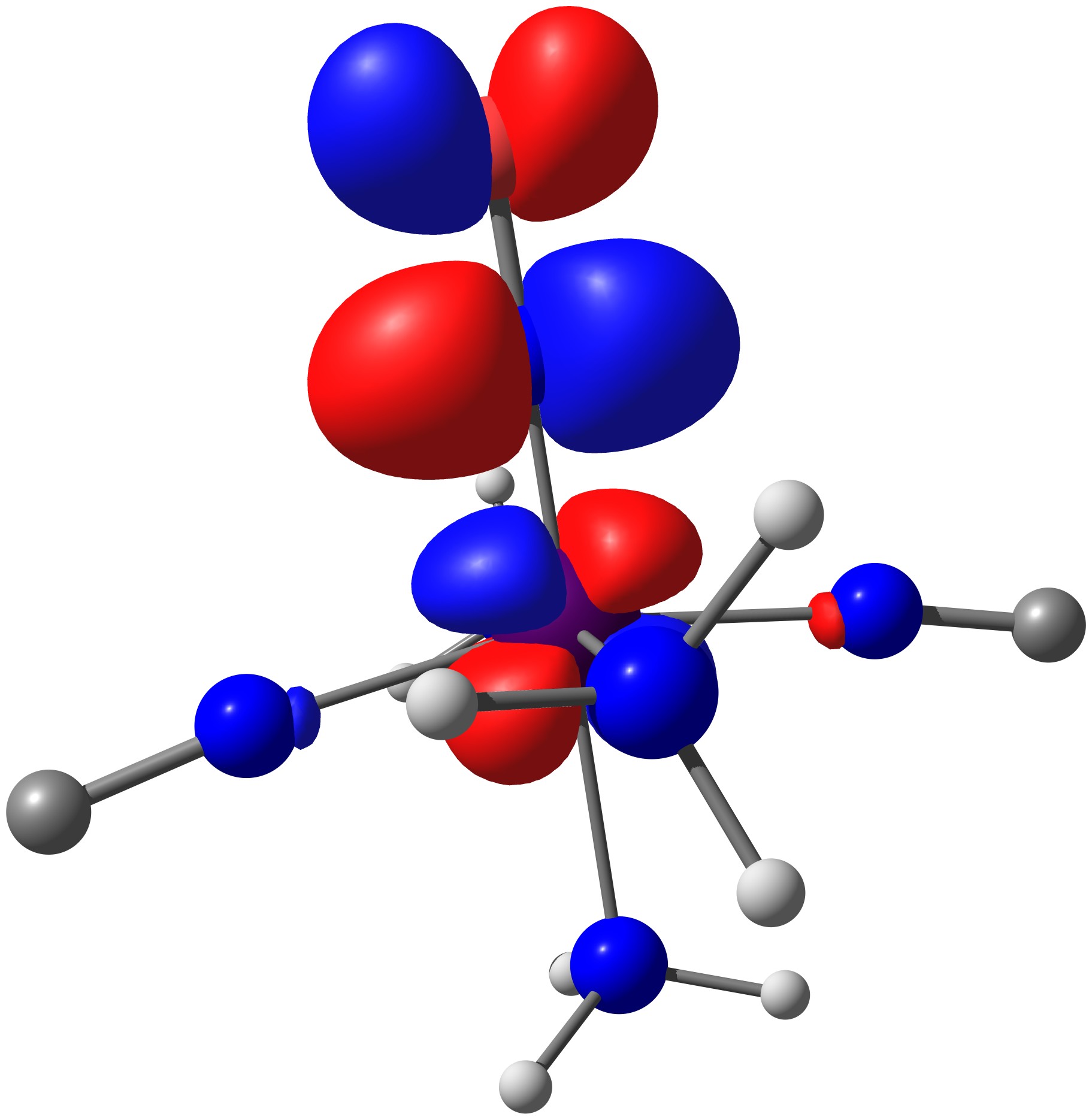}
    \caption{ }
 \end{subfigure}
     \begin{subfigure}[b]{0.2\textwidth}
    \includegraphics[width=\textwidth]{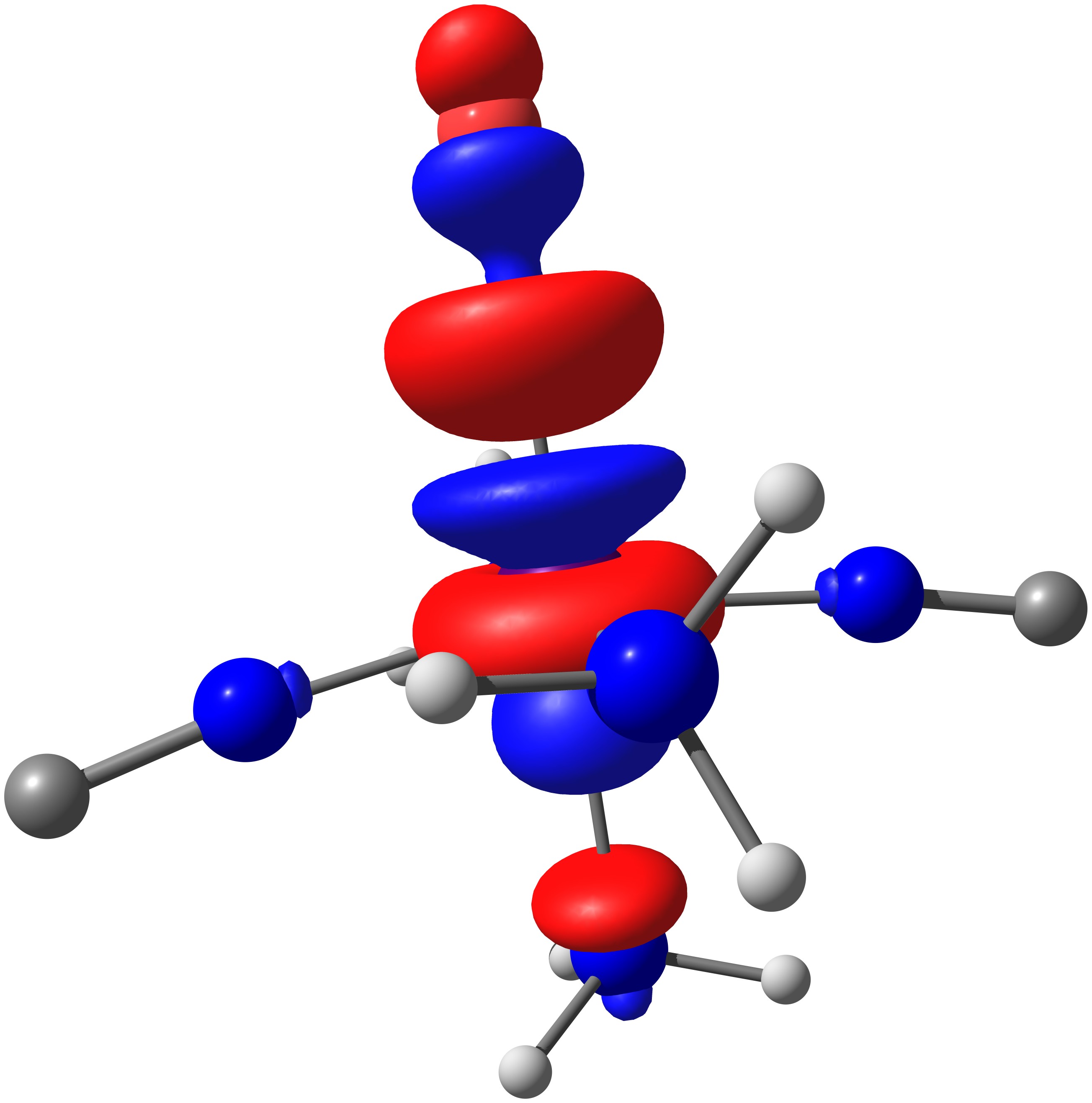}
    \caption{ }
 \end{subfigure}

      \caption{Natural orbitals of the (6e, 6o) CASSCF calculation of \ce{[Mn(CN)2(NH3)3NO]^0}.}
    \label{fig:MnNO_NOON}
\end{figure*} 

\begin{figure*}
    \centering
    \includegraphics[width=0.6\textwidth]{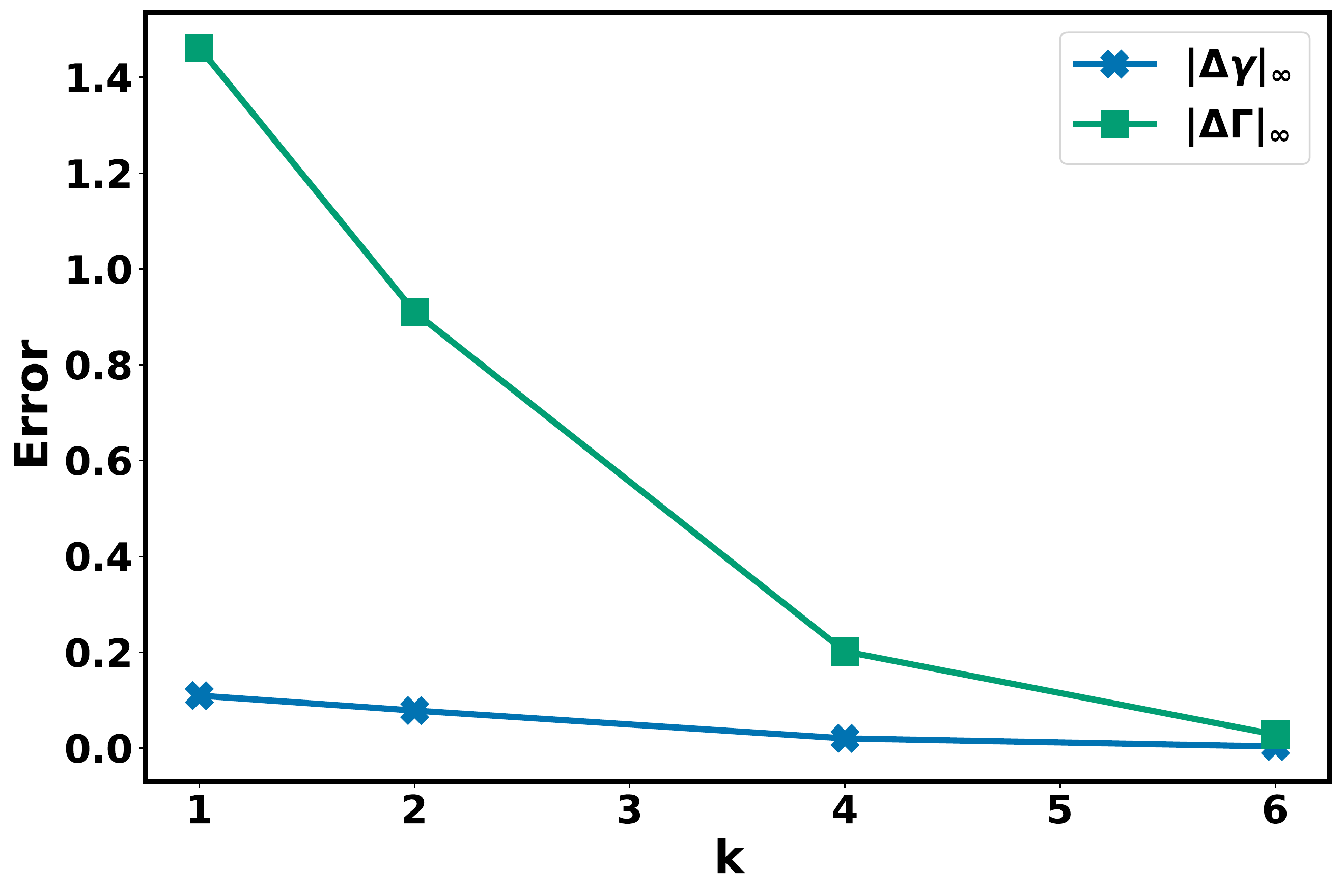}
    \caption{Error in the one and two particle density matrix as a function of the circuit repetition factor $k$ for the \ce{[Mn(CN)2(NH3)3NO]^0} monomer. The errors are defined as $|\Delta\gamma|_\infty=|\gamma_{\mathrm{CASCI}}-\gamma_{k\mathrm{-muCJ}}|_\infty$ and $|\Delta\Gamma|_\infty=|\Gamma_{\mathrm{CASCI}}-\Gamma_{k\mathrm{-muCJ}}|_\infty$. }
    \label{fig:pdm_error}
\end{figure*}

\begin{table}[htb]
    \centering
        \caption{Quantum hardware resource requirements with respect to the repetition factor $k$ (Number of qubits, number of two qubit gates (2-Q-G), number of parameter, quantum circuit depth) for the VQE simulations using the $k$-muCJ ansatz for the simulation of the \ce{[Mn(CN)2(NH3)3NO]^0} with an (6e, 6o) active space.}
    \label{tab:ressouces}
    \begin{ruledtabular}
    \begin{tabular}{l|cccccccc}
    k & 1 & 2 & 3 & 4 & 5 & 6 & 7 & 8 \\ 
    \colrule
    \# Qubits& 12 & 12& 12& 12& 12& 12& 12& 12  \\
    \# 2-Q-G & 750 & 1065 & 1380 & 1695 & 2010 & 2325 & 2640 & 2955 \\
    \# Param. & 45 &  75 & 105 & 135 & 165 & 195 & 225 & 255\\
    Depth & 216 &  378 &  540 &  702 &  864 & 1026 & 1188 & 1350\\
    \end{tabular}
    \end{ruledtabular}
\end{table}

\begin{figure*}
    \centering
    \includegraphics[width=0.75\textwidth]{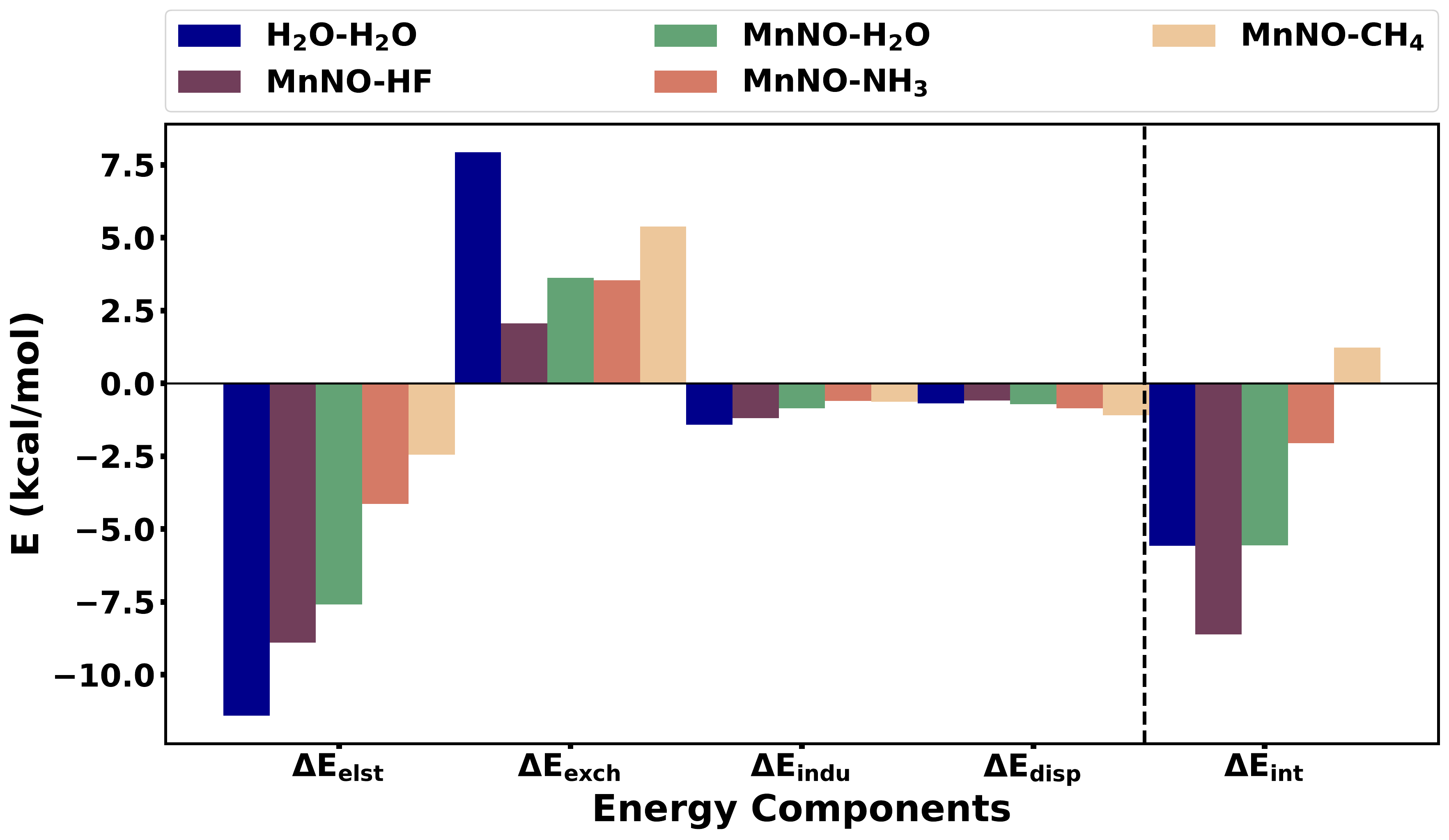}
    \caption{Term-by-term decomposition of the SAPT(4-uCJ) binding energies of each heme-nitrosyls hydrogen complex at a fixed bond distance of r(\ce{O-H})~=~2.08~\AA\  (equilibrium distance of the \ce{H2O} complex).}
    \label{fig:sapt_same_dist}
\end{figure*}

\begin{figure*}
    \centering
    \includegraphics[width=0.75\textwidth]{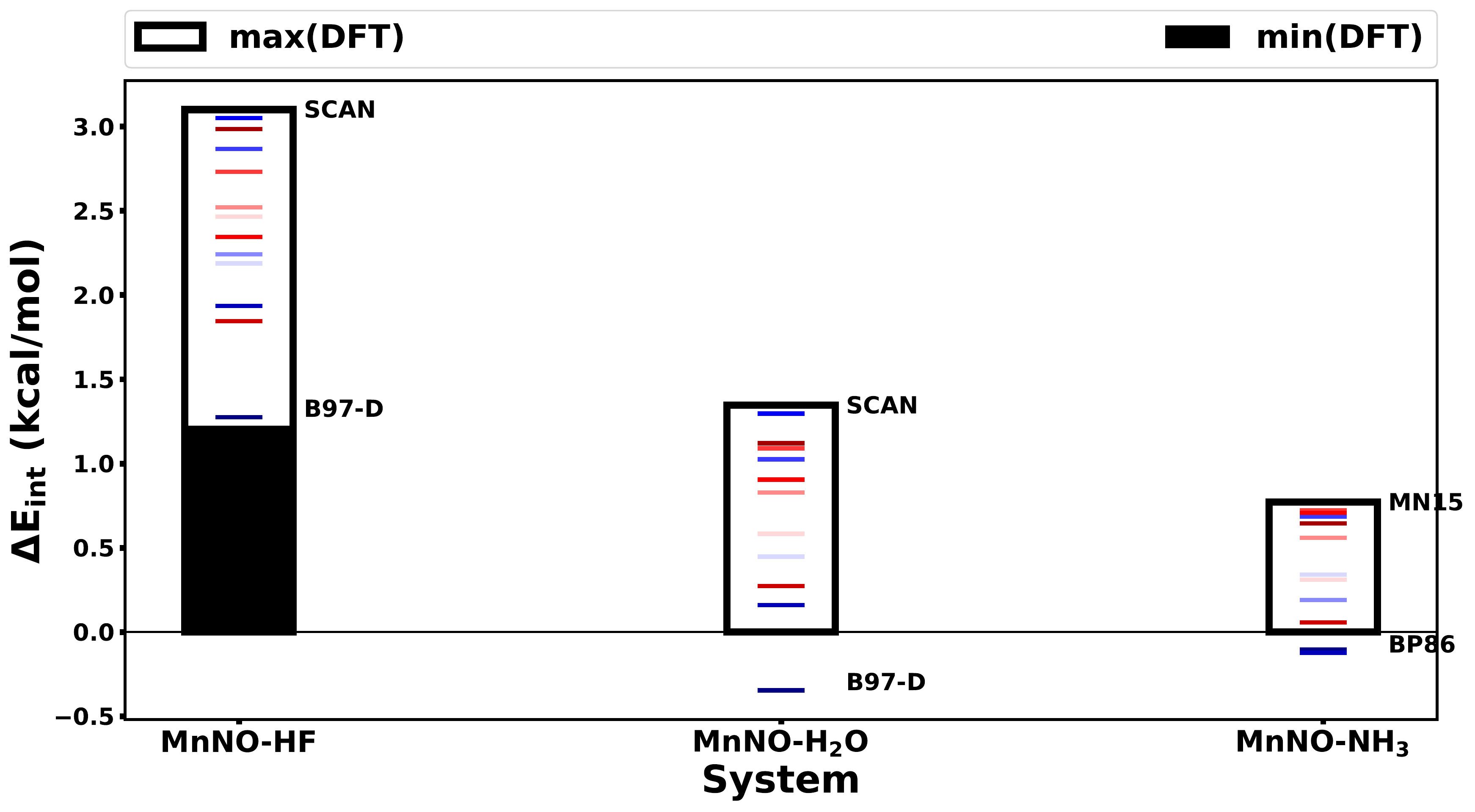}
    \caption{Supramolecuar DFT interaction energies in the def2-TZVPD\cite{weigend2005balanced} basis for the \ce{HF}, \ce{H2O} and \ce{NH3} nitrosyl hydrogen complexes. }
    \label{fig:DFT_large_basis}
\end{figure*}

\begin{figure*}
    \centering
     \begin{subfigure}[b]{0.6\textwidth}
    \includegraphics[width=\textwidth]{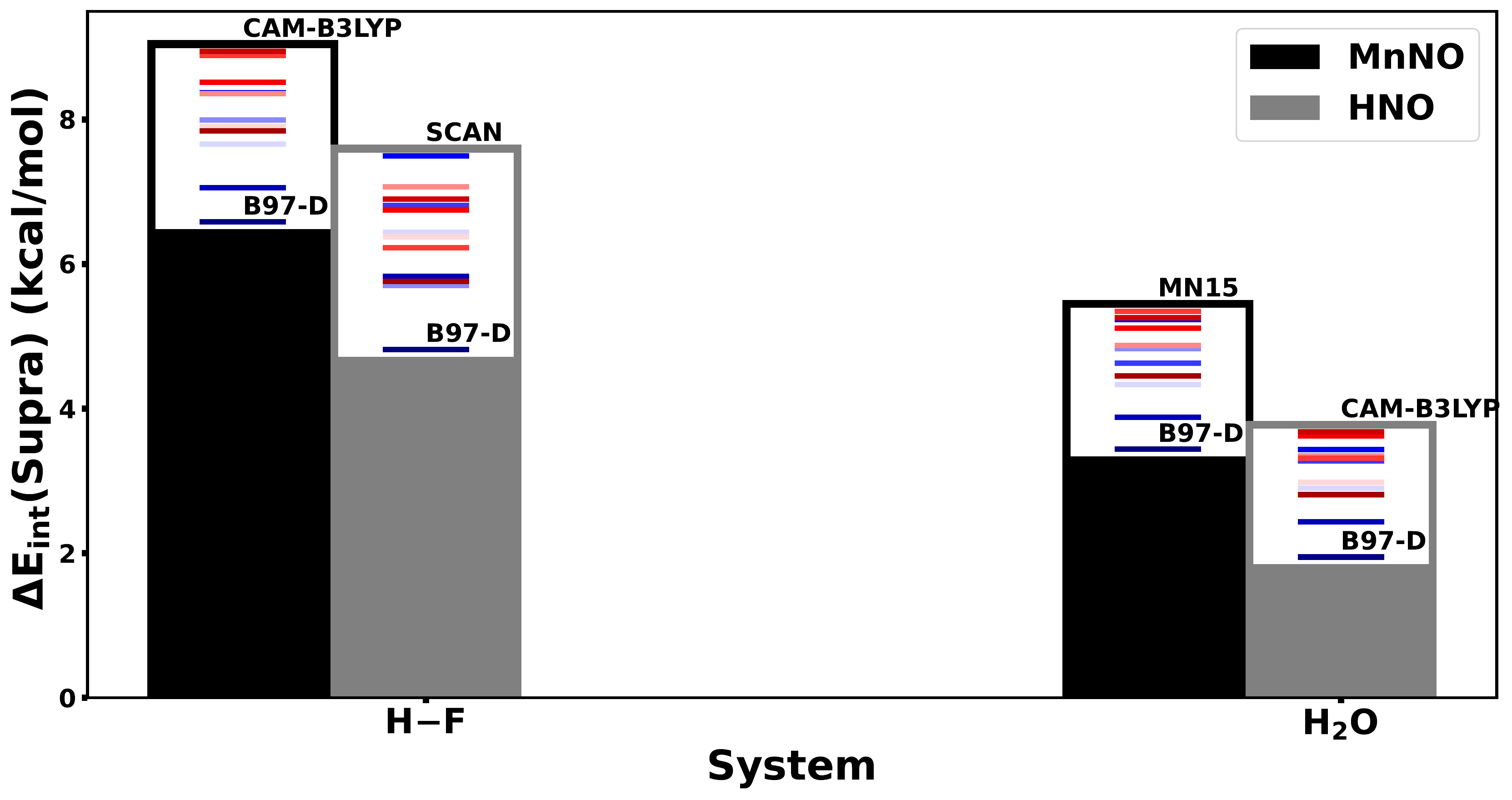}
    \caption{ }
    
 \end{subfigure}
     \begin{subfigure}[b]{0.6\textwidth}
    \includegraphics[width=\textwidth]{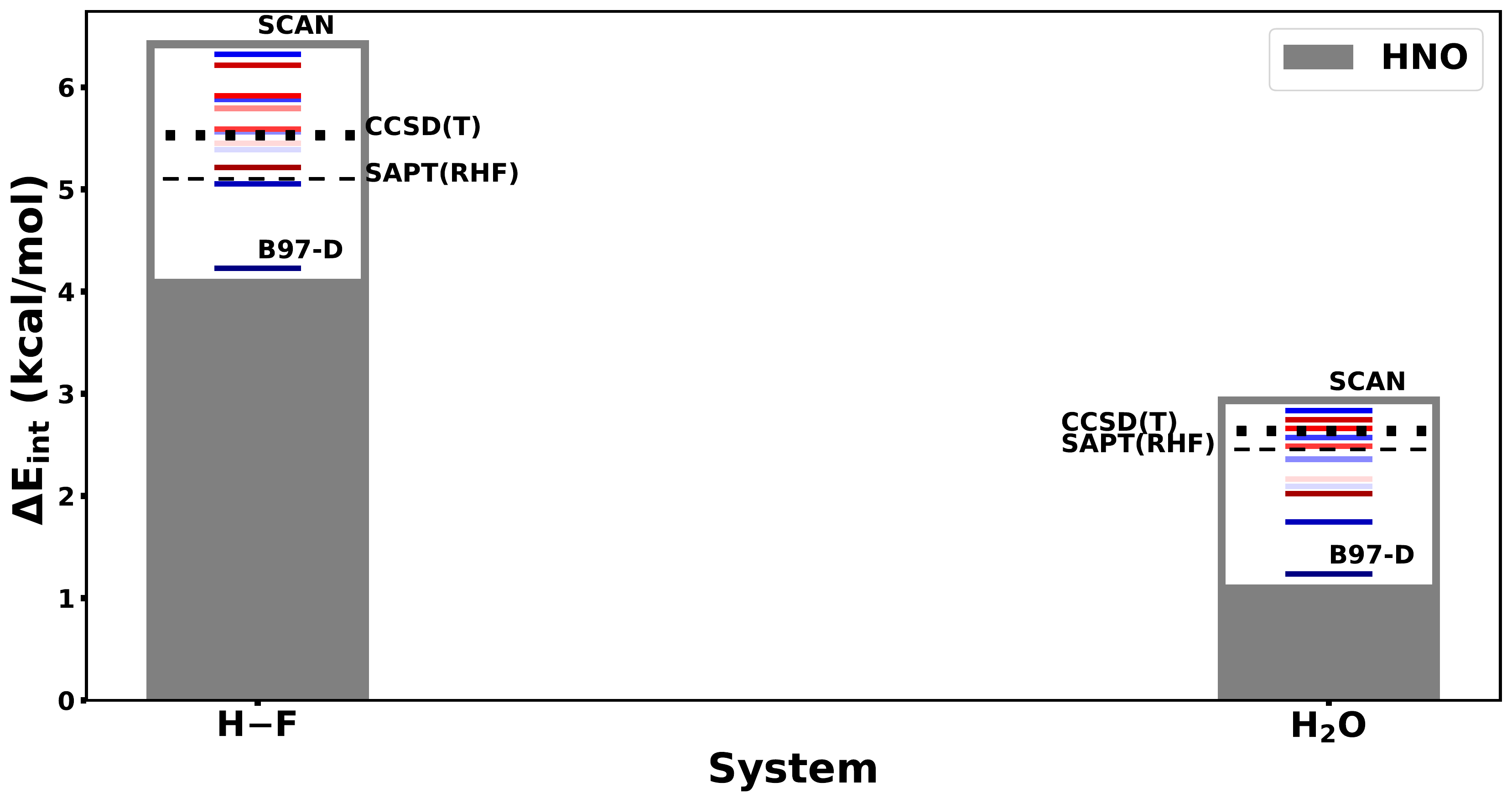}
    \caption{ }
 \end{subfigure}
 
      \begin{subfigure}[b]{0.8\textwidth}
    \includegraphics[width=\textwidth]{Figures/DFT_color_labels.pdf}
    \caption{ }
 \end{subfigure}
 
    \caption{(a) Comparison between supramolecular DFT interaction energies for the nitrosyl and nitric acid hydrogen complexes (small basis); (b) Comparison between supramolecular DFT and SAPT in their most common basis sets using the nitric acid hydrogen complexes. High level CCSD(T) reference interaction energies are also provided for reference (DFT: def2-TZVPD \& BSSE corrected; SAPT: SAPT in the jun-cc-pVDZ basis (skips the VQE calculation and uses the RHF wave function instead) and CCSD(T):  aug-cc-pVQZ \& BSSE corrected; for the DFT functional color coding see (c)).}
    \label{fig:SAPT_DFT_HNO}
\end{figure*}

\begin{sidewaystable}[htbp]
    \centering
    \caption{Detailed interaction energies from Fig.~6 (main text) using the supramolecular approach for DFT and CCSD(T) calculation (BSSE corrected). CCSD(T)/aug-cc-pVQZ can be considered very close to the true \ce{E_{int}} using both a very accurate wave function ansatz (for non strongly correlated systems) and large basis set; DFT calculations using the def2-TZVPD basis set as recommended by practical best practices from Ref.~\citenum{bursch_mewes_hansen_grimme_2022} and SAPT calculation use the jun-cc-pVDZ basis set as recommended by Ref.~\citenum{parker2014levels}.
    }
    \label{tab:dft}
    \begin{ruledtabular}
        \begin{tabular}{ll|rrrrrrrrrrrr}
    System &Basis & B97-D & BP86 & SCAN & PBE & M06-L & TPSSh & B3LYP & PBE0 & MN15 & PWB6K & \footnotesize{CAM-B3LYP} &  \footnotesize{$\omega$B97X-D}\\
    \colrule
    \mnhf & mixed & -6.58 & -7.06 & -8.37 & -7.88 & -7.99 & -7.66 & -7.91 & -8.36 & -8.89 & -8.51 & -8.94 & -7.84 \\
    \mnoh & mixed & -3.44 & -3.88 & -5.23 & -4.63 & -4.83 & -4.33 & -4.45 & -4.87 & -5.35 & -5.11 & -5.26 & -4.45 \\
    \ce{HNO\bond{...}HF}& 6-31G & -4.82 & -5.83 & -7.5 & -6.81 & -5.7 & -6.44 & -6.38 & -7.07 & -6.22 & -6.75 & -6.9 & -5.76 \\
    \ce{H2O\bond{...}H2O}& 6-31G & -1.95 & -2.43 & -3.44 & -3.28 & -2.81 & -2.9 & -2.98 & -3.34 & -3.31 & -3.62 & -3.68 & -2.81 \\
    \rule{0pt}{1ex}    \\
    \ce{HNO\bond{...}HF} & \tiny{def2-TZVPD}&-4.23 & -5.06 & -6.32 & -5.88 & -5.57 & -5.39 & -5.45 & -5.79 & -5.59 & -5.92 & -6.22 & -5.22\\
    \ce{HNO\bond{...}H2O}&\tiny{def2-TZVPD}& -1.24 & -1.75 & -2.84 & -2.57 & -2.36 & -2.09 & -2.17 & -2.49 & -2.49 & -2.66 & -2.75 & -2.02\\
    \end{tabular}
    \vspace*{0.5 cm}
    \begin{tabular}{l|rr}
    System & CCSD(T) (Basis)  & SAPT(RHF)  (Basis)  \\
    \colrule
    \ce{HNO\bond{...}HF} & -5.5 (aug-cc-pVQZ) & -5.1  (jun-cc-pVDZ)\\
    \ce{HNO\bond{...}H2O} & -2.6  (aug-cc-pVQZ) &-2.4 (jun-cc-pVDZ)\\    

    \end{tabular}
    \end{ruledtabular}
    
\end{sidewaystable}
\cleardoublepage
\bibliography{bibliography}
\end{document}